\documentclass[a4paper,11pt]{article}
\pdfoutput=1

\usepackage{jheppub}

\usepackage{amsmath,amssymb,graphicx}
\usepackage{subfigure}
\usepackage{multirow}
\usepackage{dcolumn}
\usepackage{color}
\usepackage{bm}
\usepackage{siunitx}
\usepackage{slashed}
\usepackage{natbib}
\usepackage{feynmp-auto}
\usepackage{hyperref}
\usepackage{makecell}
\usepackage{pbox}
\usepackage{array}
\newcommand{\bea}{\begin{eqnarray}}
\newcommand{\eea}{\end{eqnarray}}
\newcommand{\beq}{\begin{eqnarray}}
\newcommand{\eeq}{\end{eqnarray}}

\newcommand{\ud}{\mathrm{d}}

\newcommand{\mgamc}{Madgraph5\_aMC@NLO}
\newcommand{\whizard}{WHIZARD}

\newcommand{\bra}[1]{\langle#1|}
\newcommand{\ket}[1]{|#1\rangle}

\newcommand{\gwa}[1]{g_{W,a#1}}
\newcommand{\gwb}[1]{g_{W,b#1}}

\newcommand{\gza}[1]{g_{Z,a#1}}
\newcommand{\gzb}[1]{g_{Z,b#1}}

\newcommand{\gab}[1]{g_{A,b#1}}

\newcommand{\gva}[1]{g_{V,a#1}}
\newcommand{\gvb}[1]{g_{V,b#1}}

\newcommand{\gxb}[1]{g_{X,b#1}}

\title{\boldmath Multi-Higgs Production and Unitarity\\ in Vector-Boson Fusion
  at Future Hadron Colliders}

\author[a]{Wolfgang Kilian}
\author[b]{Sichun Sun}
\author[c,d]{Qi-Shu Yan}
\author[e]{Xiaoran Zhao}
\author[a,d]{Zhijie Zhao}

\affiliation[a]{Department of Physics, University of Siegen, 57072 Siegen, Germany}
\affiliation[b]{Department of Physics, National Taiwan University, Taipei, Taiwan}
\affiliation[c]{School of Physics Sciences, University of Chinese Academy of Sciences, Beijing 100039, China}
\affiliation[d]{Center for future high energy physics, Chinese Academy of Sciences, Beijing 100039, China}
\affiliation[e]{Centre for Cosmology, Particle Physics and Phenomenology (CP3), Universit\'{e} catholique de Louvain, Chemin du Cyclotron, 2, B-1348 Louvain-la-Neuve, Belgium}

\emailAdd{kilian@physik.uni-siegen.de}
\emailAdd{sichunssun@gmail.com}
\emailAdd{yanqishu@ucas.ac.cn}
\emailAdd{xiaoran.zhao@uclouvain.be}
\emailAdd{zhao@physik.uni-siegen.de}

\preprint{SI-HEP-2018-27,CP3-18-55,MCnet-18-21}

\abstract{
We study multi-Higgs final states in vector boson fusion (VBF) processes at
the LHC and at future proton-proton colliders, focusing on the prospects for
measurements at 27 TeV and at 100 TeV.  We use an effective Lagrangian which
includes higher-dimensional operators in the mass eigenstates which are
relevant to VBF processes and relate this to specific parameterizations and
models for new physics in the Higgs sector.  We derive theoretical constraints
on the parameter space from the unitarity of $2\to n$ scattering amplitudes
and apply the results to $V V \to hh$ and $h h h$ processes, where $V=W,Z$.
As a result, we 
present constraints on differential
distributions as appropriate to the study of $V V \to hh$ 
and $h h h$ processes.
}

\begin{document}
\maketitle
\flushbottom

\section{Introduction}

After the discovery of the Higgs boson with mass $m_h\approx 125$~GeV
at the LHC~\cite{Aad:2012tfa,Chatrchyan:2012xdj}, detailed
measurements of all of its properties have become central to any
search for new physics beyond the Standard Model~(SM).  
In the SM,
the Higgs boson has three types of interaction at tree level:
(1) the Yukawa interaction with massive fermions;
(2) the interaction with electroweak gauge bosons~($W^\pm$ and $Z$);
(3) the cubic and quartic Higgs self-interactions.
Establishing the last type of interaction is a crucial test of our current understanding of electroweak symmetry breaking (EWSB).
However,
any direct measurement of Higgs self-interactions involves producing two or more Higgs bosons in a single elementary process.

For all possible multi-Higgs production processes, the SM rates are very
small.  Beyond the SM, new effects may significantly enhance the rate,
but any such enhancement is subject to generic relations from
unitarity.  In the current paper, we investigate multi-Higgs
production in the context of vector-boson fusion topologies,
compute production rates with appropriate cuts and selection criteria,
and study the applicable unitarity constraints within an
effective-theory formalism.  The results are intended to supply future
studies of
multi-Higgs production with generic limits of event distributions and rates
that have to be considered in the physical interpretation of the analysis.

\subsection{Higgs pair-production in the SM}
The cubic Higgs self-coupling, $hhh$, contributes to processes that
involve at least two Higgs bosons in the final state.
At the LHC, the dominant process of Higgs pair-production in the SM is
gluon-gluon fusion (ggF) via a heavy top-quark loop.
Current LHC data constrain the triple Higgs self-coupling only very weakly~\cite{CMS:2017ihs}.
Multiple groups have evaluated the potential for a first meaningful measurement of the triple Higgs self-coupling at future high-luminosity runs of the LHC~\cite{Plehn:1996wb,Baur:2002rb,Li:2013flc,Bhattacherjee:2014bca,Cao:2015oaa,Cao:2016zob,Grober:2017gut}.
The considered decay channels of the Higgs pair include 
$W^+W^-W^+W^-$~\cite{Baur:2002qd,Ren:2017jbg},
$b\bar{b}\gamma\gamma$~\cite{Baur:2003gp,Yao:2013ika,Kling:2016lay,Chang:2018uwu,Kim:2018uty,He:2015spf},
$b\bar{b}W^+W^-$~\cite{Papaefstathiou:2012qe},
$b\bar{b}\tau^+\tau^-$~\cite{Baur:2003gpa,Dolan:2012rv,Barr:2013tda},
$b\bar{b}\mu^+\mu^-$~\cite{Baur:2003gp},
$W^+W^-\gamma\gamma$~\cite{Lu:2015qqa},
and $b\bar{b}b\bar{b}$~\cite{Baur:2003gpa,deLima:2014dta,Behr:2015oqq}.
It is expected that the triple Higgs self-coupling can be constrained within $40\%$ accuracy after collecting 3 ab$^{-1}$ of data at the 14 TeV LHC~\cite{Barger:2013jfa}.
Beyond the LHC,
at a future 100 TeV hadron collider,
the Higgs pair-production rate is enhanced significantly~\cite{Cao:2016zob,Kling:2016lay,Barr:2014sga,Papaefstathiou:2015iba,Li:2015yia,Zhao:2016tai,Contino:2016spe,Goncalves:2018yva},
allowing for a more accurate determination of the Higgs potential.

Vector-boson fusion (VBF), $VV\to hh$, is a subdominant process of
Higgs pair production in hadron collisions~\cite{Jones:1979bq}.  The
vector bosons $V=W^{\pm},Z$ are effectively radiated from incoming
quarks.  In addition to its dependence on the Higgs self-interaction,
this process also depends on the $hVV$ and $hhVV$ couplings.  The
single-Higgs couplings $hVV$ can be determined from the precise measurement of the
decay branching fractions $h\to WW^*$ and $h\to ZZ^*$ at the LHC, up
to a common normalization factor.  Current LHC data on the decay
branching fractions for these channels are consistent with the SM
predictions~\cite{ATLAS:2017cju,CMS:2017pzi}.  The Higgs-pair interaction
$hhVV$ has not been accessible otherwise.  In principle, the
$VV\to hh$ process allows for a simultaneous extraction of this
class of couplings and of the Higgs self-interaction $hhh$.

The VBF mode of Higgs pair-production at hadron colliders has been studied in Refs.~\cite{Dolan:2013rja,Liu-Sheng:2014gxa,Dolan:2015zja,Bishara:2016kjn,Arganda:2018ftn}.
Beyond tree level, the NLO QCD correction enhances the cross section by $\sim 7\%$~\cite{Baglio:2012np,Frederix:2014hta}.
In the high-luminosity mode of the LHC (HL-LHC) with $3$ ab$^{-1}$ at 14 TeV,
the $hhVV$ interaction can be constrained to $20\%$ \cite{Dolan:2015zja}.
A 100 TeV hadron collider has the potential to reduce the uncertainty down to $1\%$~\cite{Bishara:2016kjn}.
The $hhWW$ coupling is also accessible in the $W^{\pm} W^{\pm} h$ final state.
In Ref.~\cite{Englert:2017gdy} it is found that this particular final state can constrain this coupling to $O(100\%)$ at the HL-LHC,
and to $20\%$ at a 100 TeV collider.

There are further modes of Higgs pair-production in hadron collisions,
namely $t\bar{t}hh$ or $Vhh$ production.
The corresponding production rates are substantially smaller than in the ggF
and VBF modes~\cite{Frederix:2014hta,Baglio:2012np,Moretti:2004wa,Englert:2014uqa,Liu:2014rva,Cao:2015oxx,Barger:1988jk,Li:2016nrr,Nordstrom:2018ceg}.

\subsection{Triple Higgs production in the SM}
The quartic Higgs coupling $hhhh$ is even more elusive, and
experimentally establishing this interaction in the SM is a challenging
task.  Direct access requires processes with three Higgs bosons in the
final state.  In the dominant production channel at the LHC, $gg\to
hhh$, the total cross section at 14 TeV is only
${O}(0.01)$~fb~\cite{Plehn:2005nk,Binoth:2006ym}.
As an alternative,
the authors of Ref.~\cite{Dicus:2016rpf} have considered triple
Higgs-strahlung, $pp\to Zhhh$, but the cross section is also tiny.
At a 100 TeV hadron collider,
triple-Higgs production via ggF can become observable in principle~\cite{Maltoni:2014eza,Papaefstathiou:2015paa,Chen:2015gva,Fuks:2015hna,Kilian:2017nio,Agrawal:2017cbs,Fuks:2017zkg}.
The cross section of $gg\to hhh$ at a $100$ TeV hadron collider is estimated to $5$ fb if NLO QCD corrections are included~\cite{Maltoni:2014eza}.
Various decay channels have been investigated in some detail,
such as $hhh\to b\bar{b}b\bar{b}\gamma\gamma$~\cite{Papaefstathiou:2015paa,Chen:2015gva},
$hhh\to b\bar{b}b\bar{b}\tau\tau$~\cite{Fuks:2015hna,Fuks:2017zkg},
and $hhh\to b\bar{b}WW^*WW^*$~\cite{Kilian:2017nio}.
These results are encouraging,
but an unambiguous discovery of this process in the SM puts strong requirements on the performance of the detector and analysis.

Looking at the VBF mode of triple Higgs production, $VV\to hhh$, the
characteristics of the signal suggest a dedicated study despite the
small expected event rate.  The VBF topology, which implies forward
jets with suppressed QCD activity in the central region, improves the
signal-to-background ratio considerably.  This process is
sensitive to a $hhhVV$ interaction which does not exist in the SM but
may be expected for a strongly interacting Higgs
sector~\cite{Contino:2013gna}.  The amplitude also involves the
lower-order $hVV$ and $hhVV$ couplings, and is subject to gauge
cancellations at high energy.  For instance, an anomalous $hVV$
coupling would have a strong impact on triple Higgs production in
VBF~\cite{Belyaev:2018fky}.

At a future lepton collider, the situation is
slightly more favorable.  A high luminosity $e^+e^-$ collider in the
energy range between 500 GeV and 3 TeV gives access to the
Higgs-strahlung and VBF modes of Higgs pair-production, and allows for
a meaningful determination of the cubic Higgs
self-interaction.~\cite{Tian:2013yda,Abramowicz:2016zbo}.
Furthermore, single-Higgs production processes enable an absolute
determination of the $hVV$ couplings, which is an essential ingredient of
an unambiguous determination of the Higgs potential in pair-production
processes.  However, the SM cross section for triple Higgs production
in either $e^+e^-\to Zhhh$ or $e^+e^-\to \nu\bar{\nu}hhh$ does not
rise above $1\;\mathrm{ab}$~\cite{Djouadi:1999gv,Maltoni:2018ttu}.

\subsection{Purpose and Contents of this Paper}

For a process that is as rare as triple Higgs production, in the
presence of large background, detection becomes a challenge even if a
SM calculation predicts a sizable number of events.  Fortunately,
any disturbance of the SM interactions is likely to increase the
expected event yield, possibly by a significant amount, so even a loose
upper limit on the cross section should acquire physical meaning.  To
this end, it is important to know about model-independent upper
limits, beyond which experimentally determined bounds would lose their
immediate significance.

In this paper, we study double and triple Higgs production in VBF
processes in an effective-field theory (EFT) approach.  
In this framework,
anomalous effects are parameterized by the coefficients of higher-dimensional operators.
By investigating the consequences of S-matrix unitarity for the amplitudes $VV\to hh$ and $VV \to hhh$,
we constrain the energy-dependent parameter region where the EFT
yields a valid parameterization.
We use the packages \whizard~\cite{Kilian:2007gr} and
\mgamc~\cite{Alwall:2014hca} to compute the tree-level cross sections
including all terms linear and bilinear in the EFT parameters.
Evaluating numerical results at $14\;\mathrm{TeV}$, $27\;\mathrm{TeV}$
and at $100\;\mathrm{TeV}$, we turn the results into scale-dependent
bounds on the model parameters.  For a complete picture, it is
important to treat Higgs-pair and triple production processes on the
same footing.
By applying our methods to a detailed
study of signal and background effects which we do not attempt here,
it should become possible to properly gauge 
the achievable sensitivity of the collider to new-physics effects in the Higgs
potential, in a largely model-independent fashion.

The paper is organized as follows.
In Sec.~\ref{Sec:EFT},
we establish the framework and introduce the generic effective
Lagrangian, together with two more specific model scenarios where anomalous effects are
present in the Higgs sector.
In Sec.~\ref{Sec:unitarity},
we derive unitarity relations that arise in general, and in
particular for this class of
processes.  To this end, we generalize the well-known formalism for
quasi-elastic 
relativistic scattering processes to multi-particle final states, and
compute the specific constraints that arise within the EFT framework.  In 
Sec.~\ref{Sec:hhh-EFT}, we apply our findings to the phenomenology of
Higgs pair- and triple-production processes, considering cut
strategies, interference effects, and invariant-mass distributions.
We conclude this paper with a discussion of our results in
Sec.~\ref{Sec:conc}.

\section{Effective Lagrangian for VBF Higgs production}
\label{Sec:EFT}

For an unbiased approach to multi-Higgs production phenomenology and
constraints, we would like to employ a framework that is as
model-independent as possible.  We choose to parameterize the Higgs
interactions within an effective-theory framework.  In practice, this
may be considered as arising from the most general form-factor
approach for the interactions of interest, where all form factors are
expanded in powers of momentum.  We keep the leading terms in the
expansion.  We do not introduce new fields, and we keep the infrared
symmetries of the SM (QED, QCD) intact.  The effective Lagrangian
is expressed in terms of physical fields $W^\pm,Z,h$, etc.

It is well known that such a framework can be rendered formally
gauge-invariant by introducing a non-linear gauge representation for
the electro-weak interactions.  Vice versa, the Lagrangian in terms of
physical fields emerges from a generic non-linear representation by
selecting the unitary gauge, where Goldstone and ghost fields are
eliminated.  Furthermore, it is perturbatively equivalent to a
standard effective-theory framework with linear gauge
representation~\cite{Alonso:2016oah}. We have used the same framework
for our previous study of 
multiple Higgs production in gluon-gluon fusion~\cite{Kilian:2017nio}.

The effective Lagrangian can be written as follows:
\begin{align}
\label{eft}  
\mathcal{L}_{EFT} =& \mathcal{L}_{\overline{SM}}+\mathcal{L}_h +\mathcal{L}_{VVh}+\mathcal{L}_{Vh}, \\
\mathcal{L}_h =& -\lambda_3\frac{m_h^2}{2v}h^3-\frac{\kappa_5}{2v}h\partial^\mu{h}\partial_\mu{h} -\lambda_4\frac{m_h^2}{8v^2}h^4-\frac{\kappa_6}{4v^2}h^2\partial^\mu{h}\partial_\mu{h} + \cdots ,
  \label{eq8} \, \\
\mathcal{L}_{VVh} = & - \left(g_{W,b1} \frac{h}{v} + g_{W,b2} \frac{h^2}{2 v^2} + g_{W,b3} \frac{h^3}{6 v^3} + \cdots \right)
W^+_{\mu\nu} W^{-\, \mu\nu} \nonumber \\ &
 - \left(g_{A,b1} \frac{h}{2 v}  + g_{A,b2} \frac{h^2}{4v^2} + g_{A,b3} \frac{h^3}{12v^3} + \cdots \right)F_{\mu\nu} F^{ \mu\nu}  \nonumber \\ &
- \left(g_{X,b1} \frac{h}{ v} + g_{X,b2} \frac{h^2}{2 v^2} + g_{X,b3} \frac{h^3}{6 v^3} + \cdots \right)F_{\mu\nu} Z^{ \mu\nu} \nonumber \\ &
- \left(g_{Z,b1} \frac{h}{2 v} + g_{Z,b2} \frac{h^2}{4 v^2} + + g_{Z,b3} \frac{h^3}{12v^2} + \cdots \right)Z_{\mu\nu} Z^{ \mu\nu} \\ 
    \mathcal{L}_{VH}=&\gwa1 \frac{2m_W^2}{v} h W^{+,\mu}W^-_{\mu}
    +\gwa2 \frac{m_W^2}{v^2}h^2 W^{\mu}W_{\mu}
    +\gwa3 \frac{m_W^2}{3v^3} h^3 W^\mu W_\mu \nonumber \\ &
    + \gza1 \frac{m_Z^2}{v} h Z^{\mu} Z_{\mu}
    +\gza2 \frac{m_Z^2}{2 v^2}h^2 Z^{\mu} Z_{\mu}
    +\gza3 \frac{m_Z^2}{6 v^3} h^3 Z^\mu Z_\mu  + \cdots \,.
    \label{lvh}
\end{align}

Dots indicate higher-dimensional interactions which are not relevant for the VBF Higgs production processes that we consider.
We restrict the calculation to CP-conserving interactions and therefore omit any CP-violating operators.

In the SM at the tree level,
we have the relations $\lambda_3=\lambda_4=\gwa1=\gwa2=\gza1=\gza2=1$ and $\kappa_5=\kappa_6=\gvb1=\gvb2=\gvb3=\gwa3=\gza3=0$,
where the subscript $V$ denotes $W, A, X, Z$.
It is understood that the corresponding terms have been removed from $\mathcal{L}_{\overline{SM}}$,
such that they are not double-counted.

The higher-order operators in the kinetic-energy term (proportional to $\kappa_5$ and $\kappa_6$) are redundant and can be eliminated by applying the equation of motion of the Higgs field or by a non-linear
transformation~\cite{Giudice:2007fh}.
To eliminate $\kappa_5$,
we may replace $h\to h+\frac{a}{2v}h^2$ and get parameter shifts such as
\begin{align}
  \label{lambda3-shift-a}
    \lambda_3 &\to \lambda_3+a
    &
    \lambda_4 &\to \lambda_4+a^2+6a\lambda_3
    \\
    \kappa_5 &\to \kappa_5-2a
    &
    \kappa_6 &\to \kappa_6+5a\kappa_5-2a^2
    \\
    \gwa2 &\to \gwa2+a\gwa1
    &
    \gwa3 &\to\gwa3+3a\gwa2
\end{align}
Choosing $a=\frac{1}{2}\kappa_5$ eliminates $\kappa_5$.
Analogously,
replacing $h\to h+\frac{b}{3v^2}h^3$ results in
\begin{align}
    \lambda_4 &\to \lambda_4+4b
    \\
    \kappa_6 &\to \kappa_6-4b
    \\
    \gwa3 &\to \gwa3+2b\gwa1
  \label{gwa3-shift-b}
\end{align}
so we can eliminate $\kappa_6$ with $b=\frac{\kappa_6}{4}$.
To facilitate the comparison between parameterizations,
we choose to retain these parameters in Table~\ref{table:parameter} below (cf.\ also Ref.~\cite{Kilian:2017nio}).

The phenomenological Lagrangian~(\ref{eft}) provides a robust parameterization of new physics in the Higgs-electroweak sector,
under the condition that no new on-shell states appear in the kinematically accessible range.
In Table~\ref{thhh} we summarize the dependency of various Higgs production processes on the coefficients in the effective Lagrangian,
as can be read off from the contributing Feynman diagrams.

\subsection{Relation to the SILH effective Lagrangian}
\label{sec:silh}

As mentioned above,
the Lagrangian~(\ref{eft}) does not manifestly exhibit electroweak gauge
invariance.  To reformulate it as an equivalent $SU(2)_L\times
U(1)_Y$ invariant Lagrangian, we should apply the appropriate field
redefinitions and connect our parameters to standard parameterizations
in the literature.  In fact, one may expect a linearly gauge-invariant
Lagrangian to result from integrating out manifestly gauge-invariant
new physics at higher scales; 
for an example, cf.~Ref.~\cite{Corbett:2017ieo}.
If we then truncate the gauge-invariant expansion at some fixed order,
we obtain relations among the operator coefficients in~(\ref{eft}).

In the following,
we consider those relations for a particular version of the linear gauge
representation truncated at dimension six, the SILH
parameterization~\cite{Giudice:2007fh}.  For the translation to different
bases, cf., e.g.,
Refs.~\cite{Kaplan:1983fs,Agashe:2004rs,Contino:2013kra,Grzadkowski:2010es}, 
these parameter relations are the consequence of truncating the
gauge-invariant power-series expansion.
By allowing higher-dimensional operators ($D\geq 8$) in the EFT, we
recover the original expression~(\ref{eft}) after
fixing the gauge.

\begin{table}
\begin{center}
\begin{tabular}{|c|c|c|c|c|}
        \hline
        &  $VV \to h$ & $VV \to h h$ & $VV \to h h h$ \\
         \hline
        Parameters                     & $\gva1$, $\gvb1$ &  $\gva1$, $\gvb1$  &  $\gva1$, $\gvb1$  \\
            involved                & - &  $\gva2$, $\gvb2$, $\lambda_3$, $\kappa_5$  &   $\gva2$, $\gvb2$, $\lambda_3$, $\kappa_5$ \\
                             & - &  -  &   $\gva3$, $\gvb3$, $\lambda_4$, $\kappa_6$ \\
         \hline
    \end{tabular}
    \caption{ \label{thhh} Parameters that contribute to the
      VBF Higgs-production processes studied in this paper.}
\end{center}
\end{table}

There are various versions of the SILH effective-Lagrangian~\cite{Giudice:2007fh} parameterization.
We refer to the following definition:
\begin{align}
	\begin{split}
{\cal L}_\text{SILH} =& \frac{c_H}{2f^2}\partial^\mu \left( H^\dagger H \right) \partial_\mu \left( H^\dagger H \right)
+ \frac{c_T}{2f^2}\left (H^\dagger {\overleftrightarrow { D^\mu}} H \right)  \left(   H^\dagger{\overleftrightarrow D}_\mu H\right)
- \frac{c_6\lambda}{f^2}\left( H^\dagger H \right)^3 \\
& +\left( \frac{c_yy_f}{f^2}H^\dagger H  {\bar f}_L Hf_R +{\rm h.c.}\right) +\frac{c_g g_S^2}{16\pi^2f^2}\frac{y_t^2}{g_\rho^2}H^\dagger H G_{\mu\nu}^a G^{a\mu\nu} \\
&+\frac{ic_Wg}{2m_\rho^2}\left( H^\dagger  \sigma^i \overleftrightarrow {D^\mu} H \right )( D^\nu  W_{\mu \nu})^i
+\frac{ic_Bg'}{2m_\rho^2}\left( H^\dagger  \overleftrightarrow {D^\mu} H \right )( \partial^\nu  B_{\mu \nu})  \\
& +\frac{ic_{HW} g}{16\pi^2f^2}
(D^\mu H)^\dagger \sigma^i(D^\nu H)W_{\mu \nu}^i
+\frac{ic_{HB}g^\prime}{16\pi^2f^2}
(D^\mu H)^\dagger (D^\nu H)B_{\mu \nu}
 \\
&+\frac{c_\gamma {g'}^2}{16\pi^2f^2}\frac{g^2}{g_\rho^2}H^\dagger H B_{\mu\nu}B^{\mu\nu}.
\label{lsilh}
\end{split}
\end{align}
Regarding derivatives acting on bosonic fields, we recall that we may
apply the equations of motion
\begin{align}
	\left( D^\mu D_\mu H \right)^j =& m^2 H^j -\lambda \left( H^\dag H \right) H^j -\bar e\, \Gamma_{\!e}^\dag l^j + \epsilon_{jk} \bar{q}^k\, \Gamma_{\!u} u-
\bar d\, \Gamma_{\!d}^\dag q^j,\\
\left(D^\rho G_{\rho\mu}\right)^A =& g_s \left(\bar q \gamma_\mu T^A q
~+~ \bar u \gamma_\mu T^A u ~+~ \bar d \gamma_\mu T^A d\right)\\
\left(D^\rho W_{\rho\mu}\right)^i =&\frac{g}{2}\left( H^\dagger i \overleftrightarrow D_\mu^i H + \bar l \gamma_\mu \tau^i l~+~ \bar q \gamma_\mu \tau^i q\right)\\
\partial^\rho B_{\rho\mu} =& g'Y_HH^\dagger i \overleftrightarrow D_\mu^i H + g' \sum_{\psi \in \{l,e,q,u,d\}} Y_\psi\bar\psi\gamma_\mu\psi.
\end{align}
and trade them for terms with less derivatives and terms with
flavor-diagonal contact interactions between bosons and fermions.  
Boson-fermion contact terms yield subleading effects in dedicated VBF data
analyses, if we apply cuts that enhance the quasi on-shell
contribution for intermediate vector bosons, 
and optimize the analysis for resonant final-state vector bosons. 

Further details of deriving the relation between the SILH Lagrangian
and the effective Lagrangian~(\ref{eft}),
including conventions not listed here,
are given in Appendix~\ref{App:SILH}.  Table~\ref{table:parameter}
contains the actual translation between parameter sets.

\begin{center}
\begin{table}
	\footnotesize
  \begin{center}
  \begin{tabular}{|c|c|c|c|}
  \hline
           &  SILH                           & Higgs Inflation \\
  \hline
  $\lambda_3$         &  $(1+\frac{5}{2}c_6 v^2/f^2 )(1+\frac{3}{2}c_6 v^2/f^2 )^{-1}\zeta_h$   &  $(1+6\xi^2v^2/M_p^2)^{-1/2}$ \\
  \hline
  $\lambda_4$         &  $(1+\frac{15}{2}c_6 v^2/f^2)(1+\frac{3}{2}c_6v^2/f^2)^{-1}\zeta^2_h$        &  $(1+6\xi^2v^2/M_p^2)^{-1}$ \\
  \hline
  $\kappa_5$             &  $-2c_Hv^2/f^2\zeta^3_h$       &   $-12v^2\xi^2/M_p^2(1+6\xi^2v^2/ M_p^2)^{-3/2}$\\
  \hline
  $\kappa_6$             &  $-2c_Hv^2/f^2\zeta^4_h$      &   $-12v^2\xi^2/ M_p^2(1+6\xi^2v^2/ M_p^2)^{-2}$ \\
  \hline  
  $\gwb1$            &  $c_{HW}\frac{g^2v^2}{32\pi^2f^2}\zeta_h\zeta^2_W$   &  0\\
  \hline
  $\gwb2$            &  $c_{HW}\frac{g^2v^2}{32\pi^2f^2}\zeta^2_h\zeta^2_W$      &  0\\
  \hline
  $\gab1$           &  $-c_\gamma\frac{g^2v^2}{8\pi^2f^2}\frac{{g^\prime}^2}{g^2_\rho}\cos^2{\theta}\zeta_h\zeta_A^2$          &  0\\
  \hline
  $\gab2$           &  $-c_\gamma\frac{g^2v^2}{8\pi^2f^2}\frac{{g^\prime}^2}{g^2_\rho}\cos^2{\theta}\zeta^2_h\zeta_A^2$              & 0\\
  \hline
  \multirow{2}*{$\gxb1$}     & $\frac{g{g^\prime}v^2}{64\pi^2f^2}\left[(c_{HW}-c_{HB})+8c_\gamma\frac{g^2}{g^2_\rho}\sin^2\theta\right]\zeta_h\zeta_{A}\zeta_{Z}$ & \multirow{2}*{0} \\
                             & $+ c_\gamma\frac{g^2v^2}{4\pi^2f^2}\frac{{g^\prime}^2}{g^2_\rho}\cos^2{\theta}\zeta_h\zeta_{AZ}^2$  & \\
  \hline
  \multirow{2}*{$\gxb2$}     & $\frac{g{g^\prime}v^2}{64\pi^2f^2}\left[(c_{HW}-c_{HB})+8c_\gamma\frac{g^2}{g^2_\rho}\sin^2\theta\right]\zeta^2_h\zeta_{A}\zeta_{Z}$ & \multirow{2}*{0} \\
  & $+c_\gamma\frac{g^2v^2}{4\pi^2f^2}\frac{{g^\prime}^2}{g^2_\rho}\cos^2{\theta}\zeta^2_h\zeta_{AZ}^2  $ &     \\
  \hline
  \multirow{2}*{$\gzb1$}      &  $\frac{g^2v^2}{32\pi^2f^2}(c_{HW}+c_{HB}\tan^2{\theta})\zeta_h\zeta^2_Z - c_\gamma\frac{g^2v^2}{8\pi^2f^2}\frac{{g^\prime}^2}{g^2_\rho}\cos^2{\theta}\zeta_h\zeta_{AZ}^2$ & \multirow{2}*{0} \\
  &  $-\frac{g{g^\prime}v^2}{64\pi^2f^2}\left[(c_{HW}-c_{HB})+8c_\gamma\frac{g^2}{g^2_\rho}\sin^2\theta\right]\zeta_h\zeta_{AZ}\zeta_{Z}$  &  \\
  \hline
  \multirow{2}*{$\gzb2$}      &  $\frac{g^2v^2}{32\pi^2f^2}(c_{HW}+c_{HB}\tan^2{\theta})\zeta^2_h\zeta^2_Z- c_\gamma\frac{g^2v^2}{8\pi^2f^2}\frac{{g^\prime}^2}{g^2_\rho}\cos^2{\theta}\zeta^2_h\zeta_{AZ}^2$  & \multirow{2}*{0} \\
&$-\frac{g{g^\prime}v^2}{64\pi^2f^2}\left[(c_{HW}-c_{HB})+8c_\gamma\frac{g^2}{g^2_\rho}\sin^2\theta\right]\zeta_h\zeta_{AZ}\zeta_{Z}$  &  \\ \hline
  $\gwa1$ & $\left[1-\left(c_{W}\frac{g^2v^2}{m^2_{\rho}}+c_{HW}\frac{g^2v^2}{16\pi^2f^2}\right)\right]\zeta_h\zeta^2_W$ &$(1+6\xi^2v^2/ M_p^2)^{-1/2}$\\
  \hline
  $\gza1$ &$ \left[1-\left(c_{W}\frac{g^2v^2}{m^2_{\rho}}+c_{B}\frac{{g^\prime}^2v^2}{m^2_{\rho}}+c_{HW}\frac{g^2v^2}{16\pi^2f^2}+c_{HB}\frac{{g^\prime}^2v^2}{16\pi^2f^2}\right)\right]\zeta_h\zeta^2_Z$ &$(1+6\xi^2v^2/ M_p^2)^{-1/2}$ \\
  \hline
  $\gwa2$ &$\left[1-3\left(c_{W}\frac{g^2v^2}{m^2_{\rho}}+c_{HW}\frac{g^2v^2}{16\pi^2f^2}\right)\right]\zeta^2_h\zeta^2_W$ &$(1+6\xi^2v^2/ M_p^2)^{-1}$ \\
  \hline
  $\gza2$ & $\left[1-3\left(c_{W}\frac{g^2v^2}{m^2_{\rho}}+c_{B}\frac{{g^\prime}^2v^2}{m^2_{\rho}}+c_{HW}\frac{g^2v^2}{16\pi^2f^2}+c_{HB}\frac{{g^\prime}^2v^2}{16\pi^2f^2}\right)\right]\zeta^2_h\zeta^2_Z$ &$(1+6\xi^2v^2/ M_p^2)^{-1}$ \\
  \hline
  $\gwa3$ & $\text{From dim-8 operators or higher}$  &$\mathcal{O}(\xi)$ \\
   \hline
   $\gza3$ & $\text{From dim-8 operators or higher}$  &$\mathcal{O}(\xi)$ \\
  \hline
\end{tabular}
  \end{center}
  \caption{\label{table:parameter}  Relations between the phenomenological
    Lagrangian parameters in~(\ref{eft}-\ref{lvh}) (first column), the SILH
    effective Lagrangian~\ref{lsilh} (second column), and the free parameters
    of the Higgs-inflation model, Sec.~\ref{sec:inflation}.
    Note the extra parameters $\zeta^n_h$, $\zeta^n_W$, $\zeta^n_Z$, $\zeta^n_A$,
    $\zeta^n_{AZ}$ (defined in
    Eq.~\ref{rescale1}--\ref{rescale2}) and the $(1+6\xi^2v^2/M_p^2)^{-1/2}$
    factor, induced by the Higgs and gauge-boson wave-function
    normalization, respectively.
} 
\label{relation}
\end{table}
\end{center}

We conclude this part with a remark on oblique corrections.
According to Ref.~\cite{Barbieri:2004qk},
the parameter $\hat{S}$ is given by 
\begin{eqnarray}
\hat{S} &=& 2\frac{\cos{\theta}}{\sin{\theta}}c_{WB} \\
\hat{T} &=& -c_H
\end{eqnarray}
where $c_{WB}/v^2 g g'$ is the coefficient of the operator $H^\dagger \sigma^iHW^i_{\mu\nu}B^{\mu\nu}/gg^\prime$ for non-canonical gauge fields,
and $c_H$ is the coefficient of the operator $\vert H^\dagger D_\mu H \vert ^2$.
If we translate the basis of Ref.~\cite{Barbieri:2004qk} to our version of the SILH effective Lagrangian,
we have
\begin{eqnarray}
  \label{eq:S-T}
 \hat{S}  &=&  -(c_W+c_B)\frac{m_W^2}{m_\rho^2}\\
 \hat{T}  &=&  - \frac{v^2}{f^2}c_T
\end{eqnarray}
We recall that $\hat{S}$ is constrained by data at the $10^{-3}$ level; the
precise value depends on the variation of the other electroweak parameter
$\hat{T}$.  For our purposes, we have set $c_T$ to zero.

\subsection{Relation to models of Higgs-inflation}
\label{sec:inflation}

Higgs-inflation models~\cite{Bezrukov:2007ep,Bezrukov:2009db,Hamada:2015skp,Ren:2014sya,He:2014ora,Xianyu:2014eba,Ge:2016xcq,Ellis:2016spb,Ellis:2014dxa}
provide an interesting example of a scenario where new physics is associated with the Higgs sector,
with little impact on other SM particles.
Such models are notoriously difficult to identify,
and any possible probe of Higgs interactions should be investigated.
In the present context, this class of model provides an example of a
scenario where the effective-Lagrangian description applies, the main
effects are tied to the Higgs sector, and the parameter set is even
more restricted.  Conversely, the relations and limits that we derive
for the parameters of the generic effective Lagrangian, can be transferred
to such a restricted model in a straightforward way.

We briefly review the derivation of the phenomenological Higgs
Lagrangian for this model, where the Higgs field is coupled to gravity
in a non-minimal way.
The model is originally formulated as a Lagrangian in the Jordan frame,
\begin{align}
	\begin{split}
		S_{\mathrm{Jordan}} = \int \ud^4x \sqrt{-g} \Bigg\{
			&- \frac{M^2+2 \xi H^\dagger H}{2}R
			-\frac{1}{4} W^{a\mu\nu} W^a_{\mu\nu}-\frac{1}{4} B^{\mu\nu} B_{\mu\nu}   \\
			&  + D_\mu H^\dagger D^\mu H     -\lambda\left(H^\dagger H-\frac{v^2}{2}\right)^2
		\Bigg\}
		\;.
	\label{eq:1}
	\end{split}
\end{align}
The value of $\xi$ can vary between $1\ll\sqrt{\xi}\ll10^{17}$,
corresponding to $M \simeq M_P$.

For investigating the phenomenology,
we apply the conformal transformation from the Jordan frame to the Einstein frame
\begin{align}
  \label{eq:2}
  \hat{g}_{\mu\nu} = \Omega^2 g_{\mu\nu}
  \;,\quad
  \Omega^2 = 1 + \frac{2\xi H^\dagger H}{M_P^2}
  \;.
\end{align}

This transformation leads to a non-minimal kinetic term for the Higgs field.
In the unitary gauge $H= \frac{1}{\sqrt{2}}(0, h)^T$,
we may introduce a scalar field $\chi$ as a transformed Higgs field,
\begin{align}
  \label{eq:3}
  \ud\chi=\sqrt{\frac{\Omega^2+12\xi^2 H^\dagger H/ M_P^2}{\Omega^4}} \ud h.
\end{align}
The action in the Einstein frame is
\begin{align}
  \label{eq:4}
    S_E \supset \int d^4x\sqrt{-\hat{g}} \left\{
    - \frac{M_P^2}{2}\hat{R}
    + \partial_\mu \chi \partial^\mu \chi
    - U(\chi) 
    \right\}
 \end{align}
where $\hat{R}$ is calculated using the metric $\hat{g}_{\mu\nu}$.
We neglect any renormalization-group running effect.
The effective Higgs potential is
\begin{align}
  \label{eq:5}
  U(\chi) =
  \frac{1}{\Omega(\chi)^4}\frac{\lambda}{4}\left(h(\chi)^2-\frac{v^2}{2}\right)^2.
 \end{align}

In the context of collider physics,
we are looking at small field values $h\simeq\chi$ and $\Omega^2\simeq1$,
so the potential for the field $\chi$ is close to that of the initial Higgs field.
Inflation physics is described by the large-field behavior of the Higgs field,
the Higgs thus acting as an inflaton,
where $h\gg M_P/\sqrt{\xi}$ (or $\chi\gg\sqrt{6}M_P$).
In this range,
we can approximate
\begin{align}\label{eq:hlarge}
  h\simeq \frac{M_P}{\sqrt{\xi}}\exp\left(\frac{\chi}{\sqrt{6}M_P}\right), \quad
  U(\chi) = \frac{\lambda M_P^4}{4\xi^2}
  \left(
    1+\exp\left(
      -\frac{2\chi}{\sqrt{6}M_P}
    \right)
  \right)^{-2}.
\end{align}
The potential is exponentially flat at large $h$,
as appropriate for a model of inflation.

We are interested in collider phenomenology and thus assume small $h$ field values,
so we replace $\chi$ by $h$ again.
We plug Eq.~\eqref{eq:3} into Eq.~\eqref{eq:4} and omit higher-order terms.
After re-instating the Higgs doublet notation $H$,
we arrive at\footnote{Here we correct minor errors present in Ref.~\cite{Kilian:2017nio}. The corrections do not change the numerical results of that analysis.}
\begin{align}
	\begin{split}
  \label{eq:7}
  S_E =\int d^4x\sqrt{-\hat{g}} \Bigg\{&
    - \frac{M_P^2}{2}\hat{R}
    + \textrm{gauge kinetic terms} + \frac{ D_\mu H^\dagger  D^\mu H}{\Omega^2} +\frac{3 \xi^2}{M_p^2}\frac{ \partial_\mu (H^\dagger H ) \partial^\mu (H^\dagger H)}{ \Omega^4}  \\
	& - \frac{1}{\Omega^2}\lambda \left(H^\dagger H-\frac{v^2}{2}\right)^2 + \frac{2H^\dagger H}{\Omega^2} \left(D_\mu H^\dagger D^\mu H\right) 
   \Bigg \} \,.
   \end{split}
   \end{align}
Details regarding the gauge interaction can be found in Ref.~\cite{Obata:2014qba}.
We obtain corrections to the coefficients of the following operators:
  \begin{align}
   L_{VH}=&\gwa1 \frac{2m_W^2}{v} h W^{\mu}W_{\mu}
    +\gwa2 \frac{m_W^2}{v^2}h^2 W^{\mu}W_{\mu}
    +\gwa3 \frac{m_W^2}{3v^3} h^3 W^\mu W_\mu \nonumber \\ &
    + \gza1 \frac{m_Z^2}{v} h Z^{\mu} Z_{\mu}
    +\gza2 \frac{m_Z^2}{2v^2}h^2 Z^{\mu} Z_{\mu}
    +\gza3 \frac{m_Z^2}{6v^3} h^3 Z^\mu Z_\mu  + \cdots \,.\end{align}
 

In Table~\ref{table:parameter} we list the coefficient expressions for the
Higgs inflation model and relate them to the SILH operator basis and
to the Higgs Lagrangian that we use for our study.  It is evident that the
SILH operator basis, which is appropriate for a generic strongly interacting
model, incorporates directions in parameter space which are absent in the more
specific model of Higgs inflation.  Dedicated measurements of Higgs
self-interactions become essential if such a class of model is realized.

\section{Constraints on parameters from the unitarity of S matrix}
\label{Sec:unitarity}

When adopting an EFT for calculations, we accept a truncated Taylor
expansion as a model for any scattering amplitude, distribution or
cross section.  A term of dimension $d$ in the Lagrangian
generates uncancelled factors that rise proportional to $E^{d-4}$ in
the amplitude, where $E$ is the overall energy scale in a scattering
process~\cite{Cornwall:1974km}.  Due to this growth, the model ceases
to be weakly interacting at some scale.

The scale where the scattering strength has to saturate, can be
deduced from the optical theorem, i.e., unitarity.  Physically, the
theorem describes the effect of rescattering which occurs whenever
overlapping field amplitudes become sufficiently large.  In this
work, we want to turn the argument around and determine bounds for
EFT parameters such that sizable rescattering corrections do not occur
below some arbitrary scale~$Q$, for the concrete processes of
multi-Higgs production in VBF.

In the simplest case of spinless $2\to 2$ scattering, exploiting the optical
theorem is textbook knowledge.  A partial-wave decomposition will
reformulate the restricted $\mathcal{S}$ operator as a discrete matrix
which can be diagonalized for convenience (cf., e.g.,
Ref.~\cite{Baur:1987mt}).

The extension to relativistic $2\to n$ scattering
($n>2$), in the context of the SM and its extensions, is not as
familiar. There is no obvious decomposition of the final state.
Apparently, we have to consider $m\to n$ contributions to the
scattering operator which do not correspond to physical scattering
processes, and reformulating the problem into matrix equations is no
longer straightforward.  To arrive at useful inequalities, previous
work introduced extra
assumptions~\cite{Maltoni:2001dc,Dicus:2004rg,Dicus:2005ku}.
Neglecting spin also simplifies the problem
considerably~\cite{Yu:2013aca}.

Below, we adopt a generic approach, which we then apply to the
particular case at hand.  It turns out that the exact form of
decomposition does not matter for the purpose of deriving
phenomenologically viable parameter bounds.  In the particular case of
EFT operators, complications from kinematical dependencies are
largely absent.

\subsection{General unitarity constraints}

Unitarity is the conservation of probability in a quantum theory,
applied to the $\mathcal{S}$ operator that encodes the scattering of observable particles:
$\mathcal{S}^{\dag}\mathcal{S}=1$.
Its nontrivial part $\mathcal{T}$,
defined by $\mathcal{S}=1+i\mathcal{T}$,
satisfies the universal relation
\begin{align}
\label{eq:pwu-base}
    -i(\mathcal{T}-\mathcal{T}^{\dag})
    =
    \mathcal{T}^{\dag}\mathcal{T}
\end{align}
We are interested in unitarity conditions for matrix elements between asymptotic states which consist of a finite number $n_a$ of particles with well-defined masses.
We denote multi-particle states collectively by $\ket{\alpha,\Phi_a}$,
where $\Phi_a$ is a shorthand for the kinematical configuration of $n_a$ on-shell four-momenta (the phase-space point),
and $\alpha$ denotes the set of discrete quantum numbers such as helicity and color.
Furthermore,
we fix the total momentum of a multiparticle state $a$ to $p_a$.
With this constraint,
the manifold of configurations $(\alpha,\Phi_a)$ becomes a compact manifold for each fixed $n_a$.

Momentum conservation allows us to introduce the matrix elements of the scattering amplitude operator $\mathcal{M}$ between the initial state $\ket{\alpha,\Phi_a}$ and the final state $\ket{\beta,\Phi_b}$,
\begin{align}
    \bra{\beta,\Phi_b}\mathcal{T}\ket{\alpha,\Phi_a}
    =
    (2\pi)^4\delta^{(4)}(p_a-p_b)
    \,\bra{\beta,\Phi_b}\mathcal{M}\ket{\alpha,\Phi_a}
    \label{eq:pwu-amp-gen}
\end{align}
We take matrix elements on both sides of Eq.~\eqref{eq:pwu-base} and insert a complete set of multi-particle states $\ket{\gamma,\Phi_c}$,
\begin{align}
  \label{eq:pwu-master}  
  \lefteqn{-i\left[\bra{\beta,\Phi_b}\mathcal M \ket{\alpha,\Phi_a}
      -\bra{\alpha,\Phi_a} \mathcal M \ket{\beta,\Phi_b}^{*}\right]}
  \notag\\
  &=
  \sum_{\gamma}\int\ud\Phi_c\,
  \bra{\gamma,\Phi_c}\mathcal{M}\ket{\beta,\Phi_b}^{*}
  \bra{\gamma,\Phi_c}\mathcal{M}\ket{\alpha,\Phi_a}
\end{align}
where $\ud\Phi_c$ denotes the canonical Lorentz-invariant measure on the
phase space $\{\Phi_c\}$ constrained by $p_c=p_a=p_b$.

For calculations,
we may introduce a bijective mapping between the unit hypercube in $d_a=3n_a-4$ dimensions,
$\{x_a\in\mathbb{R}^{d_a};\ 0<(x_a)_i<1\}$ and the manifold $\{\Phi_a\}$,
for each fixed $n_a$.
For instance, we may factorize phase space as a tree consisting of $n_a-1$ momentum splittings of type $1\to2$,
with $p_a$ at the root.
There are $2(n_a-1)$ angular variables and $n_a-2$ invariant-mass variables.
This mapping introduces a Jacobian $J_a(x_a)=\ud\Phi_a/\ud x_a$,
which should incorporate symmetry factors where appropriate.
The construction provides a method of evaluating
phase-space integrals that has become standard and preserves overall
Lorentz invariance. 
If we introduce amplitude functions which include the Jacobian factors as follows,
\begin{equation}
  M_{\beta\alpha}(x_b,x_a) 
  = J^{1/2}_b(x_b)\,
  \bra{\beta,\Phi_b(x_b)}\mathcal M \ket{\alpha,\Phi_a(x_a)}
  \,J^{1/2}_a(x_a)
\end{equation}
Eq. \eqref{eq:pwu-base} takes the form
\begin{align}
  -i\left[M^{\beta\alpha*}(x_b,x_a) - M^{\alpha\beta}(x_a,x_b)\right]
  &=
  \sum_{\gamma}\int\ud x_c\,
  M^{\gamma\beta*}(x_c,x_b)\,M^{\gamma\alpha}(x_c,x_a)\label{eq:pwu-amp}
\end{align}

If massless particles are involved, the sum over intermediate states is infinite,
and the matrix elements contain non-integrable infrared,
collinear,
and Coulomb singularities,
so the integrals do not converge.
To remedy this,
we may introduce some version of phase-space slicing and sum over nearly degenerate states,
which introduces indefinite particle numbers~\cite{Kinoshita:1962ur,Lee:1964is}.
However,
in the present context where we are studying the production of massive bosons,
we focus on anomalous couplings of the Higgs boson.
As we do not change the couplings of massless particles,
we may ignore this complication,
and assume that all external and internal states are massive.
The sum over intermediate states then is a finite sum,
the matrix elements and the Jacobians are finite,
and the integration manifold (the union of the unit hypercubes for all
contributing $(n_a,\alpha)$) is compact.

In such a situation,
it is possible to introduce a scalar product of square-integrable functions on the integration manifold and to find a complete basis of functions which are mutually orthonormal with respect to this scalar product.
For instance,
choosing the canonical scalar product,
we could take a straightforward Fourier expansion.
A more physical choice could involve spherical harmonics for the normalized angular variables and an arbitrary basis for the invariant-mass variables.
In the two-particle case where there are no free invariant masses,
this becomes the standard partial-wave expansion.
We note that for each particle combination $a$,
we may choose a different kind of expansion for the corresponding phase space $\Phi_a(x_a)$.

We adopt,
for simplicity,
the canonical scalar product and a corresponding orthonormal basis $\{H^\alpha_A(x_a)\}$ on each $\alpha$ phase space,
\begin{equation}
  \int \ud x_a\, H^{\alpha*}_A(x_a)\,H^{\alpha}_B(x_a)
  = \delta_{AB},
\end{equation}
where $A$ is an appropriate (multi-)index which labels the basis functions.
We expand the amplitudes as
\begin{align}
  \label{eq:pwu-aexp}
  M^{\beta\alpha}(x_b,x_a) &=
  2\sum_{AB}a^{\alpha\beta}_{AB}H^\alpha_A(x_a)\,H^{\beta*}_B(x_b),
\end{align}
and thus express the scattering in terms of an actual matrix with elements
$a^{\alpha\beta}_{AB}$.\footnote{The factor~$2$ has been inserted for consistency with the standard two-particle partial-wave expansion.}
Explicitly,
the coefficients are
\begin{equation}
  a^{\alpha\beta}_{AB}
  = \frac12\int\ud x_a\,\ud x_b\, H^{\alpha*}_A(x_a)\,H^{\beta}_B(x_b)\,
  M^{\beta\alpha}(x_b,x_a).
\end{equation}
They take complex values and depend only on the total momentum,
$a^{\alpha\beta}_{AB} = a^{\alpha\beta}_{AB}(p_a)$,
where $p_a=p_b$.
If we choose a phase-space parameterisation which preserves Lorentz invariance,
the coefficients depend only on $s=p_a^2$.

We obtain a discrete version of Eq. \eqref{eq:pwu-amp}~\cite{Cutkosky:1960sp,Veltman:1963th},
\begin{equation}
  \label{eq:pwu-abeq}
  -i(a^{\alpha\beta}_{AB}-a^{\beta\alpha*}_{BA})
  = 2\sum_{\gamma}\sum_C a^{\alpha\gamma}_{AC}a^{\beta\gamma*}_{BC},
\end{equation}
where all coefficients are finite and the sums are convergent if the simplifications regarding massless states are applied,
as described above.

Eq.~\eqref{eq:pwu-abeq} encodes all unitarity relations of the scattering matrix in question.
To derive constraints on individual amplitudes,
we need a positivity condition.
We may diagonalize the scattering matrix and obtain exact relations for superpositions of states.
Alternatively,
we may derive less comprehensive but phenomenologically more useful relations by focusing on diagonal matrix elements,
i.e., $\alpha=\beta$ and $A=B$,
\begin{align}
  -i(a^{\alpha\alpha}_{AA}-a^{\alpha\alpha*}_{AA})
  &= 2\sum_{\gamma}\sum_C |a^{\alpha\gamma}_{AC}|^2
  \\
  &= 2|a^{\alpha\alpha}_{AA}|^2 + 2\sum_{C\neq A}|a^{\alpha\alpha}_{AC}|^2
  + 2\sum_{\gamma\neq\alpha} \sum_C |a_{AC}^{\alpha\gamma}|^2
\end{align}
To cast this in the intuitive geometry of the Argand circle,
we may express the diagonal amplitude in terms of its real and imaginary parts and write
\begin{align}
  \label{eq:pwu-constraints1}
  |\mathrm{Re}\, a^{\alpha\alpha}_{AA}|^2
  + |\mathrm{Im}\,a^{\alpha\alpha}_{AA}-\frac{1}{2}|^2
  + \sum_{C\neq A}|a^{\alpha\alpha}_{AC}|^2
  + \sum_{\gamma\neq\alpha}\sum_C |a^{\alpha\gamma}_{AC}|^2
  &= \frac{1}{4}
\end{align}
That is,
each complex-valued elastic amplitude $a_{AA}^{\alpha\alpha}(s)$ must lie on a circle with radius $r$ around $i/2$,
where the elastic radius $r=1/2$ is reduced by the total contribution of all inelastic channels.

The exact relation~\eqref{eq:pwu-constraints1} yields strict upper bounds for the elastic amplitude as well as for the total inelastic contribution,
which trivially translates into a bound for each individual final state in this representation.
We read off
\begin{equation}
  \label{eq:pwu-constraints2}
  \begin{split}
    |\mathrm{Re}\,a^{\alpha\alpha}_{AA}|^2 \le &\frac{1}{4}\\
    |\mathrm{Im}\,a^{\alpha\alpha}_{AA}-\frac{1}{2}|^2 \le &\frac{1}{4}\\
    \sum_{C\neq A}|a^{\alpha\alpha}_{AC}|^2 \le &\frac{1}{4}\\
      \sum_{\gamma\ne \alpha}\sum_C|a^{\alpha\gamma}_{AC}|^2 \le &\frac{1}{4}
  \end{split}
\end{equation}
Examples for the application of these bounds,
referring also to the treatments in Refs.~\cite{Baur:1987mt,Maltoni:2001dc,Dicus:2004rg,Yu:2013aca},
can be found in Appendix~\ref{App:pwu-further}.

The last inequality in Eq. \eqref{eq:pwu-constraints2} gives the
unitarity constraints on inelastic scattering.  The important
observation is that it is independent of the chosen basis in
multi-particle phase space, $\{H^{\gamma}\}$.
To see this,
we define the coefficients $b_{A}^{\alpha\gamma}$ as follows:
\begin{equation}
        b_{A}^{\alpha\gamma}\equiv\frac{1}{4}\int\ud x_a\ud x_b \ud x_c H_A^{\alpha *}(x_a)H_A^{\alpha}(x_b)M^{\gamma\alpha *}(x_c,x_b)M^{\gamma\alpha}(x_c,x_a)
    \label{eq:pwu-bexp}
\end{equation}
which is clearly independent of $H^{\gamma}$.
Using the expansion in Eq. \eqref{eq:pwu-aexp},
we find
\begin{align}
    b_{A}^{\alpha\gamma}=\sum_C|a_{AC}^{\alpha\gamma}|^2\ge 0
\end{align}
The unitarity constraint for inelastic scattering can be written as
\begin{align}
	\sum_{\gamma\ne \alpha}b_A^{\alpha\gamma}\le\frac{1}{4}\label{eq:pwu-bsum}
\end{align}

The problem has been reformulated as a set of discrete matrix
equations.  Diagonalization becomes a possibility also for the generic
case, although analytical solutions do not exist for realistic
applications~\cite{Corbett:2014ora}.

We are interested in inelastic scattering processes,
$VV\to hh$ and $VV\to hhh$.  For those, the
inequality~(\ref{eq:pwu-bsum}) constrains the sum over all
contributions to a particular final state.  Moreover, we read off the
weaker constraint
\begin{align}
  0 \leq b_A^{\alpha\gamma}\le\frac{1}{4}\label{eq:pwu-b}
\end{align}
for each individual contribution $\gamma$.  In fact, the
offending terms in the EFT turn out to either not depend on the phase
space point at all, or otherwise the dependence is well controlled, so
that meaningful bounds can easily be computed.  We arrive at a set of
conservative bounds on the parameter space even without an analytical
solution.  A more detailed calculation could further narrow down the
parameter space but not widen it.

\subsection{Unitarity Constraints arising from $V V \to hh$}\label{sec:pwu-hh}

We now apply the generic formalism to the two-particle inelastic
scattering process $VV\to hh$. 
In the following we derive the formulas for $W^+W^-\to hh$.
Analogous results for $ZZ\to hh$ can be obtained by replacing the couplings accordingly and taking into account the symmetry factor from two identical $Z$ boson.
We assume that the on-shell approximation is justified for the purpose
of deriving unitarity bounds,
i.e., we treat the incoming vector bosons as on-shell with
a pair invariant mass $m(W^+W^-)=m(hh)={\hat s}$.  In the actual
process, the incoming propagators are space-like with a virtuality of
$O(m_W^2)$; cf.\ Ref.~\cite{Perez:2018kav} for the possibility of
relaxing that assumption.

We are looking at an inelastic channel.
If we expand in an orthonormal basis as described in the preceding subsection,
we obtain
\begin{align}
    b_{A}^{W^+W^-\to hh}(\hat s)\equiv\sum_C|a_{AC}^{W^+W^-\to hh}(\hat s)|^2 \leq \frac14
\end{align}
where $A$ and $C$ are (multi-)indices for the initial-state and final-state basis,
respectively.
We note that the initial-state particles carry spin as well as momentum,
while the final-state phase space manifold is trivially given by the unit sphere,
for fixed energy $\sqrt{\hat s}$.

\begin{figure}[bthp]
  \centering
  \subfigure{
\label{Fig2.sub.1}   \thesubfigure
  \includegraphics[width=0.18\textwidth]{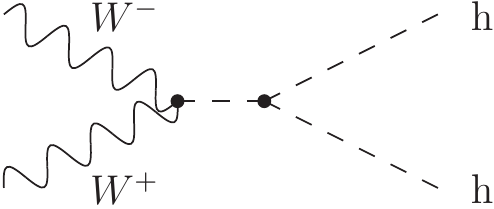}}
  \subfigure{
\label{Fig2.sub.2}   \thesubfigure
  \includegraphics[width=0.18\textwidth]{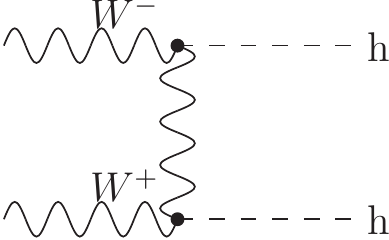}}
  \subfigure{
\label{Fig2.sub.3}   \thesubfigure
  \includegraphics[width=0.18\textwidth]{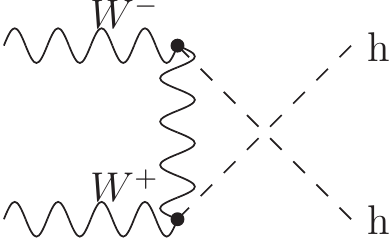}}
  \subfigure{
\label{Fig2.sub.4}   \thesubfigure
  \includegraphics[width=0.18\textwidth]{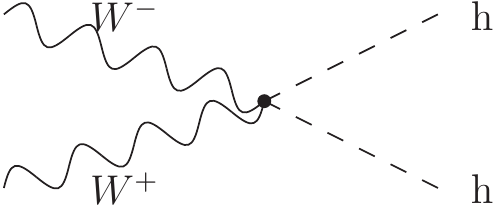}}
  \caption{Four types of Feynman diagrams which contribute to the processes $W^+ W^- \to h h $.} \label{fdvv2hh}
\end{figure} 

As shown in Fig.~\ref{fdvv2hh},
there are four distinct Feynman diagrams which contribute to the $W^+
W^- \to h h $ process in the SM, 
and this breakdown remains valid in the EFT,
\begin{align}
    M(W^+ W^-\to hh)=M_s+M_t+M_u+M_4.
\end{align}
We refer to these as the $s$-channel,
$t$-channel,
$u$-channel,
and contact-interaction amplitudes,
respectively.

In the high energy limit $s\gg m_W^2,m_H^2$,
the leading contribution in the EFT is proportional to $s$.
We thus write a series expansion as follows,
in terms of the rescaled energy $\sqrt{s/v^2}$ as a dimensionless expansion
parameter,
\begin{align}
    M(W^{+} W^{-}\to hh)= \sum_{i=0}^{+\infty}m_i (\frac{\sqrt{s}}{v})^{2-i},
\end{align}
where $m_i$ are the coefficients in the expansion.
In Table~\ref{tab-vvhh} we list the prefactors of the leading contribution for each amplitude and each one of the four independent helicity modes.
The amplitudes of all other helicity modes are related to the four modes that we include in the table.
We find that with the exception of one term ($t/u$-channel $+-$), all leading contributions are independent of the kinematics.
These table entries translate directly into bounds for amplitude
coefficients.
We also observe that only the $++$, $+-$, and $00$ modes lead to amplitudes rising proportional to $s$,
so we may focus on those when considering unitarity bounds.
Actually,
the leading contribution to the $+0$ helicity amplitude is proportional to
$\gwa1\gwb1\sqrt{s/v^2}$.
If $\gwa1$,
which is constrained via the $00$ mode,
does not deviate grossly from its SM value,
the $+0$ mode leads to a bound on $\gwb1$ which has the same $s$ dependence but is weaker than the constraint that we get from the $+-$ amplitude.
\begin{table}[h]
    \centering
\begin{tabular}{|c|c|c|c|c|c|c}
    \hline
    helicity configuration & $++$ & $+-$ & 00 & $+0$\\
    \hline
    $s$-channel & $\frac{1}{2}\kappa_5\gwb1$ & 0 & $\frac{1}{2}\kappa_5\gwa1$ & 0\\
    $t,u$-channel & $2\gwb1^2$ & $\mathcal{O}(\gwb1^2)$ & $-\gwa1^2$ & 0\\
    contact interaction & $\gwb2$ & 0 & $\gwa2$  & 0\\
    \hline
\end{tabular}
\caption{Leading contribution $m_{0}$ to the helicity amplitudes for
  $VV\to hh$, broken down by type of Feynman diagram.  The notation
  $\mathcal{O}(\gwb2^2)$ indicates that the contribution is non-zero
  but depends on the phase-space point, proportional to the coupling constants $\gwb2^2$.}\label{tab-vvhh}
\end{table}

Angular-momentum conservation directs the choice of
a convenient phase-space basis for the initial state of two vector bosons.
We couple helicity with orbital angular momentum to total angular momentum $j$,
i.e.,
adopt the Wigner D-matrix formalism~(cf.\
Appendix~\ref{App:pwu-further}) in analogy with the formalism
developed for quasi-elastic processes~\cite{Brass:2018hfw}.

We thus derive individual bounds for amplitude coefficients $b_j(h_1h_2)$,
\begin{align}
    b_j(h_1h_2)\le\frac{1}{4},
    \qquad\text{where $h_i=+-0$}.
\end{align}
The strongest bounds on the EFT coefficients that we obtain for this process,
are the following ones:
\begin{align}
    b_0(00)=&\frac{s^2}{2^9\pi^2v^4}|\gwa2-\gwa1^2+\frac{1}{2}\kappa_5\gwa1|^2\le\frac{1}{4} \label{hh1}\\
	b_0(++)=&\frac{s^2}{2^9\pi^2v^4}|\gwb2+2\gwb1^2+\frac{1}{2}\kappa_5\gwb1|^2\le\frac{1}{4} \label{hh2}\\
	b_2(+-)=&\frac{s^2}{3\times 2^{10}\pi^2v^4}\gwb1^4\le\frac{1}{4} \label{hh3}
\end{align}
In particular, the $+-$ mode contributes a bound on $\gwb1$,
i.e., the $hW^+_TW^-_T$ interaction,
which is independent of the other EFT parameters.

\subsection{Unitarity Constraints arising from $VV\to hhh$}\label{sec:pwu-hhh}

\begin{figure}[htbp]
  \centering
  \subfigure{
\label{Fig3.sub.1}   \thesubfigure
  \includegraphics[width=0.20\textwidth]{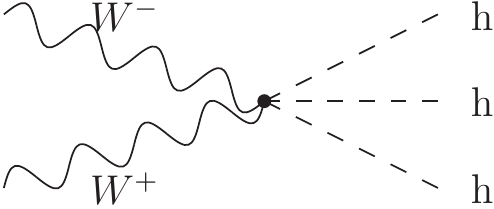}}
  \subfigure{
\label{Fig3.sub.2}   \thesubfigure
  \includegraphics[width=0.20\textwidth]{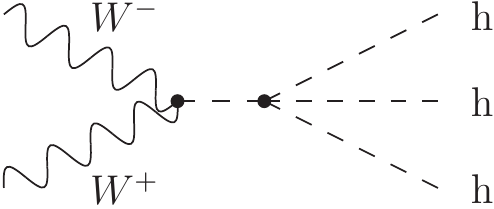}}
  \subfigure{
\label{Fig3.sub.3}   \thesubfigure
  \includegraphics[width=0.20\textwidth]{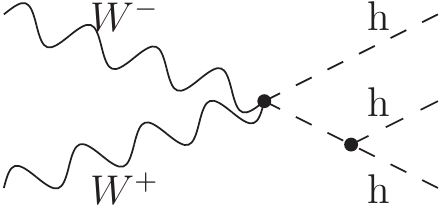}}
  \subfigure{
\label{Fig3.sub.4}   \thesubfigure
  \includegraphics[width=0.20\textwidth]{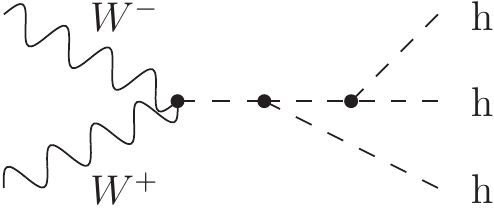}}
  \subfigure{
\label{Fig3.sub.5}   \thesubfigure
  \includegraphics[width=0.20\textwidth]{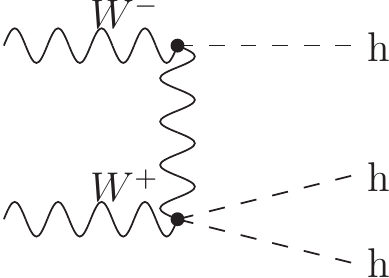}}
  \subfigure{
\label{Fig3.sub.6}   \thesubfigure
  \includegraphics[width=0.20\textwidth]{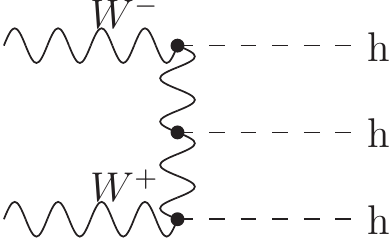}}
  \subfigure{
\label{Fig3.sub.7}   \thesubfigure
  \includegraphics[width=0.20\textwidth]{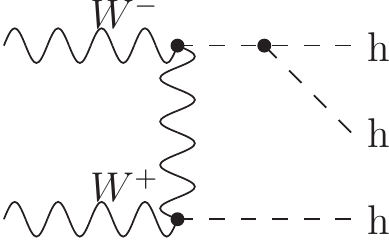}}
  \caption{Seven types of Feynman diagrams which contribute to the processes $W^+ W^- \to h h h$.} \label{fdvv2hhh}
\end{figure}

The helicity amplitudes of the process $W^+  W^- \to hhh$ are associated
with seven distinct types of Feynman diagrams, Fig.~\ref{fdvv2hhh}.
Similar to $W^+ W^-\to hh$,
in the high energy limit,
the amplitude can be expanded as a series in powers of $\sqrt{s/v^2}$,
\begin{align}
    M(W^+W^-\to hhh)=\sum_{i=0}^{+\infty} m_iv^{-1}(\frac{\sqrt s}{v})^{2-i}
\end{align}
We list the leading term $m_{0}$ in Table~\ref{ww2hhh},
for each helicity combination.
Wherever the coefficient is phase-space dependent,
we denote it as $\mathcal{O}(C)$,
where $C$ is a combination of coupling constants.
We find that the $+0$ helicity mode does not contribute to $m_0$,
and that the unitarity bounds resulting from the $m_1$ terms are weaker than the remaining ones,
as long as $\gwa1,\gwa2,\kappa_5$ are not far from their respective SM values.

\begin{center}
\begin{table}
\begin{center}
\begin{tabular}{|c|c|c|c|c|}
\hline
& $++$ & $+-$ & 00 & $+0$ \\
    \hline
    a & $\gwb3$ & 0 & $\gwa3$ & 0\\
    b & $\frac{1}{2}\gwb1\kappa_6$ & 0 & $\frac{1}{2}\gwa1\kappa_6$ & 0\\
    c & $\frac{3}{2}\gwb2\kappa_5$ & 0 & $\frac{3}{2}\gwa2\kappa_5$ & 0\\
    d & $\gwb1\kappa_5^2$ & 0 & $\gwa1\kappa_5^2$ & 0\\
    e & $6\gwb1\gwb2$ & $\mathcal{O}(\gwb1\gwb2)$ & $-4\gwa1\gwa2$ & 0\\
    f & $\mathcal{O}(\gwb1^3)$ & $\mathcal{O}(\gwb1^3)$ & $4\gwa1^3$ & 0\\
    g & $3\gwb1^2\kappa_5$ & $\mathcal{O}(\gwb1^2\kappa_5)$ & $-2\gwa1^2\kappa_5$ & 0\\
    \hline
\end{tabular}
\end{center}
\caption{Leading contribution $m_{0}$ to the helicity amplitudes in
  the high-energy limit for $VV\to hhh$, broken down by type of
  Feynman diagram. \label{ww2hhh}} 
\end{table}
\end{center}

Since this is also an inelastic channel,
we obtain unitarity bounds on the $b$-coefficients defined in Eq.~\eqref{eq:pwu-bexp},
\begin{equation}
	b_{A}^{W^+W^-\to hhh}(\hat s)\le\frac{1}{4}.
\end{equation}
Note that the $b$-coefficients are independent of the phase-space parameterisation and of the basis functions for the triple-Higgs system;
only the phase-space parameterisation and the basis functions for the $W$-boson pair do matter.
As discussed in Appendix \ref{App:pwu-further},
after we choose the Wigner D-matrix as our basis for the $W^{+} W^{-}$ state,
the $b$-coefficients are diagonal and can be denoted as $b_j(h_1h_2)$, where $j$ represents the total angular momentum,
and $h_i=+,-,0$ are the helicities of the two $W$ bosons.
We calculate the (reduced) $b$-coefficients directly according to Eq. \eqref{eq:pwu-bexp}.
Although the result is independent of the phase-space parameterisation for the triple-Higgs system,
an explicit expression is required for phase-space integration;
we adopt the form given in App.~\ref{App:pwu-rcm}.
We give the results for the three helicity modes below:

\begin{enumerate}
	\item For the $00$ helicity mode, the amplitude is constant in
          phase space.  The corresponding bound becomes
		\begin{equation}
			\begin{split}
				b_0(00)=\frac{s^3}{3\times 2^{14}\pi^4 v^6}|&\gwa3+\frac{1}{2}\gwa1\kappa_6+\frac{3}{2}\gwa2\kappa_5+\gwa1\kappa_5^2\\
															&-4\gwa1\gwa2+4\gwa1^3-2\gwa1^2\kappa_5|^2\le\frac{1}{4}
			\end{split}
			\label{3h00}
		\end{equation}

	\item For the $++$ helicity mode,
		the type-f contribution is phase-space dependent,
		and it yields a non-zero $b_j$ for $j>0$.
		However,
		we checked that the dependence is of minor importance,
		and the bounds from $b_j\le\frac{1}{4}$ with $j>0$ turn out to be much weaker than the bounds from $W^{+}W^{-}\to h h$.
		Therefore,
		we only quote $b_0$ here,
		\begin{equation}
			\begin{split}
				b_0(++)=\frac{s^3}{3\times 2^{14}\pi^4 v^6}(|&\gwb3+\frac{1}{2}\gwb1\kappa_6+\frac{3}{2}\gwb2\kappa_5+\gwb1\kappa_5^2\\
															 &+6\gwb1\gwb2+f_1\gwb1^3-3\gwb1^2\kappa_5|^2
				+f_2\gwb1^6)\le\frac{1}{4}
			\end{split}
			\label{3hpp}
		\end{equation}
		with $f_1=7.49994\pm0.00005$ and $f_2=0.0658\pm0.0006$ computed by numerical integration.
		The negligible $f_2$ reflects the fact that the dependence of $\gwb1^3$ on phase-space is small.

	\item For the $+-$ helicity mode, only $b_j$ with $j=2,4,\dots$ are non-zero,
		and among them the largest one is $b_2$, which is given by
		\begin{align}
			b_2(+-)=\frac{s^3}{3\times 2^{14}\sqrt6\pi^4 v^6}\left|\gwb1\gwb2+2\gwb1^3+\frac{1}{2}\gwb1^2\kappa_5\right|^2\le\frac{1}{4}
		\end{align}
\end{enumerate}

\section{Multi-Higgs production in VBF processes in the EFT approach}
\label{Sec:hhh-EFT}

In this section, we consider the phenomenology of multi-Higgs
production in VBF together with basic strategies to isolate the signals at a
high-energy hadron collider.  The impact of anomalous couplings,
introduced via the EFT Lagrangian, depends on the
final-state kinematics, and has to be understood in order to evaluate
the physical implications of assigning unitarity bounds to the
EFT parameter set.

We have computed the cross sections for the processes $pp\to hhjj$ and
$pp\to hhhjj$ including the full dependence on the higher-dimensional
operator coefficients, 
represented by the free parameters of the phenomenological
Lagrangian~\eqref{eft}.
To enhance the contribution of the VBF sub-process,
we apply standard VBF cuts,
as listed in Table~\ref{vbfcuts}.
We display results for the 14 TeV LHC,
for a $pp$ collider with 27 TeV c.m.\ energy,
and for a 100~TeV $pp$ collider.

\begin{figure}[hbt]
	\centering
	\includegraphics[width=0.3\textwidth]{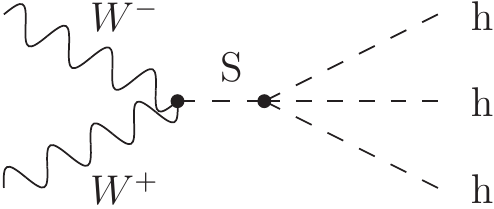}
	\caption{Triple-Higgs production diagram with a five-point vertex $WWhhh$
	effectively generated by an auxiliary field S.}
	\label{fig:hhh-eft-aux}
\end{figure}

For the numerical calculations,
we use the automatic Monte-Carlo integration and simulation packages
\whizard~2.3~\cite{Kilian:2007gr} and \mgamc~\cite{Alwall:2014hca},
where we have implemented the effective Lagrangian~\eqref{eft}.
For \mgamc~we have constructed an appropriate UFO model file.\footnote{The
current version 2.8.1 of \whizard\ does support this UFO standard.  The
calculations for the current paper relies on an earlier version which
did not support the five-point vertices in the EFT
contributions to $VV\to hhh$.}  For \whizard,
we introduced an auxiliary field $S$ with a Lagrangian
\begin{eqnarray}
	\mathcal{L}_S &=& \frac{1}{2}(\partial_\mu S)^2 - \frac{1}{2}M^2S^2 - g_{Shhh}(\partial^2 S)h^3
	+ \gwa3\frac{2m_W^2}{v^3}SW^\mu W_\mu - \frac{\gwb3}{v^3}SW^{\mu\nu} W_{\mu\nu},
\end{eqnarray}
Choosing $M=0$ and $g_{Shhh}=-1$ and restricting the calculation to the triple-Higgs production process,
$\gwa3$ and $\gwb3$ become equivalent to the parameters in our convention,
and the resulting amplitude expression is identical to the one that follows from using~\eqref{eft} directly,
cf.~Fig.~\ref{fig:hhh-eft-aux}.
The anomalous $ZZhhh$ vertex is implemented in a similar way.

We have cross-checked numerical results from
\whizard~and \mgamc,
and found mutual agreement.
As another cross-check,
we have validated selected results against the package
VBFNLO~\cite{Baglio:2014uba,Arnold:2011wj}, again
with good agreement.

For the pure SM,
we obtain the cross sections after VBF cuts as listed in Table~\ref{vbfxs}.
All numerical results are computed at leading order in the strong and electroweak perturbative expansions.

\begin{figure}
	\centering
	\subfigure[14 TeV]{
		\label{etaj014}
	\includegraphics[width=0.4\textwidth]{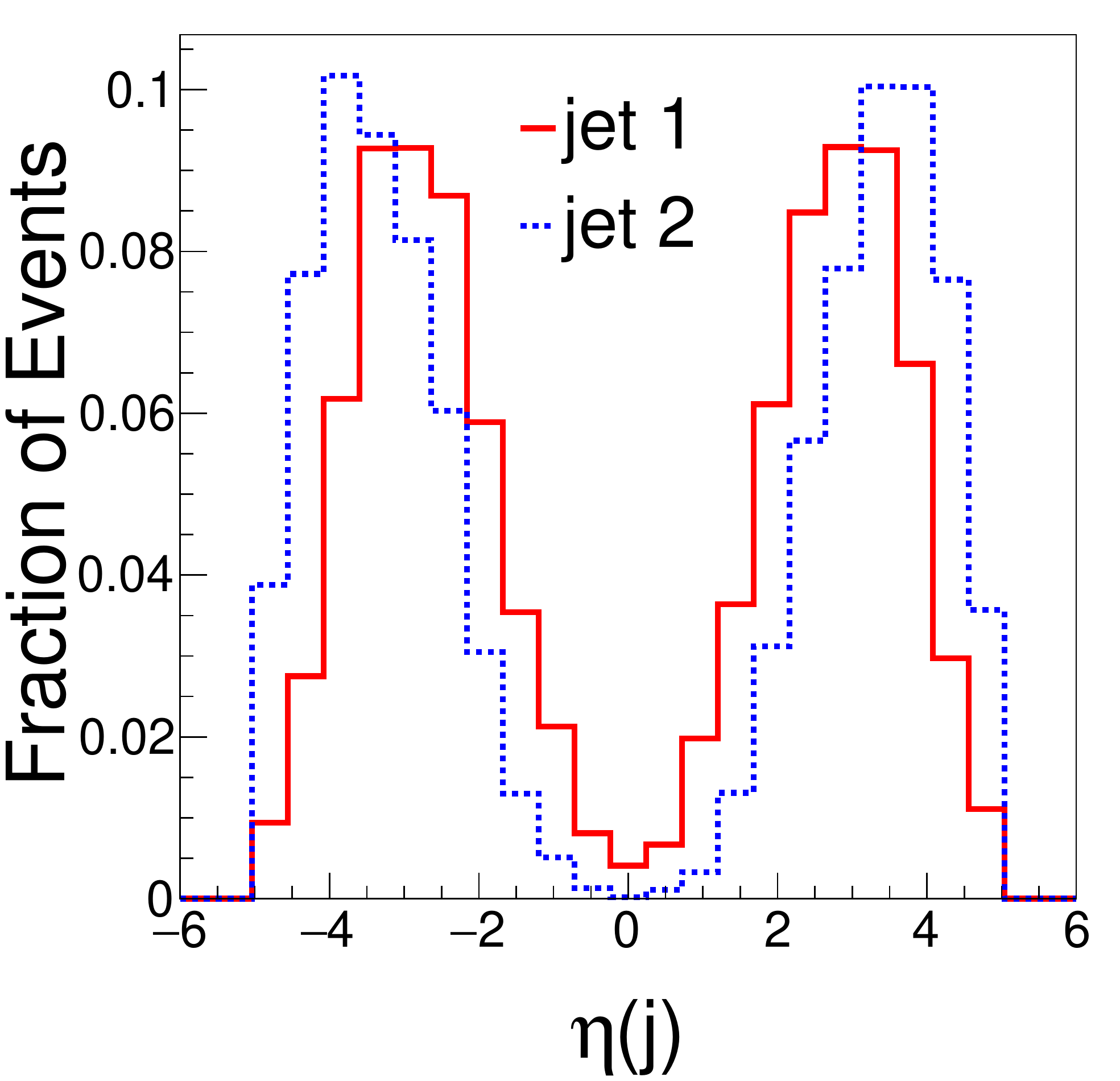}}
	\subfigure[100 TeV]{
		\label{etaj100}
	\includegraphics[width=0.4\textwidth]{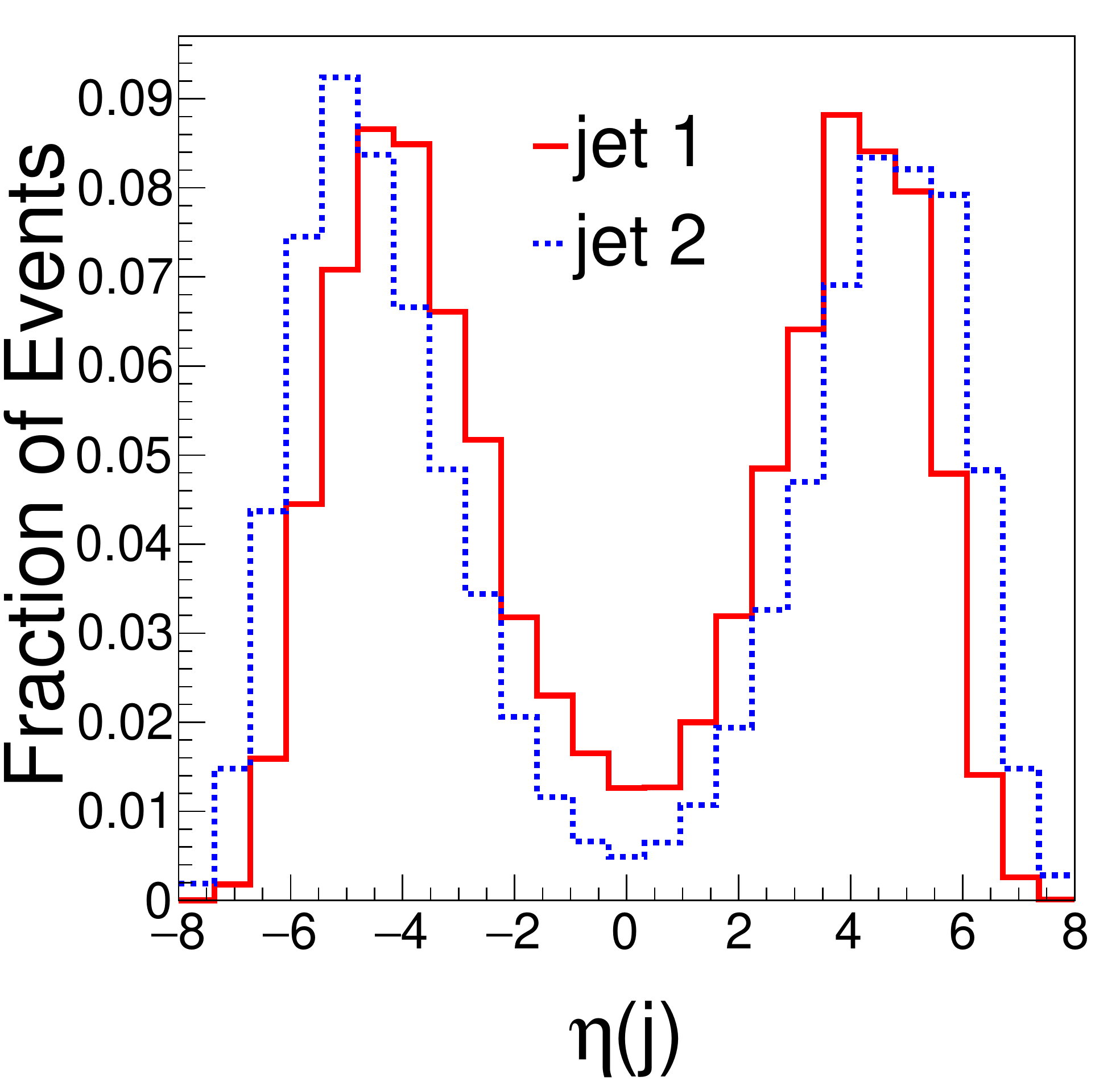}}
	\caption{Rapidity distribution ($\eta$) of the forward tagging jets at (a)
		14 TeV and (b) 100 TeV.  Jet 1 (2) labels the harder (softer) jet,
	respectively.}\label{fig_etaj}
\end{figure}

The VBF cuts in Table~\ref{vbfcuts} force the remnant jets to a forward/backward configuration,
with high energy and momentum,
as it is expected from $q\to Wq'$ splitting in the VBF signal region.
We require $p_T(j)>20\;\mathrm{GeV}$ for 14 and 27~TeV,
and $30\;\mathrm{GeV}$ for
100~TeV,
respectively.
Regarding the transition from LHC kinematics to a 100~TeV collider,
our numerical results demonstrate that the forward jets can acquire significantly larger rapidity than at lower energy~(Fig.~\ref{fig_etaj}).
Therefore, we assume a better rapidity coverage for the detector at 100~TeV and have adapted our cuts in Table~\ref{vbfcuts} accordingly.

\begin{table}[ht]
	\centering
	\begin{tabular}{|c|c|c|c|}
		\hline
		Cuts           &  $\sqrt{s} = 14$ TeV    &  $\sqrt{s} = 27$ TeV       &  $\sqrt{s}=100$ TeV \\ 
		\hline
		$P_t(j)$       &  $>20$ GeV  &  $>20$ GeV  &     $> 30$ GeV \\
		\hline
		$\Delta R (j,j)$      &  $>0.8$   &  $>0.8$               &    $> 0.8$ \\
		\hline
		$|\eta(j)|$     &  $<5.0$   &  $<5.0$            &   $<8.0$ \\ 
		\hline
		$\Delta \eta(j,j)$ & $>3.6$ & $>3.6$  &  $> 4.0$ \\
		\hline
		$M(j,j)$ & $>500$ GeV  & $>500$ GeV &  $>800$ GeV \\
		\hline
	\end{tabular}
	\caption{Acceptance cuts used for the calculation of VBF (multi-)Higgs
		production in $pp$ collision 
	(VBF cuts), for three different collider energies.\label{vbfcuts}} 
\end{table}

\begin{table}[ht]
	\centering
	\begin{tabular}{|c|c|c|c|}
		\hline
		Process           &  $\sigma(14 \textrm{TeV})$ [fb]  & $\sigma (27 \textrm{TeV})$ [fb]  &  $\sigma (100 \textrm{TeV})$ [fb] \\
		\hline
		$pp\to hjj$       &  $1.64\times{10^3}$   &  $4.87\times{10^3}$&  $2.60\times{10^4}$ \\
		\hline
		$pp\to hhjj$      &  $1.10$               &  $4.32$             &  $41.2$  \\
		\hline
		$pp\to hhhjj$     &  $2.73\times{10^{-4}}$   &     $1.73\times{10^{-3}}$     &  $4.50\times{10^{-2}}$ \\
		\hline
	\end{tabular}
	\caption{SM values for the cross sections of the processes $pp\to hjj$,
		$pp\to hhjj$ and $pp\to hhhjj$ with VBF cuts, at three different
	collider energies.\label{vbfxs}}
\end{table}

\subsection{Higgs pair-production}

The analysis of Higgs pair-production, $VV\to hh$, aims at
a determination of the $VVhh$ and $hhh$ couplings.  Anomalous lower-order
couplings such as $VVh$ would affect the amplitudes but can be
measured in lower-order processes, with considerably higher event
rates.  We 
assume that such measurements yield better precision.  For the
sake of simplicity, we thus fix the $VVh$ couplings to their SM values,
$\gva1=1$ and $\gvb1=0$.  In addition, we assume the
custodial-symmetry relations $\gwa2=\gza2=\gva2$ and
$\gwb2=\gzb2=\gvb2$ whenever contributions of the $Z$ boson are considered.
Furthermore, we introduce the shifts
$\delta \gva2=\gva2-1,\delta\lambda_3=\lambda_3-1$ which multiply
the deviation with respect to the SM, as parameters in our
calculation.

By construction, the tree-level result for the cross section in the
EFT depends linearly and bilinearly on the free parameters, and can be
cast into the form
\begin{align}
	\label{pphh}
	\sigma(pp \to hhjj) &= \sum_{i+j+k\le2}\sigma^{hh}_{ijk}(\delta \gva2)^i \gvb2^j (\delta \lambda_3)^k\\
	&=\sigma^{hh}_{000} + \sigma^{hh}_{100} \delta\gva2 + \sigma^{hh}_{200} (\delta\gva2)^2 + 
	\sigma^{hh}_{010} \gvb2  + \sigma^{hh}_{110} \delta \gva2\gvb2  + \sigma^{hh}_{020} \gvb2^2  \nonumber \\
	   &\quad +  \sigma^{hh}_{001} \delta\lambda_3
	+ \sigma^{hh}_{101} \delta \gva2 \delta\lambda_3 + \sigma^{hh}_{011} \gvb2 \delta\lambda_3
	+ \sigma^{hh}_{002} (\delta\lambda_3)^2.
\end{align}

In Table~\ref{vbf2heft},
we display the values of the SM cross section
$\sigma^{hh}_{000}$ and of all EFT coefficients
$\sigma^{hh}_{ijk}$, evaluated for the three collider energies of
$14\;\mathrm{TeV}$, $27\;\mathrm{TeV}$, and $100\;\mathrm{TeV}$. 
As discussed in Sec.~\ref{sec:pwu-hh},
the amplitude contributions proportional to $\delta\gva2$ and $\gvb2$
grow linearly with~$s$.  For any nonzero values of $\delta\gva2$ or
$\gvb2$, the resulting contributions to the cross section will break
perturbative unitarity 
as $s$ increases.   If such terms are present, we should expect a
rapidly growing enhancement in the Higgs-pair invariant mass distribution
$m_{hh}$, eventually dampened 
by non-perturbative rescattering corrections.  We will discuss this
property below.

The amplitude contribution proportional to
$\delta\lambda_3$ does not grow with energy relative to the SM cross
section.  The dependence on $\delta\lambda_3$ is dominated by the
low-energy region, and constraining $\delta\lambda_3$ is much more
challenging than constraining $\delta\gva2$ and $\gvb2$.  Conversely,
the uncertainty on $\delta\lambda_3$ is not essential for studying the
dependence on $\delta\gva2$ and $\gvb2$.  In the following, we
focus on the effects induced by $\delta\gva2$ and $\gvb2$.

	\begin{table}
		\begin{center}
			\begin{tabular}{|c|c|c|c|c|c|c|c|c|c|c|}
				\hline
				[fb] & $\sigma^{hh}_{\textrm{SM}}=\sigma^{hh}_{000}$ & $\sigma^{hh}_{100}$ & $\sigma^{hh}_{200}$ & $\sigma^{hh}_{010}$ & $\sigma^{hh}_{110}$ & $\sigma^{hh}_{020}$ & $\sigma^{hh}_{001}$ & $\sigma^{hh}_{101}$ & $\sigma^{hh}_{011}$ & $\sigma^{hh}_{002}$ \\
				\hline
				14 TeV & 1.10 & -3.51 & 11.0 & 1.31 & 1.7 & 87.8 & -0.81 & 3.6 & 0.35 & 0.66\\
				\hline
				27 TeV & 4.32 & -15.0 & 61.1 & 6.91 & 9.6 & 957 & -2.89 & 14.1 & 1.4 & 2.3\\
				\hline
				100 TeV & 41.2 & -158 & 1302 & 79.2 & 95 & $4.80\times10^4$ & -21.8 & 123 & 11.2 & 16.9 \\
				\hline
			\end{tabular}
		\end{center}
		\caption{Coefficients $\sigma^{hh}_{ijk}$ (in $\mathrm{fb}$) in the
			expression~\eqref{pphh} for VBF $hh$ at three different collider energies.}\label{vbf2heft}
	\end{table}

When adopting the EFT approach for an analysis, there is an underlying
assumption that the included terms are dominant, and higher-order terms
can be dropped.  In a cross-section calculation, the formally leading
term is the interference of the linear EFT contribution with the SM
part, which should be larger than the quadratic EFT contribution.
In the present case, the linear coefficient $\sigma^{hh}_{010}$ is 
much smaller than the naive expectation
$\mathcal{O}(\sqrt{\sigma^{hh}_{000}\sigma^{hh}_{020}})$.  The
interference effect is suppressed, and the squared term
$\sigma^{hh}_{020}$ is more important even for small values of $\gvb2$.

In fact, there is a cancellation among regions in phase space caused
by a helicity mismatch between the SM and the EFT terms.  The parameter
$\gvb2$ multiplies the coupling between the transverse vector bosons
and Higgs bosons, while $\delta \gva2$ and the SM contribution describe
the coupling between longitudinal vector bosons and
Higgs bosons.  Consequently, different azimuthal distributions reflect
the different polarisations of intermediate vector bosons.  The finite
vector-boson masses induce some level of mixing, so the cancellation
is not exact.

\begin{figure}[ht]
	\centering
	\includegraphics[width=0.49\textwidth]{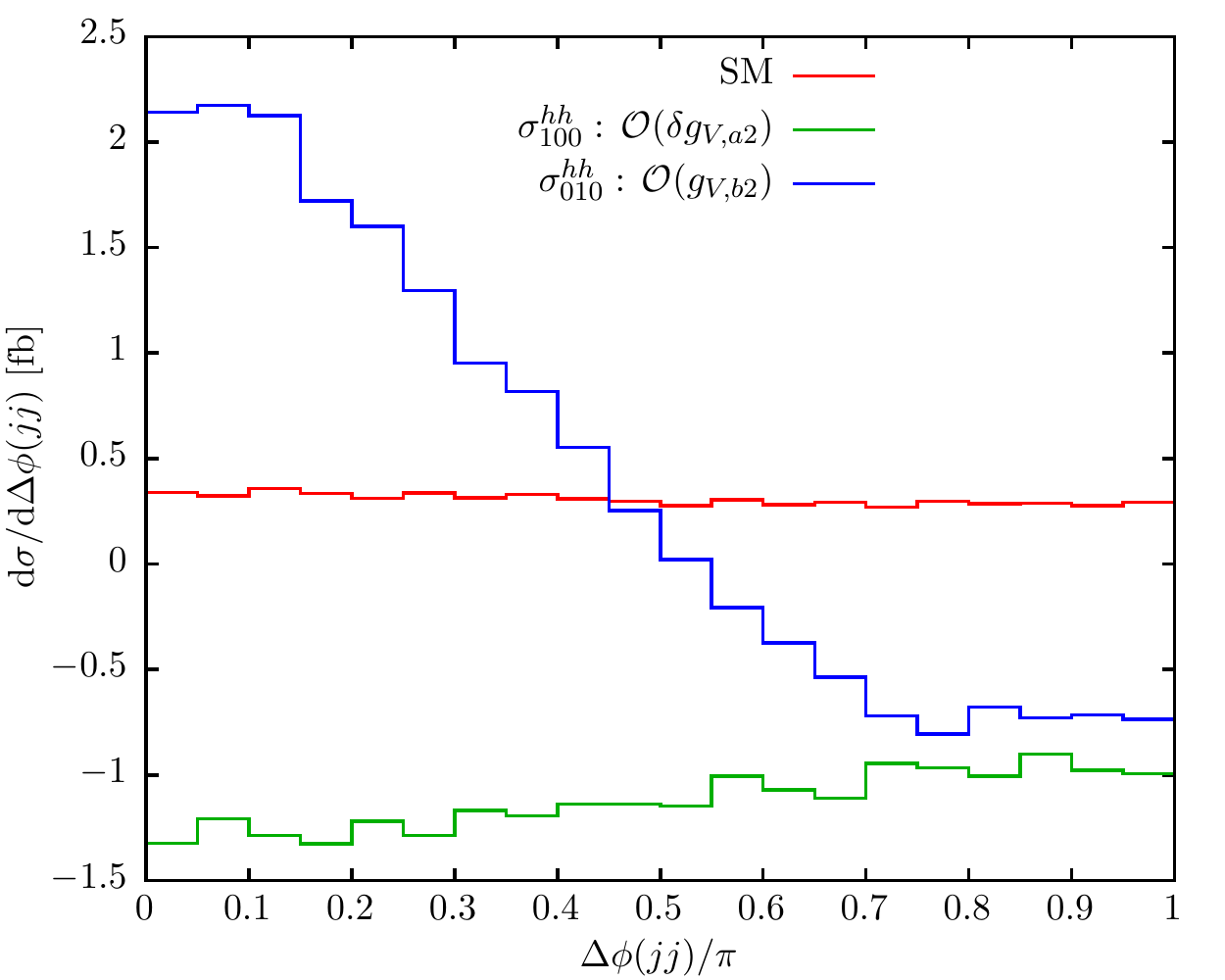}
	\includegraphics[width=0.49\textwidth]{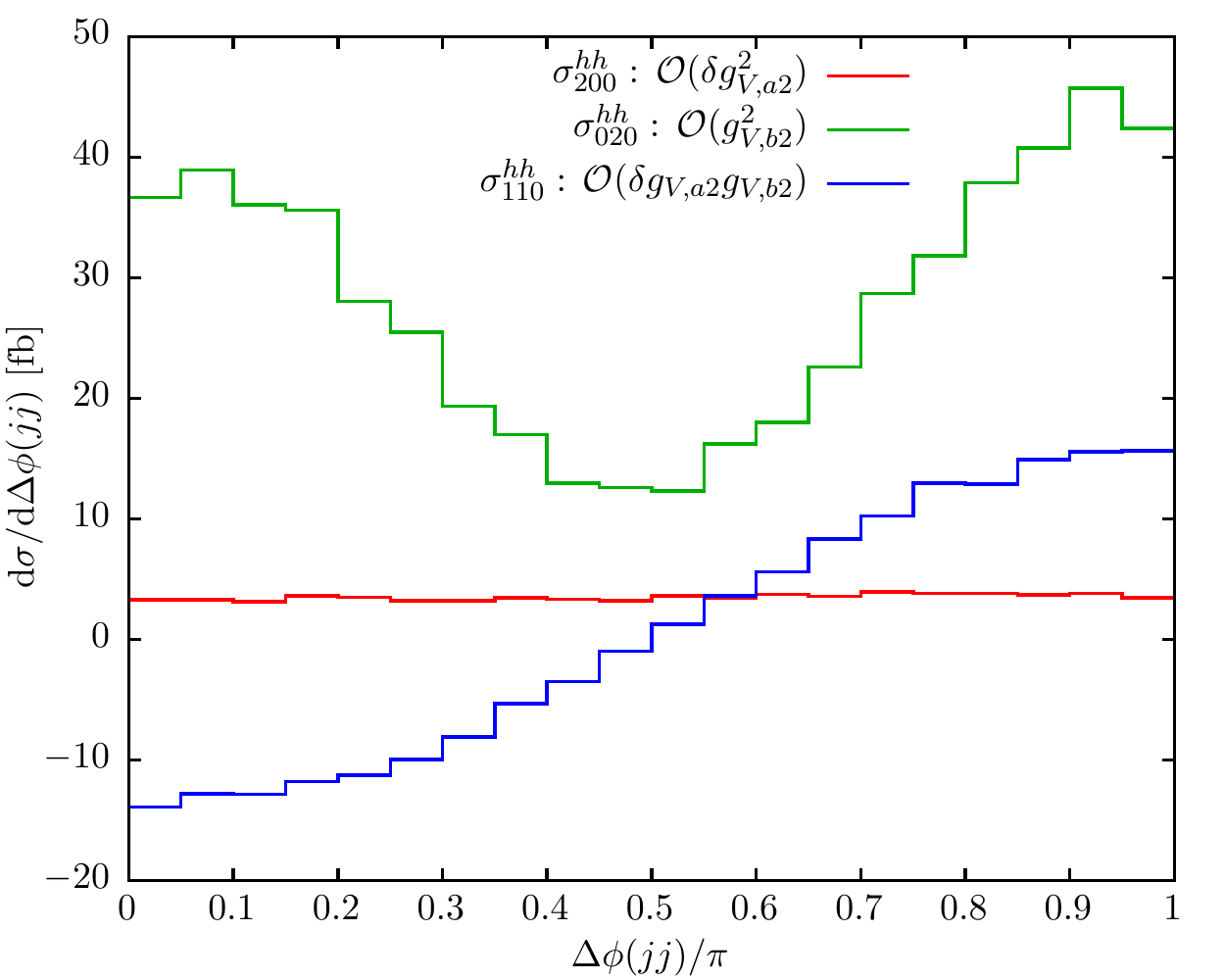}
	\caption{Distribution of $\Delta\phi(jj)$ for various
          modifications of VBF $hh$ production in the EFT, at the 14
          TeV LHC.}\label{fig:2h-dphijj} 
\end{figure}

To illustrate this property, in Fig.~\ref{fig:2h-dphijj} we show the
distribution of $\Delta\phi(jj)$ at the 14 TeV LHC,  
i.e., the azimuthal-angle correlation of the two VBF jets. 
For the $\sigma^{hh}_{000}$, $\sigma^{hh}_{100}$, and
$\sigma^{hh}_{200}$ terms, the distribution of $\Delta\phi(jj)$ is almost flat.
By contrast, for $\sigma^{hh}_{010}$, $\sigma^{hh}_{110}$, and
$\sigma^{hh}_{020}$, the differential cross section depends
on $\Delta\phi(jj)$ with a sign flip near $\pi/2$. 

We conclude that the interference contribution of $\gvb2$, for
instance, can significantly be enhanced by constructing an appropriate
observable, weighing events by azimuthal distance.
Furthermore, the azimuthal dependence in Fig.~\ref{fig:2h-dphijj} discriminates between the two
different EFT parameters, and should definitely be accounted for in an
analysis.  Since helicity-mixing effects disappear with increasing
energy, the discrimination becomes even clearer for the higher
collider energies of 27 TeV or 100 TeV.

We now turn to the distribution in the observable $m(hh)$, the
invariant mass of the final state as the total c.m.\ energy of the
elementary $VV\to hh$ scattering.  In this observable, effects growing
with energy are manifest, and we should study the behavior of the terms
proportional to $\delta \gva2$ and $\gvb2$.

\begin{figure}[ht]
  \centering
  \includegraphics[width=0.49\textwidth]{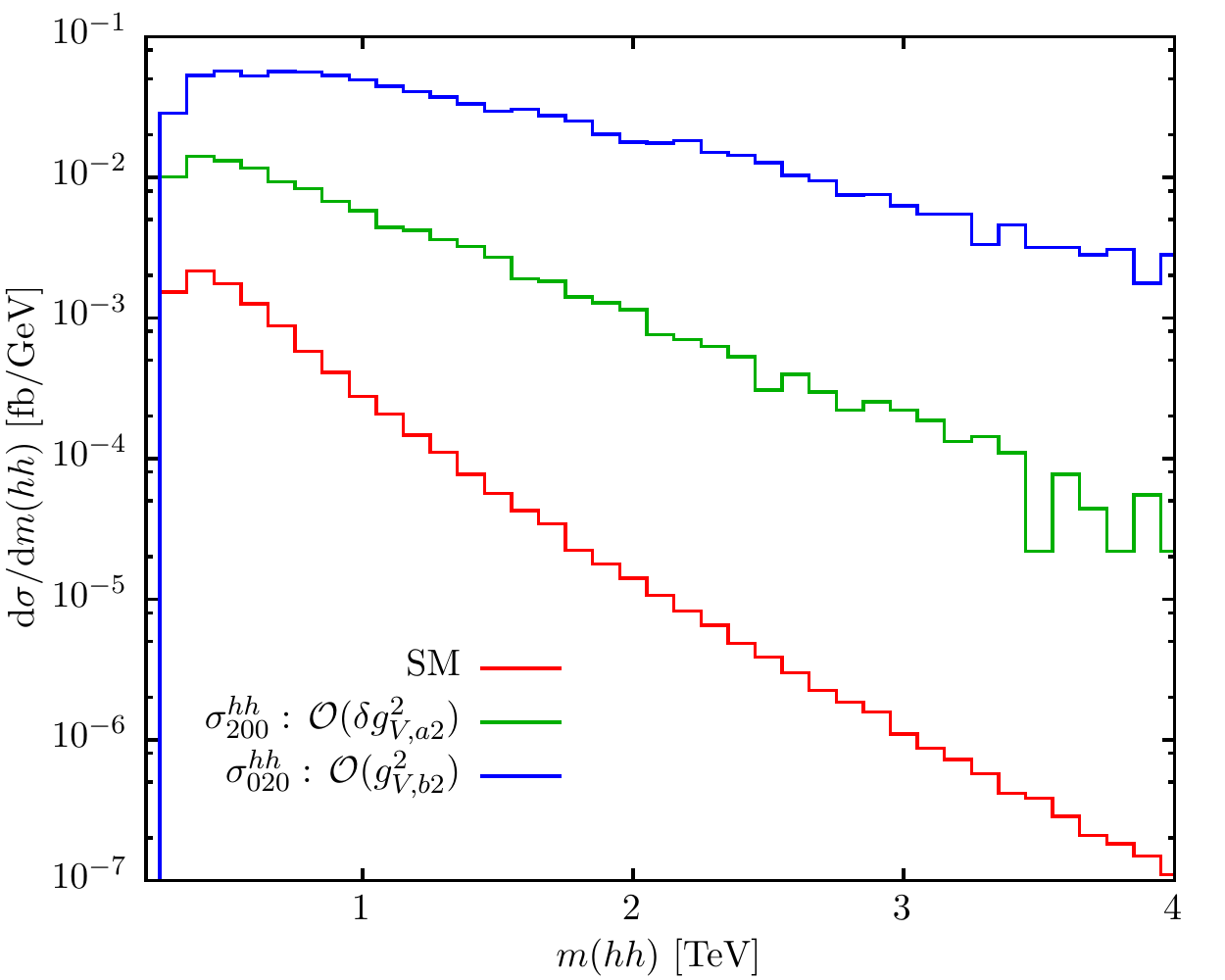}
  \includegraphics[width=0.49\textwidth]{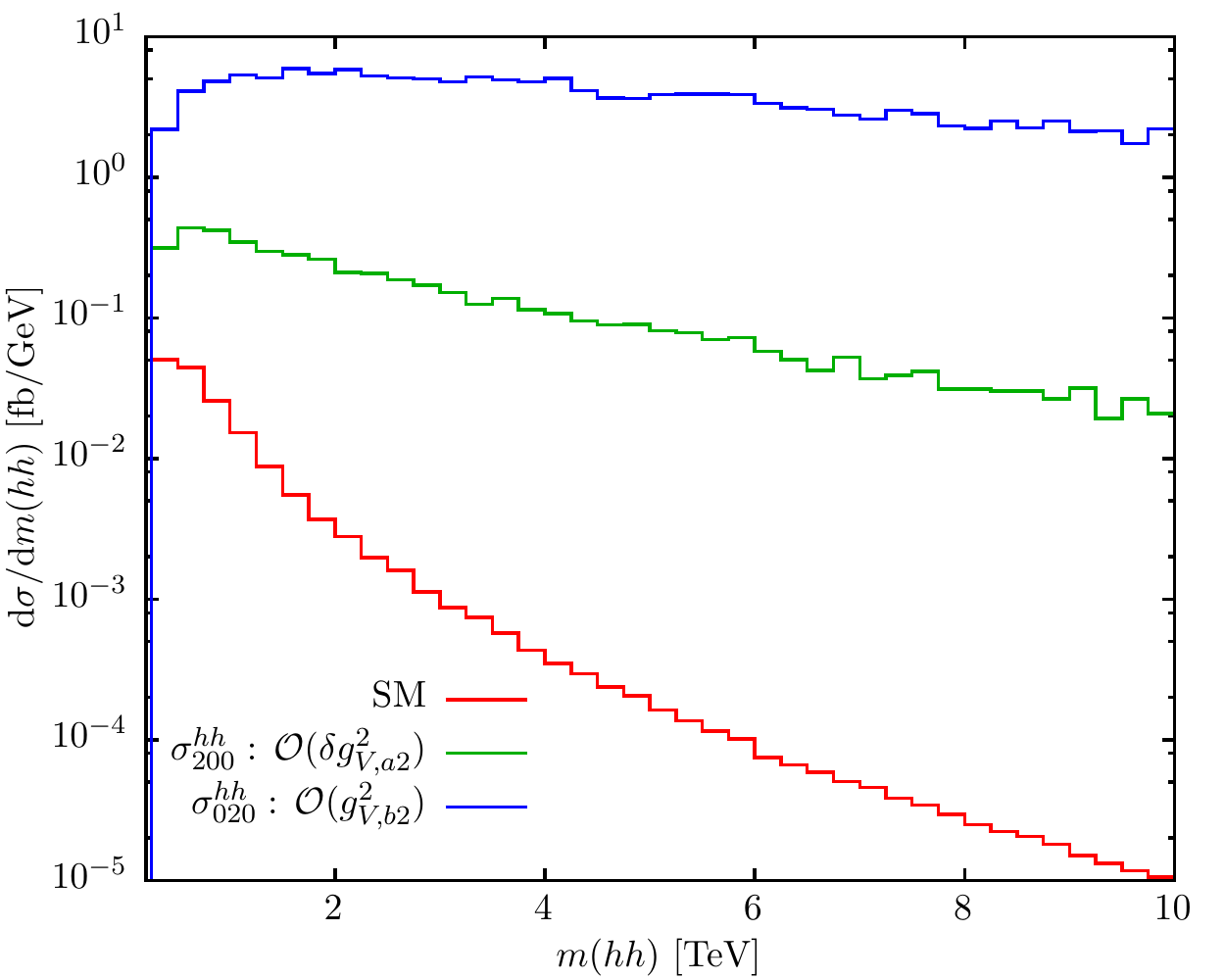}
  \caption{Higgs-pair invariant-mass distribution for the coefficients
    of the
    leading power corrections to VBF $hhh$ production in $pp$
    collisions.  The SM baseline is drawn in red.  We
    display results for 14 TeV (left) and 
    for 100 TeV (right).}\label{fig:2h-invm}
\end{figure}

In Fig.~\ref{fig:2h-invm}, we show the $m(hh)$ distribution of the
$\sigma^{hh}_{200}$ and $\sigma^{hh}_{020}$ terms (green and blue,
respectively) and the SM distribution (red).  As discussed
above, the EFT distributions decrease much slower with increasing
$m(hh)$, compared to the SM curve.  $\delta \gva2$ and $\gvb2$ lead to
similar $VV\to hh$ sub-amplitudes where the leading contribution for
$\delta \gva2$ corresponds to longitudinal vector bosons, while for
$\gvb2$ it corresponds to transverse vector bosons.  The dominant
contribution to the complete off-shell process originates from
quasi-on-shell collinear splitting $q\to Vq^{\prime}$. The
emission of longitudinal vector boson from a quark is kinematically
suppressed in relation to transverse vector
bosons~\cite{Dawson:1984gx}.  Consequently, for comparable values of
the EFT coefficients $\gvb2$ and $\delta \gva2$, the contribution from
$\gvb2$ is much larger than $\delta \gva2$ and exhibits a slower
decrease as a function of $m(hh)$.

The power-like increasing EFT distributions will leave the
perturbative regime beyond some point in the high-energy region that
depends on the parameters,
invalidating the EFT as a systematic approximation to the (unknown)
true amplitudes.  Given numerical values for the EFT parameters,
perturbative unitarity thus provides an upper bound on the range of
validity of the EFT.  We may require that the EFT should be valid up
to some scale $Q$, in the range $m(hh)\le Q$, The perturbative
unitarity constraints derived in Sec. \ref{sec:pwu-hh} then provide
bounds on the values of $\delta \gva2$ and $\gvb2$.
Restricting $\delta \gva2$ and $\gvb2$ to vary only within this
parameter range, we can compute the differential cross section
$\frac{\ud \sigma}{\ud m(hh)}$ in the region $m(hh)\le Q$, and derive
the maximally allowed deviation at any point within that region.

Explicitly, we model the deviation from the SM by linear and quadratic
terms in the EFT parameters, dropping any higher-dimensional
contributions.  Allowing a single parameter, say, $\delta \gva2$, to
deviate from zero and imposing the unitarity bound for the entire
$m(hh)$ distribution below the scale~$Q$, we obtain the formal
condition
\begin{align}
  \label{eq:ds-dmhh}
	\max_{\delta \gva2}\left.\frac{\ud \sigma^{\textrm{NP}}}{\ud m(hh)}\right|_{m(hh)=Q}
		=
		&\max_{\delta \gva2}\left|\delta \gva2\left(\left.\frac{\ud \sigma^{hh}_{100}}{\ud m(hh)}\right|_{m(hh)=Q}
		+\delta \gva2\left.\frac{\ud\sigma^{hh}_{200}}{\ud m(hh)}\right|_{m(hh)=Q}\right)\right|.
\end{align}
Here, we can take the unitarity bounds from
Sec. \ref{sec:pwu-hh},
\begin{align}
  \label{eq:dgva2-bounds}
  -\frac{8\sqrt{2}\pi v^2}{Q^2}\le \delta \gva2\le\frac{8\sqrt{2}\pi v^2}{Q^2}.
\end{align}
Analogous results hold for $\gvb2$ if varied on its own.  Regarding
the general case of both $VVhh$ couplings non-zero, we recall that
their interference is small and asymptotically suppressed, as
discussed above.  In effect, the overall deviation of the distribution
is approximately the sum of the maximally allowed deviations from
$\delta \gva2$ and $\gvb2$ being varied separately.

\begin{figure}[ht]
  \centering
  \includegraphics[width=0.49\textwidth]{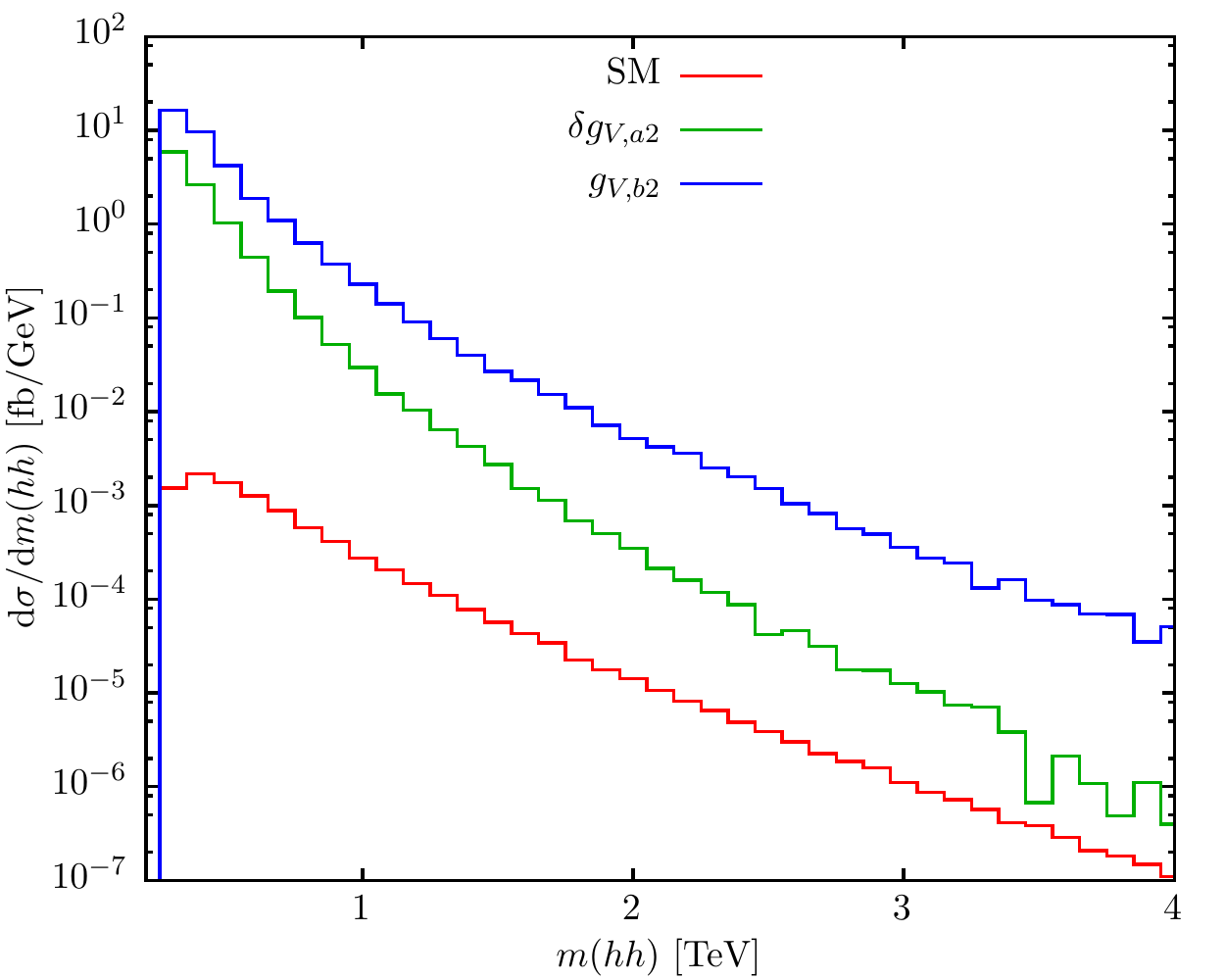}
  \includegraphics[width=0.49\textwidth]{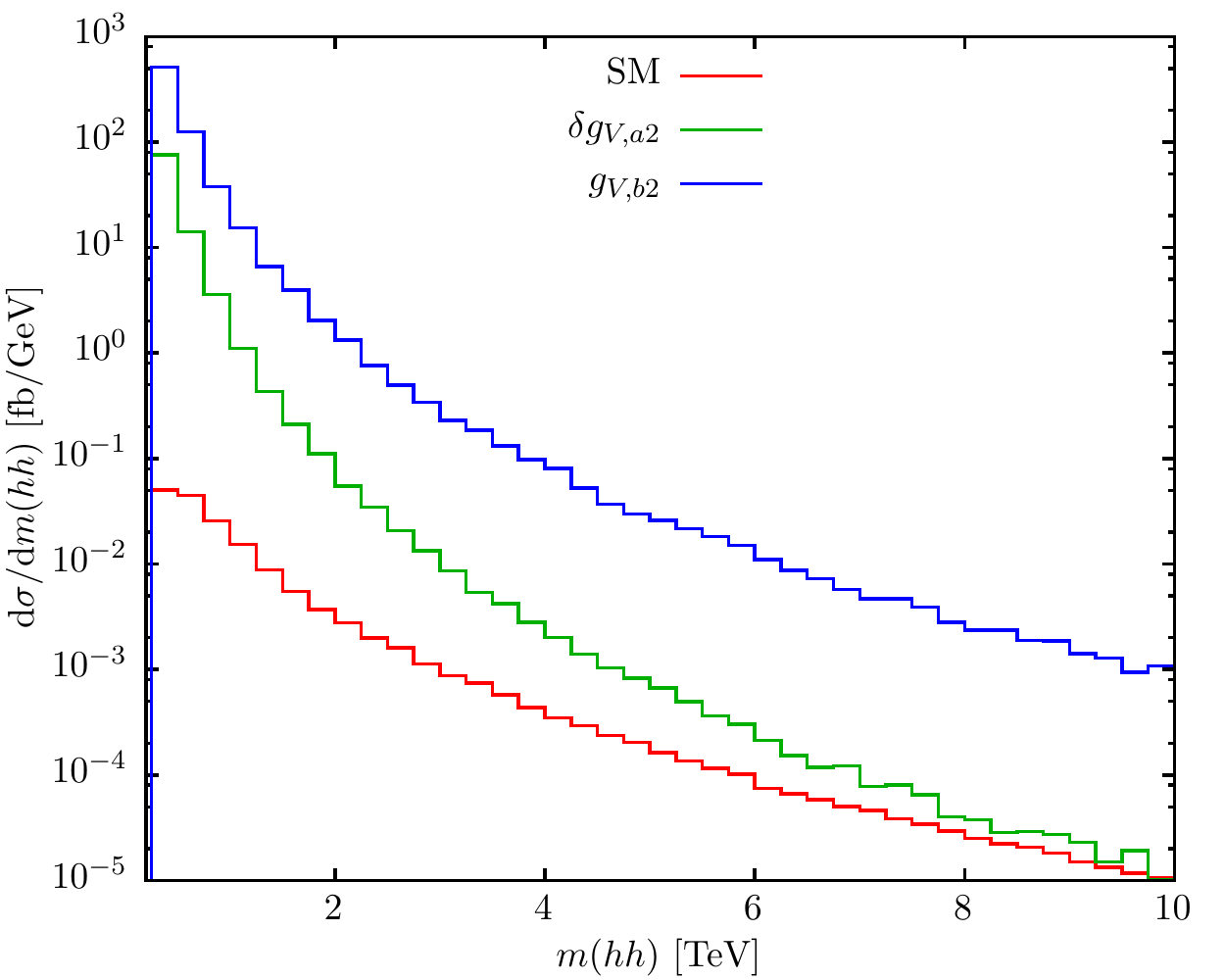}
  \caption{Maximally allowed enhancement of the differential cross
    section of VBF $hh$ production in $pp$ collisions, as a function of
    the Higgs-pair 
    invariant mass.   We display results for 14 TeV (left) and
    for 100 TeV (right), identifying $m(hh)$ with the cutoff scale $Q$.} 
		\label{fig:2h-maxm}
\end{figure}

In Fig.~\ref{fig:2h-maxm}, we display the results following from that
argument, namely the curves of maximal event rate.
Taking the conditions~(\ref{eq:ds-dmhh}, \ref{eq:dgva2-bounds})
literally, we obtain an upper bound on the $m(hh)$
distribution for all energies, where we identify $m(hh)$ with $Q$.  We
observe that there is more freedom for 
enhancement in the direction of~$\delta \gvb2$, which corresponds to
transverse gauge bosons interacting with a Higgs pair, than
for~$\delta \gva2$ which parameterizes the longitudinal coupling.

\subsection{Triple Higgs production}

We now consider the $hhhjj$ final state which contains $VV\to hhh$ as the
relevant sub-amplitude, on the same footing as the $hhjj$ final state
in the previous subsection.  The Feynman diagrams are displayed in
Fig.~\ref{fdvv2hhh}.   In analogy with $VV\to hh$,  we assume that $g_{W,a1},g_{W,a2},\lambda_3$ and $g_{W,b1},g_{W_b2},\kappa_5$ 
have been sufficiently well determined from lower-multiplicity measurements,
single-Higgs and double-Higgs production in particular.   For simplicity
we set these parameters to their SM values $1$ and $0$, respectively. 

Expanding the full dependence on the remaining free parameters, we can
write the cross section as 
\begin{align}
  \label{pphhh}
  \sigma(pp \to hhhjj) &= \sum_{i+j+k\le2}\sigma^{hhh}_{ijk}\gva3^i \gvb3^j (\delta\lambda_4)^k\\
  &=\sigma^{hhh}_{000} + \sigma^{hhh}_{100} \gva3 + \sigma^{hhh}_{200} (\gva2)^2 + 
     \sigma^{hhh}_{010} \gvb3  + \sigma^{hhh}_{110}  \gva3\gvb3 + \sigma^{hhh}_{020} \gvb3^2 \nonumber \\
  &\quad+ \sigma^{hhh}_{001} \delta\lambda_4
       + \sigma^{hhh}_{101}  \gva3 \delta\lambda_4 + \sigma^{hhh}_{011} \gvb3 \delta\lambda_4
       + \sigma^{hhh}_{002} (\delta\lambda_4)^2,
\end{align}
We have computed the $\sigma^{hhh}_{ijk}$ coefficients by numerical integration using \whizard.  The results are listed in Table~\ref{vbf3heft}.

\begin{table}
		\centering
		\begin{tabular}{|c|c|c|c|c|c|}
			\hline
			[fb] & $\sigma^{hhh}_{\textrm{SM}}=\sigma^{hhh}_{000}$ & $\sigma^{hhh}_{100}$ & $\sigma^{hhh}_{200}$ & $\sigma^{hhh}_{010}$ & $\sigma^{hhh}_{110}$ \\
			\hline
			14 TeV  & \num{2.792e-4} & \num{-5.21e-4} & 0.146 & \num{2.44e-3} & \num{1.55e-2} \\
			27 TeV  & \num{1.66e-3}  & \num{-1.45e-3}   & 2.30 & \num{2.00e-2} & 0.195         \\
			100 TeV & \num{3.10e-2}  & \num{5.18e-2}    & 495   & 0.398         & \num{11}    \\
			\hline
			[fb] & $\sigma^{hhh}_{020}$ & $\sigma^{hhh}_{001}$ & $\sigma^{hhh}_{101}$ & $\sigma^{hhh}_{011}$ & $\sigma^{hhh}_{002}$ \\
			\hline
			14 TeV  & 3.29         & \num{-8.32e-5} & \num{7.24e-3} & \num{4.7e-4} & \num{2.09e-4}\\
			27 TeV  & 121          & \num{-3.55e-4} & \num{5.28e-2} & \num{2.9e-3} & \num{1.06e-3}\\
			100 TeV & \num{6.67e4} & \num{-3.07e-3} & 1.61         & \num{4.6e-2}   & \num{1.38e-2}\\
			\hline
		\end{tabular}
		\caption{Coefficients $\sigma_{ijk}^{hhh}$ (in $\mathrm{fb}$) in the expression~\eqref{pphhh} for the process VBF $hhh$ at three
	different collider energies.}\label{vbf3heft}
	\end{table}

Only the couplings to vector bosons, $\gva3$ and $\gvb3$,
lead to unitarity violation at high energy in the EFT calculation.  Here,
$\gva3$ couples a pair of longitudinal vector bosons to the Higgs
triplet, while $\gvb3$ multiplies 
the triple-Higgs coupling of transverse vector bosons.   By contrast,
a deviation in the quartic Higgs coupling $\delta\lambda_4$ does not
lead to a power-like enhancement.  Asymptotically, its effect is
subleading compared to $\gva3$ and $\gvb3$, and we expect
significantly more resolution power for the latter parameters.

Of course, the main goal of such a measurement would be to get a
handle on the quartic Higgs self-coupling.  The apparent dominance of
the other EFT couplings makes it even more important to understand
their impact on the process.
\begin{figure}[ht]
  \includegraphics[width=0.49\textwidth]{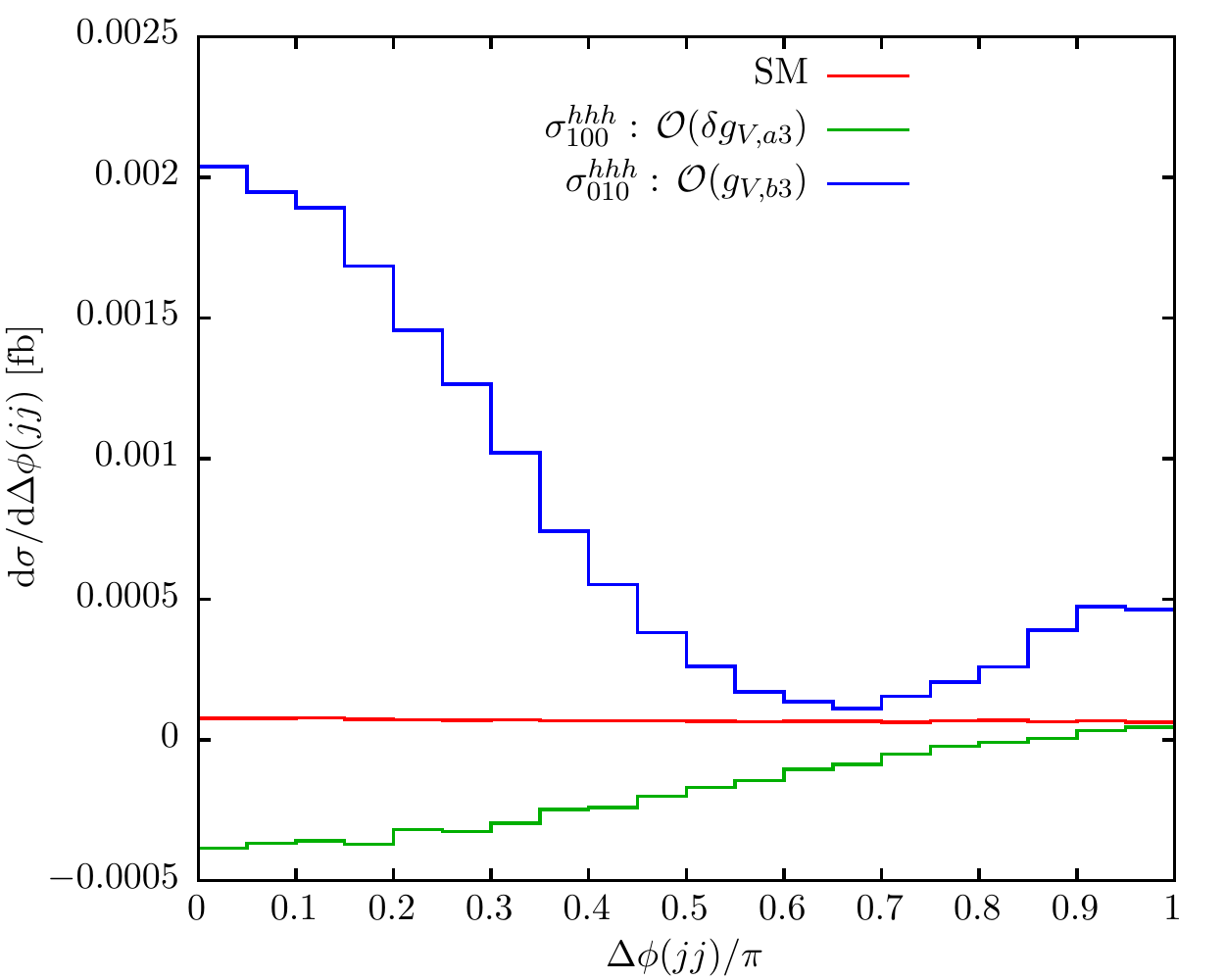}
  \includegraphics[width=0.49\textwidth]{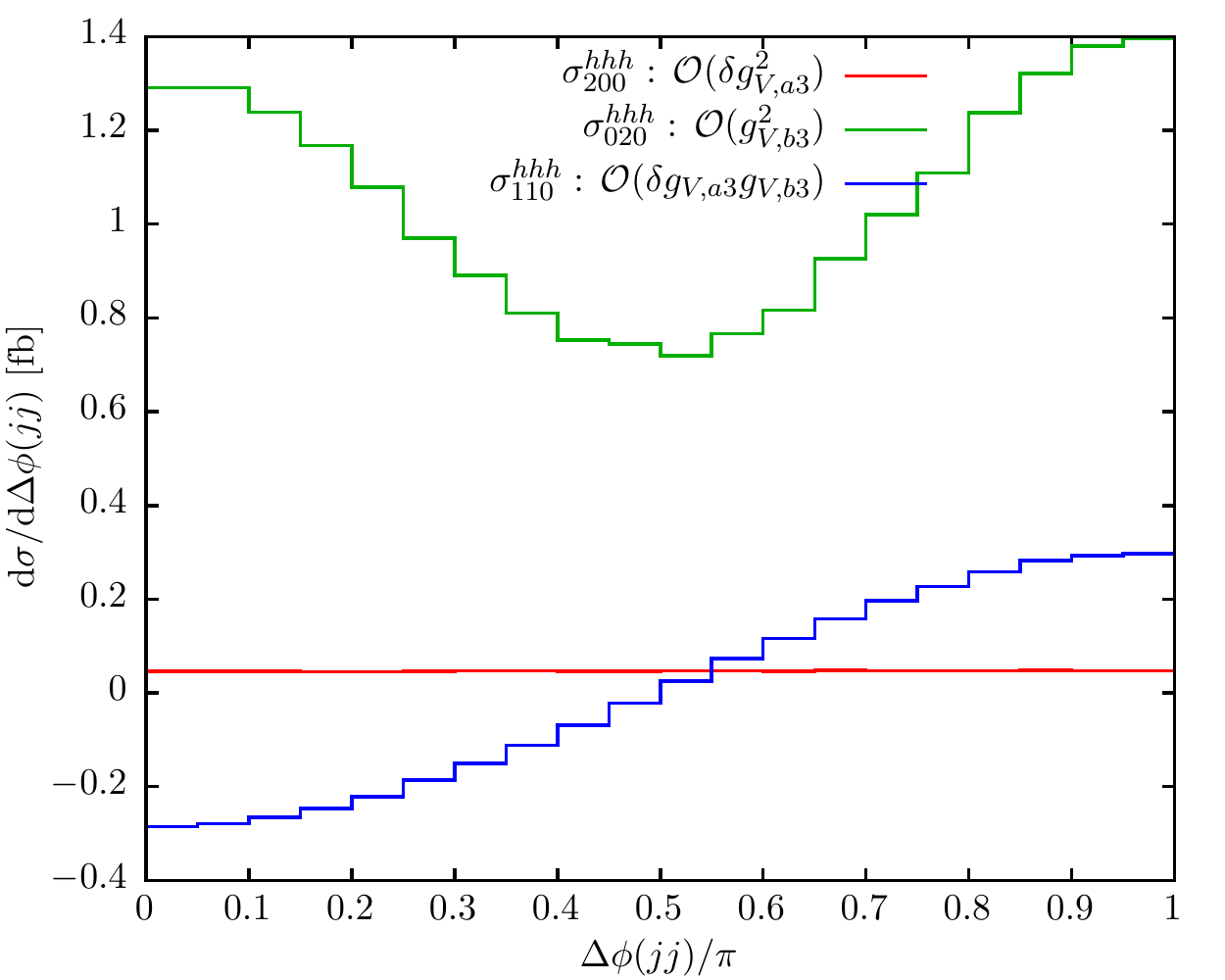}
  \caption{Distribution of $\Delta\phi_{jj}$ for various modifications
    of VBF $hhh$ production, at the 14 TeV LHC.}\label{fig:3h-dphijj}
\end{figure}

In Fig.~\ref{fig:3h-dphijj}, we show the azimuthal $\Delta\phi_{jj}$
distributions for VBF to $hhh$ at the 14 TeV LHC.   For the 27 TeV and
100 TeV colliders, the results are similar.  The
behavior is superficially analogous to the double-Higgs case, taking
finite-mass effects into account.  Apart from the much smaller
absolute values, there also are significant
differences.
The Lorentz structure of the
interference terms is more complicated.  This is reflected in
the  $\sigma^{hhh}_{100}$ and $\sigma^{hhh}_{010}$ distributions.

\begin{figure}[htbp]
  \centering
  \includegraphics[width=0.49\textwidth]{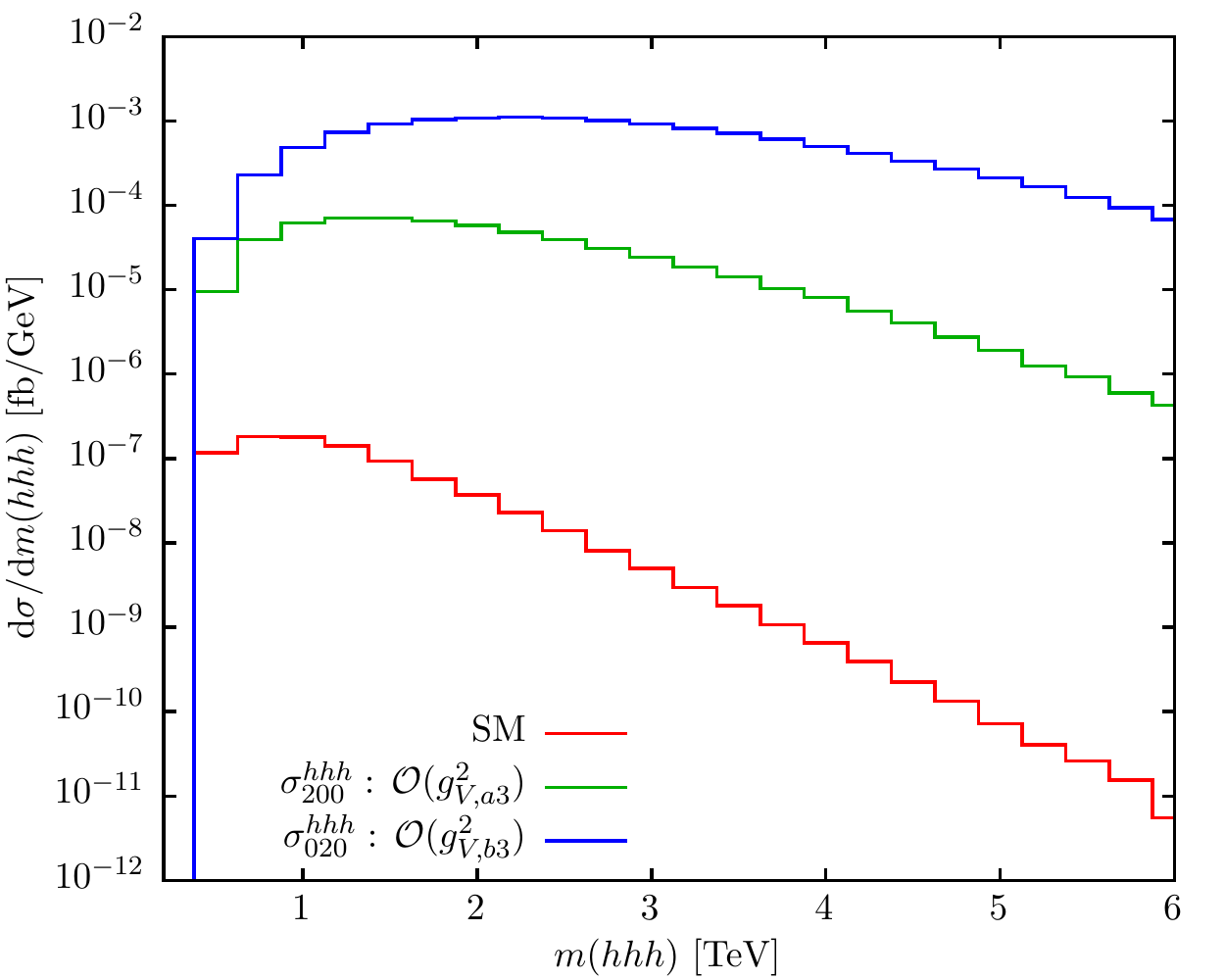}
  \includegraphics[width=0.49\textwidth]{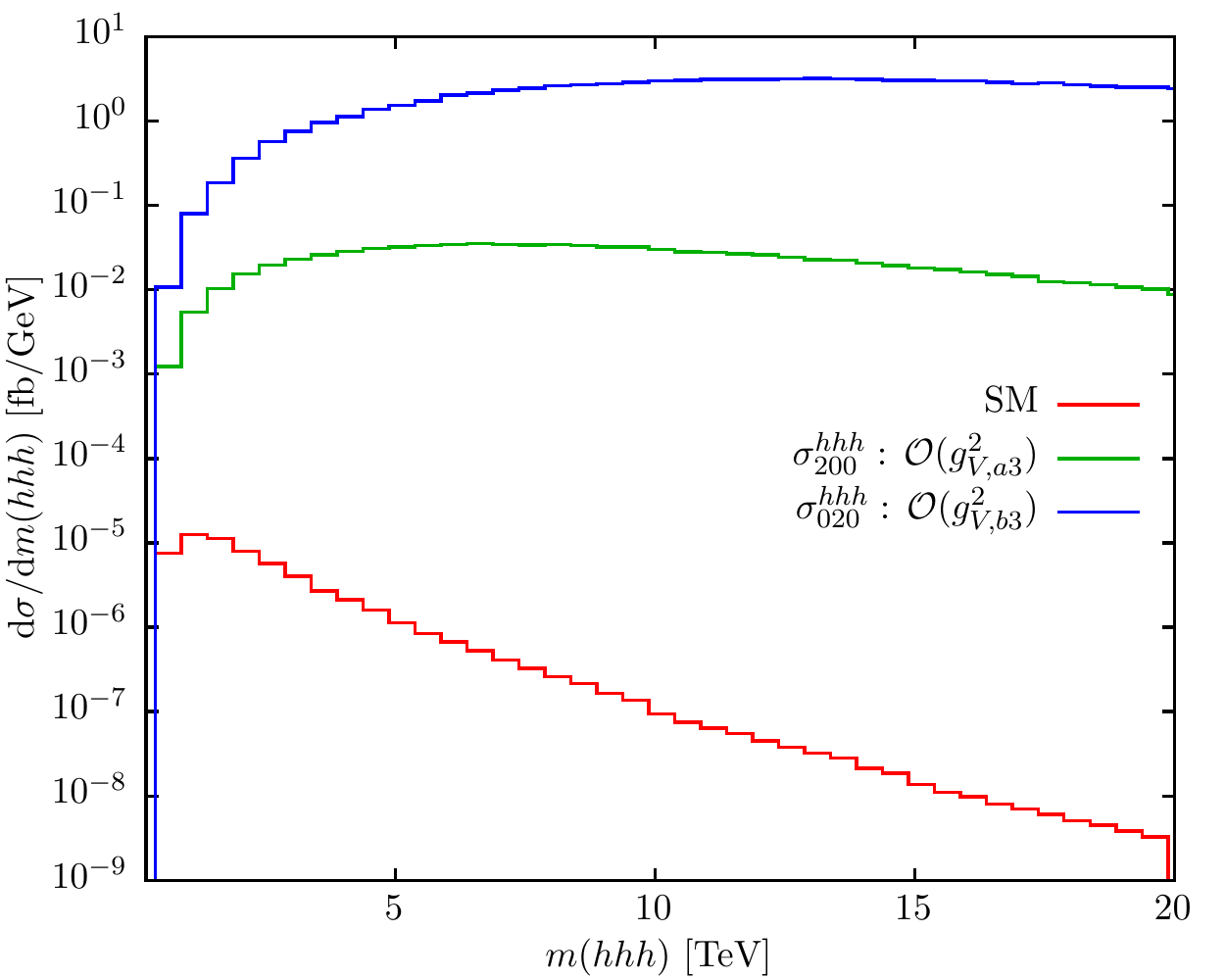}
  \caption{Triple-Higgs invariant mass distribution for the
    coefficients of the leading power corrections to VBF $hhh$
    production in $pp$ 
    collisions.  We display results for 14 TeV (left) and for 100 TeV
    (right).}\label{fig:3h-invm} 
\end{figure}

To examine the energy dependence of the $\gva3$ and $\gvb3$ contributions,
in Fig.~\ref{fig:3h-invm} we plot the appropriate $m(hhh)$
distributions.  While the SM curve displays the expected decrease with
increasing $m(hhh)$, the
$\mathcal{O}(\gva3^2)$ and $\mathcal{O}(\gvb3^2)$ coefficients exhibit
a wide plateau with a very slow falloff for high masses.

\begin{figure}[htbp]
  \centering
  \includegraphics[width=0.49\textwidth]{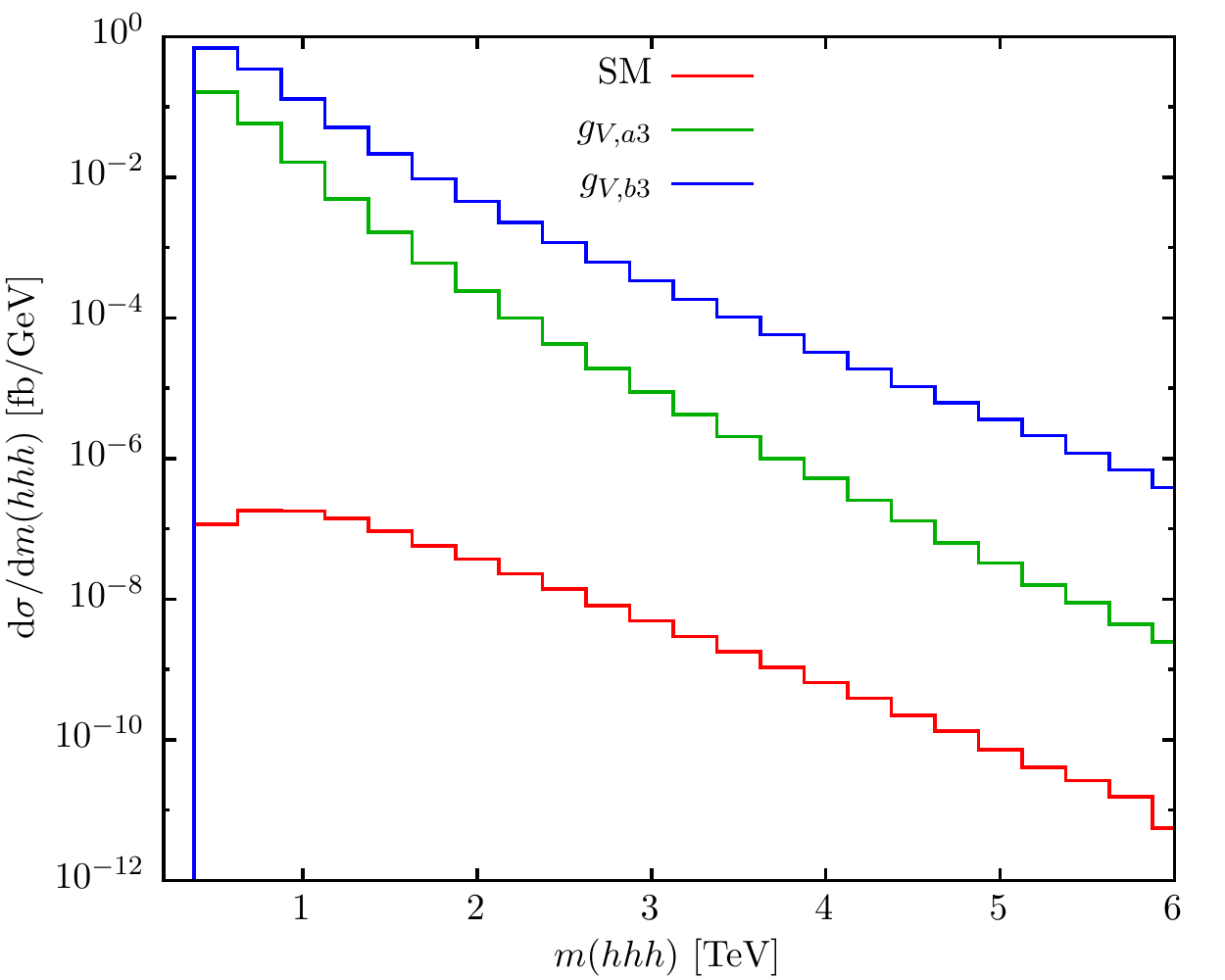}
  \includegraphics[width=0.49\textwidth]{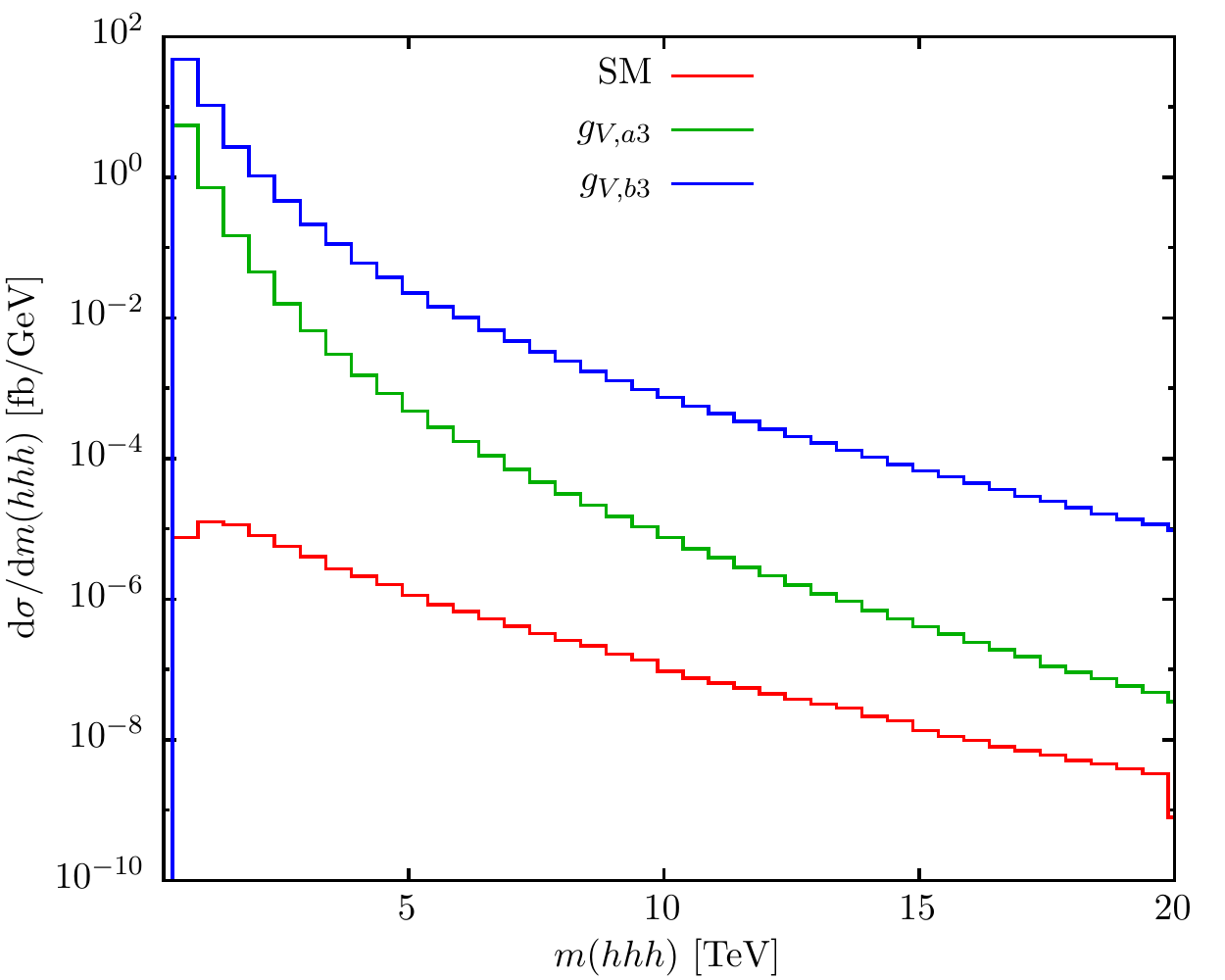}
  \caption{Maximally allowed enhancement of the differential cross
    section of VBF $hhh$ production in $pp$ collisions, as a function
    of the triple-Higgs invariant mass.  We display results for 14 TeV
    (left) and for 100 TeV (right), identifying $m(hhh)$ with the
    cutoff $Q$.\label{fig:3h-maxm}}
\end{figure}

This behavior demonstrates the necessity of considering
unitarity constraints on the EFT parameters.  Based on formulas given
in Sec.~\ref{sec:pwu-hhh} with a cutoff $Q$, $\gva3$ and $\gvb3$ are
restricted to the range
\begin{align}
  |\gva3|,|\gvb3|<\frac{64\sqrt{3} v^3 \pi^2}{Q^3}.
\end{align}
In analogy with the double-Higgs case discussed before, we use this relation
to deduce the maximally allowed enhancement of the differential cross
section as a function of $m(hhh)$, varying either $\gva3$ or $\gvb3$.
The results are presented in Fig.~\ref{fig:3h-maxm}.  As the
interference is negligible, we may sum the two sources
incoherently to obtain an absolute upper bound.  We conclude that the
potential enhancement is substantial in the transverse mode,
somewhat less so in the longitudinal mode.  This
conclusion is qualitatively similar to the double-Higgs case, cf.\ Fig.~\ref{fig:2h-maxm}. 

While we do not attempt a full study of the collider sensitivity to
the final state in this work, we may remark that the presence of one
extra Higgs boson relative to $hh$, reduces the signal
reconstruction efficiency but also helps to to suppress backgrounds.
Despite the tiny cross section, obtaining some experimental
limits on the parameters $\gva3$ and $\gvb3$ may become feasible.
This, in turn, would help to set meaningful bounds 
on the actual quartic Higgs self-coupling.

\subsection{Bounds on multi-Higgs production in the SILH and Higgs-inflation models}
After discussing unitarity bounds in the generic EFT context, we apply
the formalism to the more specific SILH (Sec.~\ref{sec:silh}) and
Higgs-inflation (Sec.~\ref{sec:inflation}) models.  We recall from
Sec.~\ref{sec:silh} that the SILH model describes a truncation of the
EFT expansion in a linear gauge representation, such that interactions
with different multiplicities become related to each other.  The
Higgs-inflation model is even more restricted, and all corrections
depend on just a single parameter.

For the SILH model we simplify the treatment by considering only
$c_W$ and $c_B$ as non-zero parameters, and furthermore impose
the relation $c_B=-c_W$ to ensure that the $\hat{S}$ parameter is zero at
tree-level, cf.\ Eq.~\eqref{eq:S-T}.  Denoting
$\hat{c}_W=c_W\frac{m_W^2}{m_{\rho}^2}$ and
$\hat{c}_Z=(1-\tan^2\theta_W)\hat{c}_W\approx0.71\hat{c}_W$, we have
\begin{align}
  \delta\gwa2 &= 3\delta\gwa1=-3\hat{c}_W,
  &
    \delta\gza2 &= 3\delta\gza1=3\hat{c}_Z.
\end{align}
Evaluating the unitarity constraints given in
Sec.~\ref{Sec:unitarity} for the $W$ fusion sub-amplitudes in this
parameterization, we obtain
\begin{align}
  b_0^{W^+W^-\to hh}(00)&=\frac{s^2}{2^9\pi^2 v^4}(\hat{c}_W-\hat{c}_W^2)^2\le\frac{1}{4}\label{eq:bcoef-2h-silh}\\
  b_0^{W^+W^-\to hhh}(00)&=\frac{s^3}{3\times 2^{10}\pi^4 v^6}(1+\hat{c}_W)^2(\hat{c}_W-\hat{c}_W^2)^2\le\frac{1}{4}\label{eq:bcoef-3h-silh}
\end{align}
If we assume that $\delta\gwa1\ll 1$, we can ignore higher-order
terms in $\delta\gwa1$ and reduce this to
\begin{align}
  b_0^{W^+W^-\to hh}(00)&\approx\frac{s^2}{2^9\pi^2 v^4}\hat{c}_W^2\le\frac{1}{4}\\
  b_0^{W^+W^-\to hhh}(00)&\approx\frac{s^3}{3\times 2^{10}\pi^4 v^6}\hat{c}_W^2\le\frac{1}{4}
\end{align}
Comparing the constraints from the double- und triple-Higgs production
processes, we find that once $\sqrt{s}\ge \sqrt{6}\pi v\approx
1.9~\textrm{TeV}$, 
the $b_{0}$ value that results from VBF triple-Higgs production is
larger than its double-Higgs counterpart, and thus provides a stronger
constraint.  In addition, considering that the differential cross
section is proportional to $b$~(see Eq. \eqref{eq:pwu-xs}), this observation
suggests that the cross section deviation from the SM prediction would
be larger for triple-Higgs than for double-Higgs production.   The
triple-Higgs production process should be considered as supplementing
relevant information, provided the difficulties of isolating the final
state can be overcome.

We expand the cross section for VBF double-Higgs production in the form
\begin{align}
  \sigma(pp\to hhjj)
  &=\sum_{j=0}^{4} \sigma_j^{hh}\hat{c}_W^j\label{eq:xs-2h-silh}
\end{align}
and list the coefficient values in Table~\ref{tab:xs-2h-silh}.  In
contrast to the generic EFT~\eqref{eft} where all distinct
interactions have distinct coefficients, the SILH parameters enter the
amplitude quadratically, and thus appear with up to fourth power in
the cross section.

\begin{table}[ht]
  \centering
  \begin{tabular}{|c|c|c|c|c|c|c}
    \hline
    [fb] & $\sigma_{SM}^{hh}=\sigma_{0}^{hh}$ & $\sigma_1^{hh}$ & $\sigma_2^{hh}$ & $\sigma_3^{hh}$ & $\sigma_4^{hh}$ \\
    \hline
    14 TeV & 1.10 & 0.90 & 10.6 & -9.3 & 16.1 \\
    \hline
    27 TeV & 4.32 & 3.0 & 56.4 & -64.5 & 78.1\\
    \hline
    100 TeV & 41.2 & 14.3 & \num{1.13e3} & \num{-1.79e3} & \num{1.29e3}\\
    \hline
  \end{tabular}
  \caption{Coefficients $\sigma_j^{hh}$(in fb) in
    Eq.~\eqref{eq:xs-2h-silh} for VBF $hh$ in SILH with $c_B=-c_W$.
    We show values for $pp$ collisions at 14 TeV, 27 TeV, and 100 TeV,
    respectively.}\label{tab:xs-2h-silh}
\end{table}

We observe that $\sigma_2^{hh}\gg\sigma_1^{hh}$, in line with the
discussion in the preceding subsections.  On the other hand,
$\sigma_2^{hh},\sigma_3^{hh},\sigma_4^{hh}$ are all of the same order;
they are accounted for in the $b_0$ term in
Eq.~\eqref{eq:bcoef-2h-silh}.  As long as $\hat{c}_W\ll 1$, the
contributions from $\sigma_3^{hh}$ and $\sigma_4^{hh}$ are much
smaller than the one from $\sigma_2^{hh}$.

Similarly, for VBF triple-Higgs production, the cross section becomes
\begin{align}
  \sigma=\sum_{j=0}^{6}\sigma^{hhh}_j \hat{c}_W^2
\end{align}
and the numerical results are shown in Table~\ref{tab:xs-3h-silh}. As
expected, from the leading contribution 
represented by Eq.~\eqref{eq:bcoef-3h-silh}, we find that
($|\sigma_{0}^{hhh}|, |\sigma_{1}^{hhh}|, |\sigma_{3}^{hhh}|,
|\sigma_{5}^{hhh}|) \ll (\sigma_2^{hhh}|, |\sigma_4^{hhh}|,
|\sigma_6^{hhh}|$). Therefore, in the case $\hat{c}_W\ll 1$,
$\sigma^{hhh}_j$ contributions with $j>2$ are negligible with respect
to the $\sigma^{hhh}_2$ contribution. 

	\begin{table}[ht]
		\centering
		\footnotesize
		\begin{tabular}{|c|c|c|c|c|c|c|c|c|}
			\hline
			[fb] & $\sigma_{SM}^{hhh}=\sigma_{0}^{hhh}$ & $\sigma_1^{hhh}$ & $\sigma_2^{hhh}$ & $\sigma_3^{hhh}$ & $\sigma_4^{hhh}$ & $\sigma_5^{hhh}$ & $\sigma_6^{hhh}$ \\
			\hline
			14 TeV & \num{2.792e-4} & \num{1.61e-3} & 1.89 & 0.198 & -3.85 & -0.218 & 2.09 \\
			\hline
			27 TeV & \num{1.66e-3} & \num{2.52e-3} & 30.1 & 1.87 & -58.3 & -1.97 & 29.6\\
			\hline
			100 TeV & \num{3.10e-2} & -0.28 & \num{6.58e3} & 93.6 & \num{-1.22e4} & -93.2 & \num{5.87e3} \\
			\hline
		\end{tabular}
		\caption{Coefficients $\sigma_j^{hhh}$(in fb) for
                  different contributions to VBF $hhh$ in SILH with
                  $c_B=-c_W$.  We show values for $pp$ collisions at 14 TeV, 27 TeV, and 100 TeV,
                  respectively.}\label{tab:xs-3h-silh}
	\end{table}

\begin{figure}[ht]
  \includegraphics[width=0.49\textwidth]{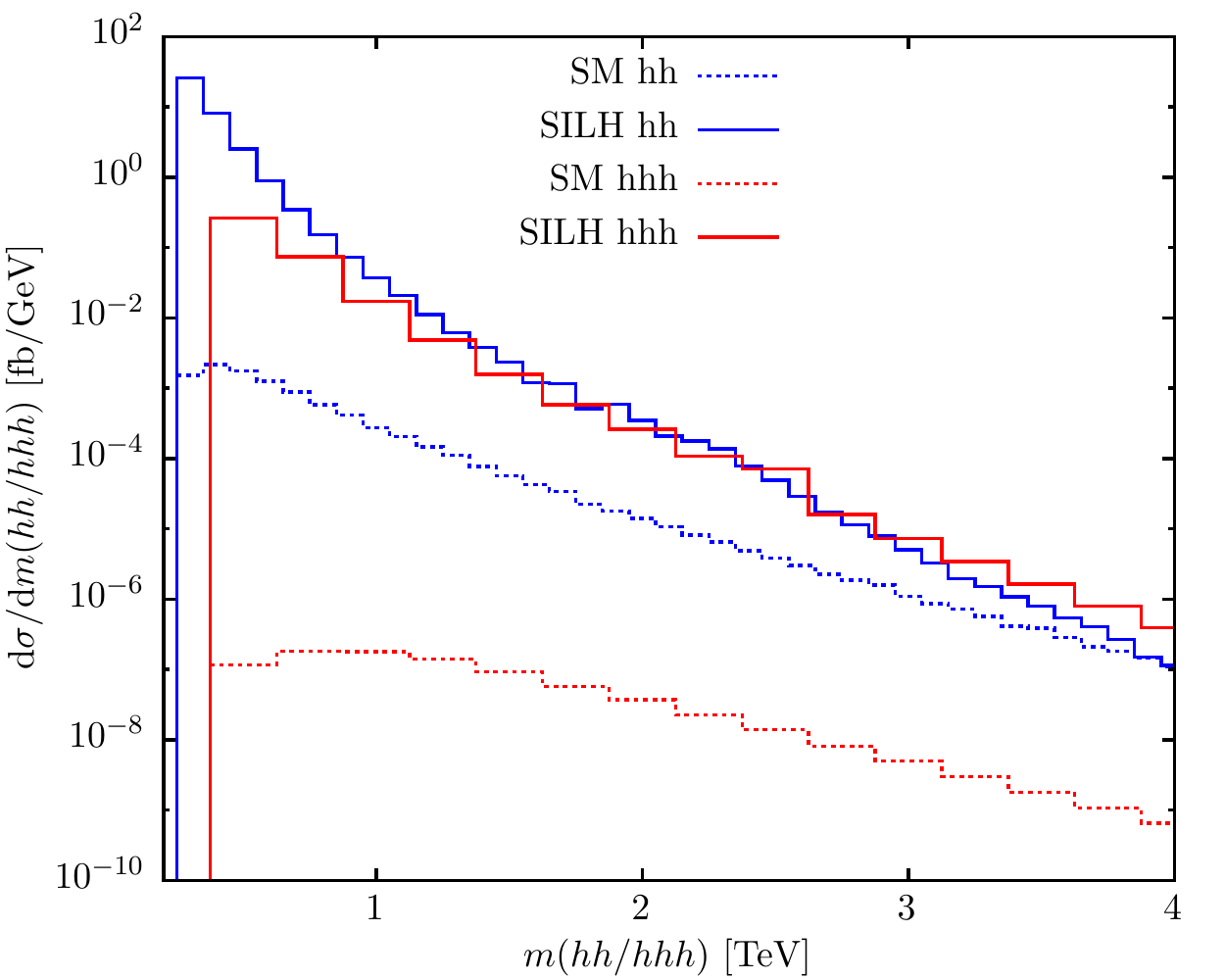}
  \includegraphics[width=0.49\textwidth]{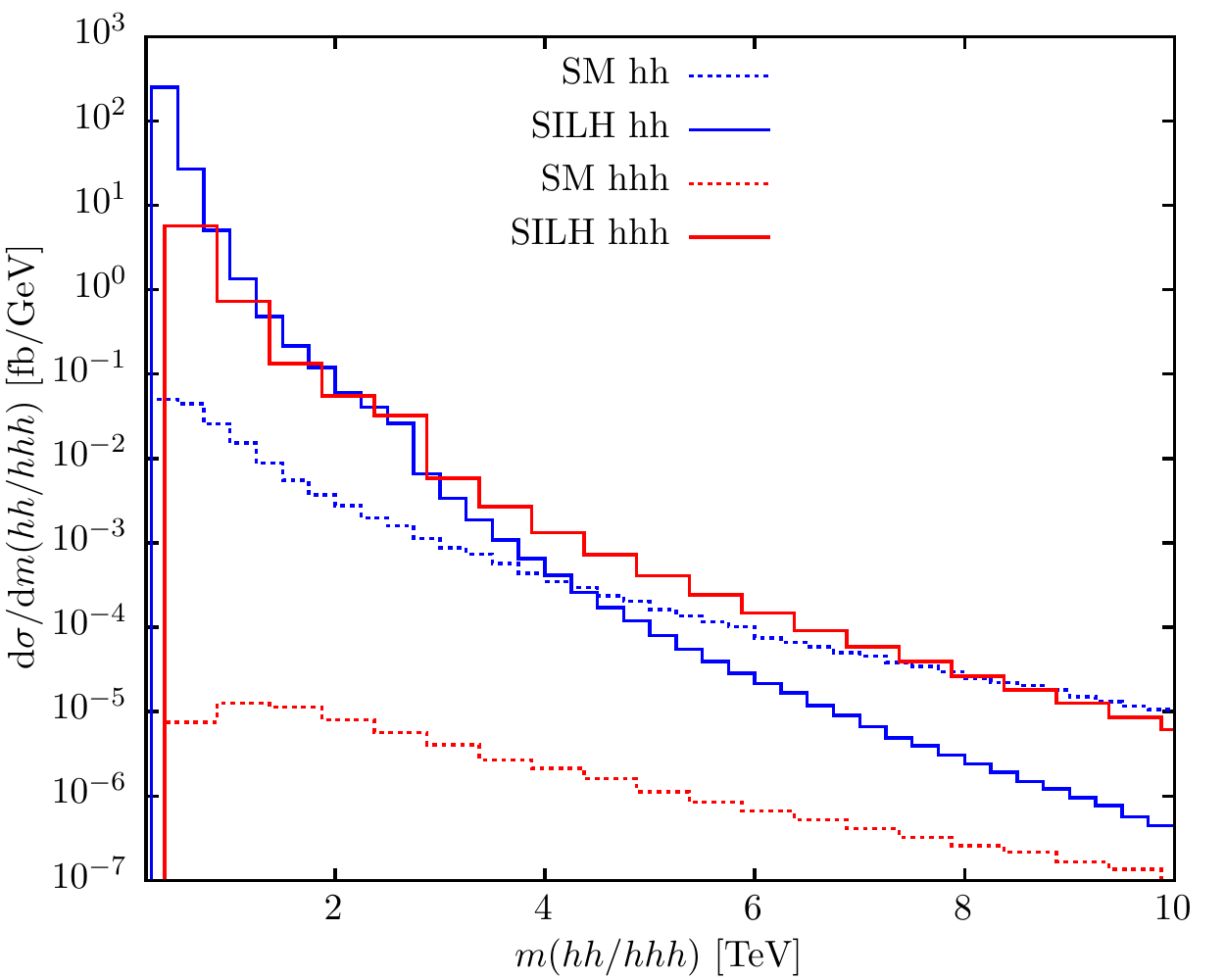}
  \caption{Maximally allowed enhancement of the differential cross
    section of VBF $hhh$ production in $pp$ collisions in the SILH
    model, as a function 
    of the triple-Higgs invariant mass.  We display results for 14 TeV
    (left) and for 100 TeV (right), identifying $m(hhh)$ with the
    cutoff $Q$.}\label{fig:silh-maxm}
\end{figure}

In Fig. \ref{fig:silh-maxm}, we present the maximally allowed
enhancement with respect to the SM differential cross section for VBF
double-Higgs and triple-Higgs production, respectively, after taking
into account all 
unitarity constraints for $W^+W^-\to hh$, $W^+W^-\to hhh$, $ZZ\to
hh$, and $ZZ\to hhh$. 
The SM cross section for VBF triple-Higgs production is indeed tiny,
2--3 orders 
smaller than the cross section for VBF double-Higgs production. 
However, if we push the SILH parameters to their unitarity limits,
the differential cross section for both processes becomes
much larger than the SM cross section.  In fact, the enhancement of
VBF triple-Higgs production can amount to 3 to 6 orders of magnitude,
becoming comparable to VBF double-Higgs production.   In the
region of large multi-Higgs invariant mass, the former rate can even surpass
VBF double-Higgs production.  This is due to the fact that the
$VV\to hhh$ cross section grows faster than $VV\to hh$, 
as indicated by the $b_0$ values given in Eq. \eqref{eq:bcoef-2h-silh}
and Eq. \eqref{eq:bcoef-3h-silh}. 

It is obvious that in such a region of strong interactions, the SILH
model as a truncated EFT will most likely be inappropriate, and
higher-order terms and non-perturbative effects will dominate the
rates.  Nevertheless, this particular example should serve to
illustrate the generic property of strongly interacting quantum field
theories, that multiple production of particles is no more suppressed
relative to low-multiplicity processes, and only inclusive observables
are under control.  This observation holds as soon as the masses of
the involved particles -- in our case, the Higgs boson -- become
negligible.  If strong interactions are a possibility, setting bounds
on high-multiplicity final states is of major physical relevance even
if the process itself cannot be detected.

Turning to the one-parameter Higgs-inflation model as described in
Sec.~\ref{sec:inflation}, we denote $\hat{x}=6\xi^2v^2/M_p^2$ and
express the unitarity constraints given in Sec.~\ref{Sec:unitarity} as
\begin{align}
  b_0^{W^+W^-\to hh}(00)&=\frac{s^2}{2^9\pi^2 v^4}(1+\hat{x})^{-4}\hat{x}^2\le\frac{1}{4}\label{eq:bcoef-2h-hinf}\\
  b_0^{W^+W^-\to hhh}(00)&=\frac{s^3}{3\times2^{10}\pi^4 v^6}(1+\hat{x})^{-7}\hat{x}^4\le\frac{1}{4}\label{eq:bcoef-3h-hinf}
\end{align}
In the limit of small $\hat{x}$, VBF triple-Higgs boson production
provides a weaker bound if $\sqrt{s}\ge \frac{16\sqrt3}{3}v\approx2.3
$TeV. 

The cross sections for VBF double-Higgs and triple-Higgs production
take the form
\begin{align}
  \sigma(hh)&=(1+\hat{x})^{-6}\sum_{j=0}^{4}\sigma_j^{hh}\hat{x}^j\label{eq:xs-2h-hinf}\\
  \sigma(hhh)&=(1+\hat{x})^{-9}\sum_{j=0}^{6}\sigma_j^{hhh}\hat{x}^j\label{eq:xs-3h-hinf}
\end{align}
We list the coefficients in Table~\ref{tab:xs-2h-hinf} and
Table~\ref{tab:xs-3h-hinf}, respectively.  We can verify that the
magnitude of the coefficients is consistent with
Eq.~\eqref{eq:bcoef-2h-hinf} and Eq.~\eqref{eq:bcoef-3h-hinf}: 
$|\sigma_2^{hh}|,|\sigma_3^{hh}|,|\sigma_4^{hh}|\gg|\sigma_{0}^{hh}|,|\sigma_1^{hh}|$,
and $|\sigma_{4}^{hhh}|,|\sigma_5^{hhh}|,|\sigma_6^{hhh}|\gg|\sigma_{0}^{hhh}|,|\sigma_{1}^{hhh}|,|\sigma_2^{hhh}|,|\sigma_3^{hhh}|$.

	\begin{table}
		\centering
		\begin{tabular}{|c|c|c|c|c|c|}
			\hline
			[fb] & $\sigma_{\textrm{SM}}^{hh}=\sigma_0^{hh}$ & $\sigma_1^{hh}$ & $\sigma_2^{hh}$ & $\sigma_3^{hh}$ & $\sigma_4^{hh}$ \\
			\hline
			14 TeV & 1.10 & 8.07 & 33.7 & 47.1 & 20.3 \\
			27 TeV & 4.32 &33.6 & 154 & 223 & 98.9 \\
			100 TeV & 41.2 & 331 & \num{2.20e3} & \num{3.57e3} & \num{1.64e3} \\
			\hline
		\end{tabular}
		\caption{Coefficients  $\sigma_j^{hhh}$ (in fb) in
                  Eq.~\eqref{eq:xs-2h-hinf}. The values correspond to
                  the Higgs-inflation model and apply to $pp$
                  collisions at 14 TeV, 27 TeV, and 100 TeV, respectively.}\label{tab:xs-2h-hinf}
	\end{table}
	\begin{table}
		\centering
		\footnotesize
		\begin{tabular}{|c|c|c|c|c|c|c|c|}
			\hline
			[fb] & $\sigma_{\textrm{SM}}^{hhh}=\sigma_0^{hhh}$ & $\sigma_1^{hhh}$ & $\sigma_2^{hhh}$ & $\sigma_3^{hhh}$ & $\sigma_4^{hhh}$ & $\sigma_5^{hhh}$ & $\sigma_6^{hhh}$ \\
			\hline
			14 TeV & \num{2.792e-4} & \num{3.72e-3} & \num{4.07e-2} & -0.132 & 2.49 & 5.51 & 2.86 \\
			27 TeV & \num{1.66e-3} & \num{2.47e-5} & 0.327 & -1.06 & 37.3 & 79.0 & 40.3\\
			100 TeV & \num{3.10e-2} & 0.55 & 11.6 & -46.0 & \num{7.77e3} & \num{1.57e4} & \num{7.89e3} \\
			\hline
		\end{tabular}
		\caption{Coefficients $\sigma_j^{hhh}$ (in fb) in
                  Eq.~\eqref{eq:xs-3h-hinf}.  The values correspond to
                  the Higgs-inflation model and apply to $pp$
                  collisions at 14 TeV, 27 TeV, and 100 TeV, respectively.}\label{tab:xs-3h-hinf}
	\end{table}

	\begin{figure}[ht]
		\includegraphics[width=0.49\textwidth]{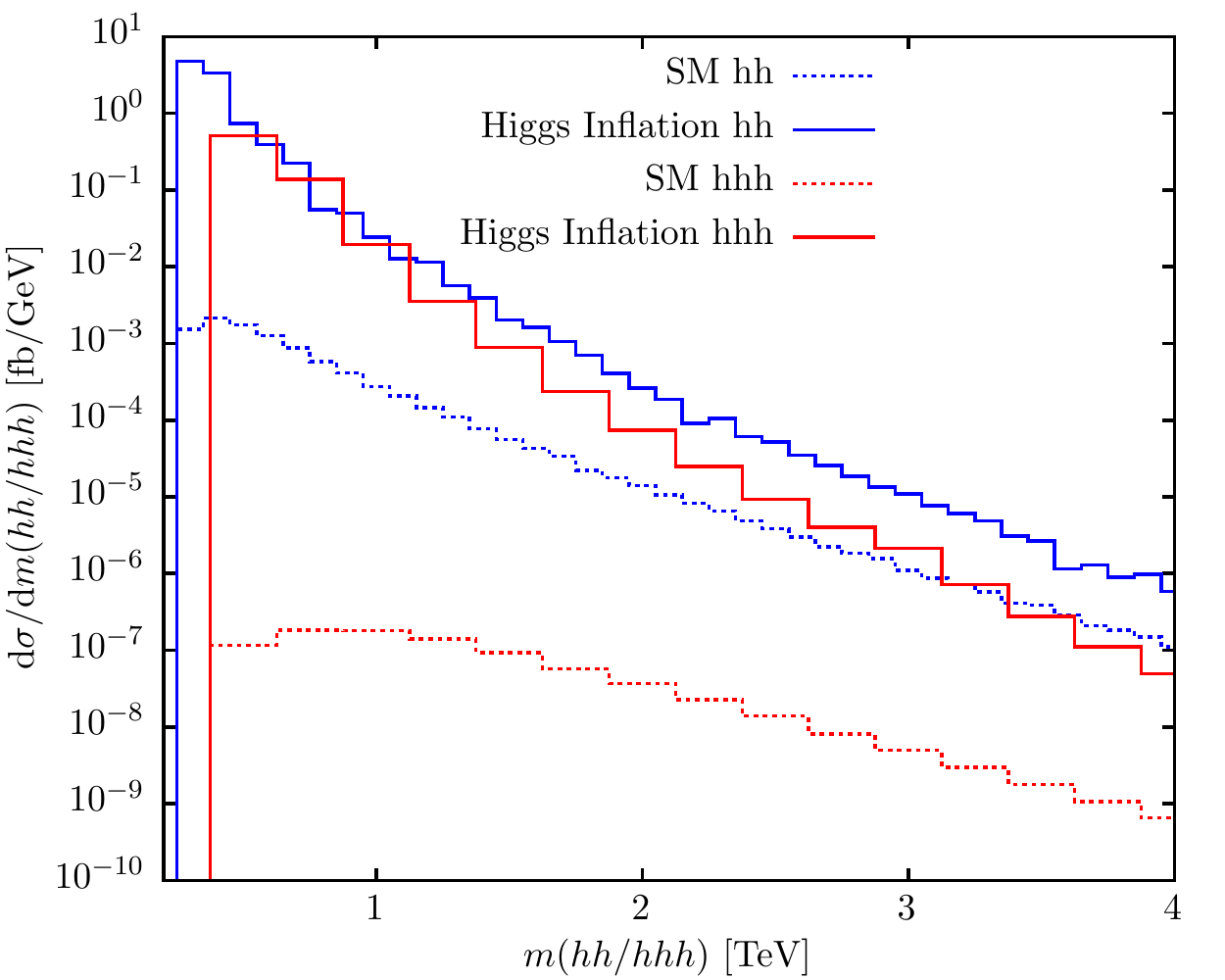}
		\includegraphics[width=0.49\textwidth]{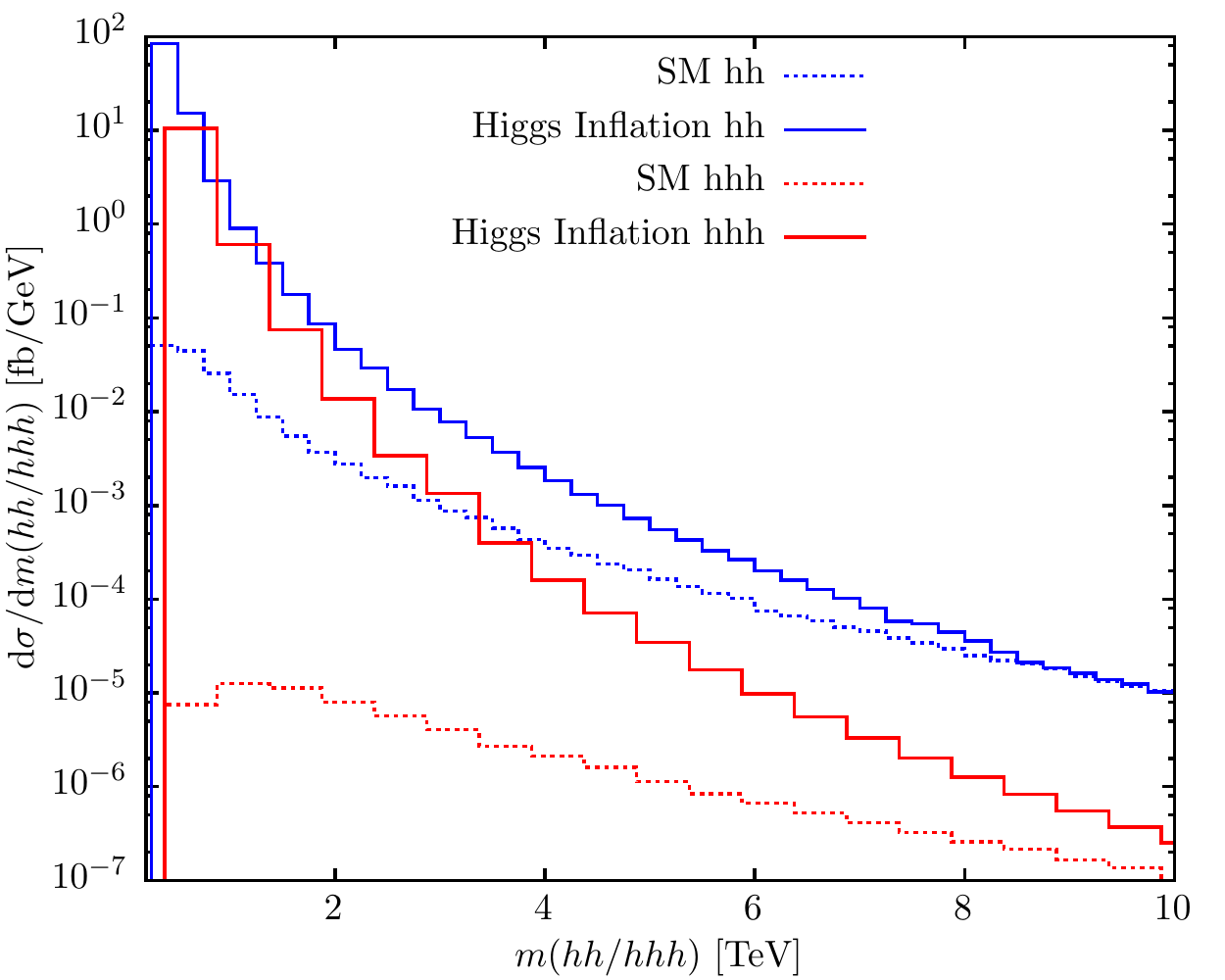}
		\caption{Maximally allowed enhancement of the differential cross
    section of VBF $hhh$ production in $pp$ collisions in the Higgs-inflation
    model, as a function 
    of the triple-Higgs invariant mass.  We display results for 14 TeV
    (left) and for 100 TeV (right), identifying $m(hhh)$ with the
    cutoff $Q$.}\label{fig:hinf-maxm}
	\end{figure}

In Fig.~\ref{fig:hinf-maxm}, we show the maximally allowed enhancement
of the differential cross section in this restricted model.
Again, the maximal deviations allowed by unitarity lift both
multi-Higgs processes by orders of magnitude, and the rates become
comparable to each other.  In contrast to the SILH case, the
triple-production process approaches double production in the
low-energy part of phase space, but falls off faster at high energy.
This can be understood from Eq. \eqref{eq:bcoef-2h-hinf} and
Eq. \eqref{eq:bcoef-3h-hinf}.

Comparing the unitarity bounds within the three models or parameterizations
that we have considered, Figs.~\ref{fig:2h-maxm}, \ref{fig:3h-maxm},
Fig.~\ref{fig:silh-maxm}, and Fig.~\ref{fig:hinf-maxm}, we note that in the
models with additional relations, there are regions where the $hh$ and $hhh$
cross sections become both large and of comparable magnitude.  Individual
bounds such as (\ref{eq:xs-2h-hinf}, \ref{eq:xs-3h-hinf}) then have to be
combined to a common bound according to~\eqref{eq:pwu-bsum}.  We should also
account for multiple $W$ and $Z$ bosons in the final state.  Since we have
introduced the specific models for illustrative purposes, not expected to be
complete or realistic in such an extreme parameter range, we did not attempt
a more complete calculation.   We expect combined bounds to be
more restrictive, reduced by an effective number of contributing final states
which depends on the details of the model in the strongly interacting regime.

\section{Summary and Discussion}\label{Sec:conc}

We have studied multi-Higgs boson production via VBF processes at the
LHC and at future hadron colliders.  While our emphasis lies on the
rare triple-Higgs production mode, we have treated double- and
triple-Higgs production processes in a common framework based on a
generic Higgs-sector effective Lagrangian.

The generic effective Lagrangian given in Eq.~(\ref{eft}) can be related to
more restricted scenarios beyond the SM.  Specifically, we have
investigated two examples, 
the SILH model -- linearly realized gauge symmetry, truncated at
dimension six --, and a Higgs-inflation model.  Various
further models with potentially strong interactions in the Higgs sector
have been proposed in the literature, such
as the minimal composite Higgs model~\cite{Agashe:2004rs}, the
composite twin-Higgs model~\cite{Geller:2014kta,Barbieri:2015lqa,Low:2015nqa},
or the composite minimal neutral naturalness model~\cite{Xu:2018ofw},
cf.\ also~\cite{Li:2019ghf}.
Such models
can be related to the generic effective Lagrangian in
a similar way.  

It is not surprising that even at a $100\;\mathrm{TeV}$ proton-proton collider,
observing the VBF multi-Higgs final state is very difficult and challenging if the SM is correct.
Beyond the SM, models such as the ones mentioned
above have revived 
the interest in new strong interactions with little impact on
observables outside the Higgs sector, but potentially striking effects
on the Higgs itself.

Anomalous interactions short of a new complete, weakly interacting
theory, spoil the delicate gauge cancellations of the SM, leading to
amplitudes which grow rapidly with energy and eventually saturate the
unitarity limits.  In this paper, we have derived unitarity limits for
inelastic two- and three-particle production in the Higgs sector, and
investigated their impact on the various possible contributions and
form factors.  It turns out that the unitarity limits are rather weak,
and allow for an enhancement of double and triple Higgs production
with rates that likely can be observed at a hadron collider, either
the high-luminosity LHC or at a future high-energy $pp$ machine.  In
principle, the triple-Higgs rate can surpass the double-Higgs rate in
part of the phase space, where the latter could also be enhanced by a large
factor.  We have computed the maximally possible enhancement in the
context of either the generic EFT, or of the specific SILH and
Higgs-inflation models.

For a more realistic study of the collider sensitivity to multi-Higgs
production, one has to find strategies that enhance the signal
in a difficult experimental environment, beyond the VBF cuts that we apply on
the parton level.  Regarding the theoretical models
with which to compare, unitarity bounds have to be incorporated in a way that
accounts for phase space in detail.  There are generic algorithms such as in
Refs.~\cite{Kilian:2014zja,Brass:2018hfw,Perez:2018kav}, where the amplitude is projected onto the
submanifold of amplitudes consistent with unitarity.  Such a formalism should
be applied to the case of inelastic and multi-particle production, to yield
more specific limits, and proper matching to global-fit data obtained within the
low-energy EFT.  We defer this program to future work.

Our results confirm that despite the tiny rates for VBF multi-Higgs
processes in the SM, substantial enhancements of multi-Higgs processes
are a real possibility that is not in
conflict with unitarity.  Such final states should
definitely be searched for in dedicated analyses.

	\begin{acknowledgments}
		We thank Kaoru Hagiwara for useful discussions.
		W.K.\ thanks the CLICdp and Theory Groups for their hospitality at CERN,
		where part of this work was completed.  W.K.\ is
                supported by the DFG Collaborative Research Center TRR
                257 ``Particle Physics Phenomenology after the Higgs Discovery''.
                Z.J. Zhao has been partially supported by a
		Nikolai Uraltsev Fellowship of the Center for Particle Physics,
		University of Siegen, and partially supported by the Natural Science
		Foundation of China under the grant No. 11875260. 
		S.C. Sun is supported by the MOST of Taiwan
		under grant number of 105-2811-M-002-130 and the CRF Grants of the
		Government of the Hong Kong SAR under HUKST4/CRF/13G.  Q.S. Yan 
		is supported by the Natural Science Foundation of China
		under the grant No.  11475180 and No. 11875260.
		X.R. Zhao has received funding from the European Union's Horizon 2020 research 
		and innovation programme as part of the 
		Marie Sk{\l}odowska-Curie Innovative Training Network MCnetITN3 (grant agreement no. 722104).
		We would like to acknowledge the Mainz Institute for Theoretical Physics(MITP) for enabling us to complete this work.

	\end{acknowledgments}

	\appendix

	\section{Relating the SILH parameterization to the Higgs EFT Lagrangian}
	\label{App:SILH}

	In our notation, the field strength tensors of the $U(1)$ and $SU(2)$ gauge
	groups are defined as
	\begin{align}
		B_{\mu\nu} &= \partial_\mu B_\nu -\partial_\nu B_{\mu}, \\
		W^{i}_{\mu\nu} &= \partial_\mu W^i_\nu -\partial_\nu W^i_{\mu} - g\epsilon^{ijk}W^j_\mu W^k_{\nu},
		\label{eq1}
	\end{align}
	respectively.  The mass eigenstates of the gauge bosons are
	\begin{align}
		B_\mu &= \cos{\theta}A_\mu - \sin{\theta}Z_\mu, \\
		W^{1}_{\mu} &= \frac{1}{\sqrt{2}}\left(W^{-}_\mu + W^{+}_\mu\right), \\
		W^{2}_{\mu} &= \frac{1}{\sqrt{2}}\left(W^{-}_\mu - W^{+}_\mu\right), \\
		W^{3}_\mu &= \cos{\theta}Z_\mu + \sin{\theta}A_\mu.
		\label{eq3}
	\end{align}
	In unitary gauge, the Higgs doublet is given by
	\begin{align}
		H &= \frac{1}{\sqrt{2}}
		\begin{pmatrix}
			0 \\
			v+h
		\end{pmatrix}
	\end{align}
	By using the equation of motion of $W_{\mu\nu}$ and
	$B_{\mu\nu}$ (cf.\ \cite{Grzadkowski:2010es}),
	\begin{align}
		\left(D^\rho W_{\rho\mu}\right)^i 
		&= 
		\frac{g}{2}\left( H^\dagger i \sigma^i\overleftrightarrow D_\mu H 
		+ \bar l \gamma_\mu \sigma^i l + \bar q \gamma_\mu \sigma^i q\right), 
		\\
		\partial^\rho B_{\rho\mu} 
		&= 
		g'YH^\dagger i \overleftrightarrow D_\mu H 
		+ g' \sum_{\psi \in \{l,e,q,u,d\}} Y_\psi\bar\psi\gamma_\mu\psi,
	\end{align}
	we obtain the following expressions for the operators with coefficients $c_W$
	and $c_B$:
	\begin{align}
		\frac{ic_Wg}{2m_\rho^2}
		\left( H^\dagger  \sigma^i \overleftrightarrow {D^\mu} H \right )
		( D^\nu  W_{\mu \nu})^i 
		&= 
		\frac{ic_W}{m_\rho^2}\frac{g^2}{4}
		\left( H^\dagger \sigma^i\overleftrightarrow D^\mu H \right)
		\left( -H^\dagger i \sigma^i\overleftrightarrow D_\mu H \right)
		+ \dots
		\notag\\
		&= \frac{c_W}{m_\rho^2}\frac{g^2}{4}
		\bigg[-\frac{g^2}{4\cos^2{\theta}}Z^\mu Z_\mu (v+h)^4 
		-\frac{g^2}{2}W^{+\mu} W^-_\mu(v+h)^4 \bigg] 
		\notag \\
		&\quad 
		+ \dots 
		\label{ow} \\
		\frac{ic_Bg'}{2m_\rho^2}
		\left( H^\dagger  \overleftrightarrow {D^\mu} H \right)
		( \partial^\nu  B_{\mu \nu}) 
		&=
		\frac{ic_B}{m_\rho^2}\frac{{g^\prime}^2}{4}
		\left( H^\dagger  \overleftrightarrow {D^\mu} H \right)
		(-H^\dagger i \overleftrightarrow D_\mu H)
		+\dots
		\notag \\
		&= \frac{c_B}{m_\rho^2}\frac{{g^\prime}^2}{4}
		\left[-\frac{g^2}{4\cos^2{\theta}}Z^\mu Z_\mu (v+h)^4\right] 
		+ \dots
		\label{ob}
	\end{align}
	To obtain expressions for the operators with coefficients $c_{HW}$ and
	$c_{HB}$, we write the relations
	\begin{align}
		2(D^\mu H)^\dagger\sigma^i(D^\nu H)W^i_{\mu\nu} 
		&= 
		H^\dagger\sigma^i\overleftrightarrow{D^\mu}H(D^\nu W^i_{\mu\nu}) 
		- H^\dagger\sigma^i(D^\mu D^\nu H)W^i_{\mu\nu} 
		\notag \\
		&\quad
		- (D^\nu D^\mu H)^\dagger \sigma^iHW^i_{\mu\nu} 
		+ \text{total derivative} 
		\notag \\
		&= 
		H^\dagger\sigma^i\overleftrightarrow{D^\mu} H(D^\nu W_{\mu\nu})^i 
		+ i\frac{g}{2}H^\dagger HW^{i\mu\nu}W^i_{\mu\nu} 
		\notag \\
		&\quad  +i\frac{g^\prime}{2} H^\dagger \sigma^iHB^{\mu\nu}W^i_{\mu\nu}
		+ \text{total derivative}
		\label{ohw}\\
		2(D^\mu H)^\dagger(D^\nu H)B_{\mu\nu} 
		&= 
		H^\dagger\overleftrightarrow{D^\mu}H(\partial^\nu B_{\mu\nu}) 
		- H^\dagger(D^\mu D^\nu H)B_{\mu\nu} 
		\notag \\
		&\quad
		-(D^\nu D^\mu H)^\dagger HB_{\mu\nu} 
		+ \text{total derivative}
		\notag \\
		&=
		H^\dagger\overleftrightarrow{D^\mu} H\partial^\nu B_{\mu\nu} 
		+ i\frac{g}{2}H^\dagger\sigma^i HW^{i\mu\nu}B_{\mu\nu}
		\notag \\
		&\quad
		+ i\frac{g^\prime}{2} H^\dagger HB^{\mu\nu}B_{\mu\nu}
		+ \text{total derivative}
		\label{ohb}
	\end{align}
	Here we used that $D^\mu D^\nu = \frac{1}{2}[D^\mu, D^\nu] +
	\frac{1}{2}\{D^\mu, D^\nu\}$ and $[D^\mu, D^\nu] =
	-ig\frac{\sigma^i}{2}W^{i\mu\nu}-ig^\prime Y B^{\mu\nu}$. 
	Also note that $\{D^\mu, D^\nu\}$ vanishes when it is being contracted with an
	anti-symmetric tensor $W_{\mu\nu}$, $B_{\mu\nu}$.
	Eq.~\ref{ohw} and Eq.~\ref{ohb} correspond to the analogous relations in
	Ref.~\cite{Contino:2013kra}.  Expanding in unitary gauge, we obtain
	\begin{align}
		\frac{ic_{HW}g}{16\pi^2f^2}( D^\mu H)^\dagger  \sigma^i ({D^\nu} H)W_{\mu \nu}^i &= 
		\frac{ic_{HW}g}{16\pi^2f^2}\bigg[\frac{1}{2}\left( H^\dagger  \sigma^i \overleftrightarrow {D^\mu} H \right )( D^\nu  W_{\mu \nu})^i  \nonumber \\
		&\quad -i\frac{g}{4}H^\dagger H W^i_{\mu\nu}W^{i\mu\nu} - i\frac{g^\prime}{4}H^\dagger \sigma^iH W^i_{\mu\nu}B^{\mu\nu} \bigg]  \nonumber \\
		&= \frac{c_{HW}g^2}{64\pi^2f^2}\bigg[-\frac{m_Z^2}{v^2}Z^\mu Z_\mu (v+h)^4 -\frac{2m_W^2}{v^2}W^{+\mu} W^-_\mu(v+h)^4 \bigg] \nonumber \\
		&\quad +\frac{c_{HW}g^2}{128\pi^2f^2}(v+h)^2W^i_{\mu\nu}W^{i\mu\nu} +\frac{c_{HW}g{g^\prime}}{64\pi^2f^2}H^\dagger \sigma^iH W^i_{\mu\nu}B^{\mu\nu} \nonumber \\
		&\quad +\dots \label{ohw2} \\
			\frac{ic_{HB}g^\prime}{16\pi^2f^2}( D^\mu H)^\dagger({D^\nu} H)B_{\mu \nu}
			&= \frac{ic_{HB}g^\prime}{16\pi^2f^2}\bigg[\frac{1}{2}\left( H^\dagger \overleftrightarrow {D^\mu} H \right )( \partial^\nu  B_{\mu \nu})  \nonumber \\
			&\quad -\frac{g^\prime}{4}H^\dagger H B_{\mu\nu}B^{\mu\nu} - i\frac{g}{4}H^\dagger \sigma^iH W^i_{\mu\nu}B^{\mu\nu} \bigg]  \nonumber \\
			&= \frac{c_{HB}{g^\prime}^2}{64\pi^2f^2}\bigg[-\frac{m_Z^2}{v^2}Z^\mu Z_\mu (v+h)^4 \bigg] \nonumber \\
			&\quad +\frac{c_{HB}{g^\prime}^2}{128\pi^2f^2}(v+h)^2B_{\mu\nu}B^{\mu\nu} +\frac{c_{HB}g{g^\prime}}{64\pi^2f^2}H^\dagger \sigma^iH W^i_{\mu\nu}B^{\mu\nu} \nonumber \\
			&\quad +\dots \label{ohb2}
		\end{align}
		We arrive at the following kinetic part of the Lagrangian which includes the
		anomalous contributions,
		\begin{align}
			\mathcal{L}_{\mathrm{kin}}
			&= \frac{1}{2}\left(1+c_H\frac{v^2}{f^2}\right)\partial^\mu h \partial_\mu h-\frac{1}{2}\left(1+c_{HW}\frac{g^2v^2}{32\pi^2f^2}\right)W^{+\mu\nu}W^{-}_{\mu\nu} \nonumber \\
			&\quad -\frac{1}{4}\left[1+\frac{g^2v^2}{32\pi^2f^2}\left(c_{HW}+c_{HB}\tan^2{\theta}-4c_\gamma\frac{{g^\prime}^2}{g^2_\rho}\sin^2{\theta}\right)\right]Z_{\mu\nu}Z^{\mu\nu} \nonumber \\
			&\quad -\frac{1}{4}\left[1-\frac{g^2v^2}{32\pi^2f^2}\left(4c_\gamma\frac{{g^\prime}^2}{g^2_\rho}\cos^2{\theta}\right)\right]A_{\mu\nu}A^{\mu\nu} \nonumber \\
			&\quad -\frac{1}{4}\left[\frac{g{g^\prime}v^2}{32\pi^2f^2}(c_{HW}-c_{HB})+\frac{gg^\prime v^2}{16\pi^2f^2}\left(4c_\gamma\frac{g^2}{g^2_\rho}\sin^2\theta\right)\right]Z_{\mu\nu}A^{\mu\nu}
			\label{kinetic}
		\end{align}
		The fields may be rescaled by
		\begin{align} 
			h &= \left(1+c_H\frac{v^2}{f^2}\right)^{-\frac{1}{2}}h'=\zeta_hh' 
			\label{rescale1}
			\\
			W^{\pm}_{\mu} &= \left(1+c_{HW}\frac{g^2v^2}{32\pi^2f^2}\right)^{-\frac{1}{2}}W'^{\pm}_{\mu} = \zeta_WW'^{\pm}_{\mu} \\
			Z_{\mu} &=\left[1+\frac{g^2v^2}{32\pi^2f^2}\left(c_{HW}+c_{HB}\tan^2{\theta}-4c_\gamma\frac{{g^\prime}^2}{g^2_\rho}\sin^2{\theta}\right)\right]^{-\frac{1}{2}}Z'_\mu = \zeta'_ZZ'_{\mu} \\
			A_{\mu} &= \left[1-\frac{g^2v^2}{32\pi^2f^2}\left(4c_\gamma\frac{{g^\prime}^2}{g^2_\rho}\cos^2{\theta}\right)\right]^{-\frac{1}{2}}A'_\mu =\zeta_AA'_\mu
		\end{align}
		to rewrite the Lagrangian in terms of normalized fields as
		\begin{align}
			\mathcal{L}_{kin} &= \frac{1}{2}\partial^\mu h' \partial_\mu h'-\frac{1}{2}W'^{+\mu\nu}W'^{-}_{\mu\nu}  -\frac{1}{4}Z'_{\mu\nu}Z'^{\mu\nu} -\frac{1}{4}A'_{\mu\nu}A'^{\mu\nu} \nonumber \\
							  &\quad -\frac{1}{4}\left[\frac{g{g^\prime}v^2}{32\pi^2f^2}(c_{HW}-c_{HB})+\frac{gg^\prime v^2}{16\pi^2f^2}\left(4c_\gamma\frac{g^2}{g^2_\rho}\sin^2\theta\right)\right]\zeta_A\zeta'_Z Z'_{\mu\nu}A'^{\mu\nu}\\
							  &= \frac{1}{2}\partial^\mu h' \partial_\mu h'-\frac{1}{2}W'^{+\mu\nu}W'^{-}_{\mu\nu}  -\frac{1}{4}Z'_{\mu\nu}Z'^{\mu\nu} -\frac{1}{4}A'_{\mu\nu}A'^{\mu\nu} -\frac{1}{4}y_{ZA}\zeta_A\zeta'_Z Z'_{\mu\nu}A'^{\mu\nu}
		\end{align}
		To eliminate the $ZA$ mixing term, we introduce a linear shift as
		follows, 
		\begin{align} 
			A''_{\mu} &= A'_\mu +y_{ZA}\zeta_A\zeta'_Z Z'_{\mu}/2 
			\\
			Z''_{\mu} &= \sqrt{1+y^2_{ZA}\zeta^2_A\zeta'^2_Z/4} Z'_{\mu} = \zeta_Z^{-1} Z _\mu
			\label{rescale2}
		\end{align}
		This leads to
		\begin{align}
			A_\mu = \zeta_A A''_\mu - \frac{y_{ZA} \zeta_A^2 \zeta'_Z}{4} Z''_\mu = \zeta_A A''_\mu - \zeta_{AZ}  Z''_\mu
		\end{align} 
		In the final result, all electroweak gauge bosons are canonically normalized, and
		we may omit the primes from the redefined fields.  The factors $\zeta_h$,
		$\zeta_W$, $\zeta_Z$,$\zeta'_{Z}$ $\zeta_A$ and $\zeta_{AZ}$ are introduced
		for convenience.

		From Eq.~\ref{ow} and Eq.~\ref{ob}, we also have the mass terms
		\begin{align}
			\mathcal{L}_{mass} &= -\lambda v^2\left(1+\frac{3}{2}c_6\frac{v^2}{f^2}\right)\zeta_h^2h^2
			+\frac{g^2v^2}{4}\left(1-c_W\frac{g^2v^2}{2m^2_\rho}-c_{HW}\frac{g^2v^2}{32\pi^2f^2}\right)\zeta_W^2W^{+\mu}W^{-}_{\mu} \nonumber \\
			&\quad +\frac{g^2v^2}{8\cos^2{\theta}}\left(1-c_W\frac{g^2v^2}{2m^2_\rho}-c_B\frac{{g^\prime}^2v^2}{2m^2_\rho}-c_{HW}\frac{g^2v^2}{32\pi^2f^2}-c_{HB}\frac{{g^\prime}^2v^2}{32\pi^2f^2}\right)\zeta^2_ZZ_{\mu}Z^{\mu} \nonumber \\
			\label{mass}
		\end{align}
		There are shifts in the $W$ mass and $Z$ mass given by
		\begin{align}
			m^2_W &= \frac{g^2v^2}{4}\left(1-c_W\frac{g^2v^2}{2m^2_\rho}-c_{HW}\frac{g^2v^2}{32\pi^2f^2}\right)\zeta_W^2 \\
			m^2_Z &= \frac{g^2v^2}{4\cos^2\theta}\left(1-c_W\frac{g^2v^2}{2m^2_\rho}-c_B\frac{{g^\prime}^2v^2}{2m^2_\rho}-c_{HW}\frac{g^2v^2}{32\pi^2f^2}-c_{HB}\frac{{g^\prime}^2v^2}{32\pi^2f^2}\right)\zeta^2_Z 
		\end{align}
		After rescaling the fields, we read off the parameter relations that are
		listed in Table ~\ref{table:parameter}

		\section{Details for the derivation of unitarity
		constraints}\label{App:pwu-further} 
		\subsection{$2\to2$ scattering}
		The application of unitarity conditions to elastic $2\to 2$ scattering is well
		known.  In this subsection, for completeness, we provide the explict
		derivation and its connection 
		to the generic formulas in Sec.~\ref{Sec:unitarity}.  The derivation is not
		restricted to elastic scattering; it applies to any combination of
		two-particle initial and final states $a$ and $b$, respectively.

		For a two-particle state vector $\ket{\alpha,\Phi_a}$, working in the center
		of mass frame, 
		it is convenient to choose the polar angle $\theta_{a}$ and azimuthal angle
		$\phi_{a}$ as phase-space parameters,
		or correspondingly, the normalized kinematics variables are
		$\vec{x}_a=(\frac{1}{2}(\cos\theta_a+1),\frac{\phi_a}{2\pi})$.  The 
		Jacobian determinant is given by 
		\begin{align}
			J_{a}=&\frac{1}{8\pi}\frac{1}{S_{\alpha}}s^{-1}
			\sqrt{[s-(m_{a 1}+m_{a 2})^2]
			[s-(m_{a 1}-m_{a 2})^2]}
		\end{align}
		where $m_{a 1},m_{a 2}$ are the masses of particles in $a$.
		$S_{\alpha}$ is the symmetry factor that accounts for identical particles in
		$a$ with quantum-number combination $\alpha$:
		if the two particles are identical then $S_{\alpha}=2$, otherwise
		$S_{\alpha}=1$. 

		Following Ref.~\cite{Jacob:1959at}, in the center of mass frame, the
		scattering matrix 
		from the two-particle state $\ket{\alpha,\Phi_a}$ to $\ket{\beta,\Phi_b}$ can
		be expressed 
		as follows:\footnote{To be consistent with the explicit choice of
			polarization vector in Eq.~\eqref{eq:pwu-epsvector}, our phase convention
		differs from Ref.~\cite{Jacob:1959at}.} 
		\begin{align}
			\begin{split}
				M^{\beta\alpha}(x_b,x_a)\equiv&J_{b}^{\frac{1}{2}}\bra{\beta,\theta_b,\phi_b}\mathcal{M}\ket{\alpha,\theta_a,\phi_a}J_{a}^{\frac{1}{2}}\\
				=&
				2
				\sum_{j} (2j+1)a_j^{\alpha\beta}
				D_{\lambda_{\alpha}\lambda_{\beta}}^{j}(\zeta_1,\zeta_2,\zeta_3)\\
				=&
				2\sum_{j,m}(2j+1)a_j^{\alpha\beta}
				D_{m\lambda_{\alpha}}^{j *}(\phi_a,\theta_a,0)
				D_{m\lambda_{\beta}}^{j}(\phi_b,\theta_b,0)
				\label{eq:pwu-ddexp}
			\end{split}
		\end{align}
		where $\theta_a$ ($\theta_b$) are the polar angles and $\phi_a$ ($\phi_b$) are
		the 
		azimuthal angles for the states $\ket{\alpha,\Phi_a}$ ($\ket{\beta,\Phi_b}$),
		respectively.  $\zeta_1,\zeta_2,\zeta_3$ denote corresponding Euler angles
		which represent the rotation from direction $(\theta_a,\phi_a)$
		to direction $(\theta_b,\phi_b)$.
		We use the standard convention for parameterizing four-momenta
		in terms of polar and azimuthal angles,
		\begin{align}
			p^{\mu}=(E, |\vec{p}|\sin\theta\cos\phi, |\vec{p}|\sin\theta\sin\phi, |\vec{p}|\cos\theta)
		\end{align}
		If the particle is a massive vector boson, we define the polarization states
		as follows:
		\begin{align}
			\ket{p,+}=&\frac{1}{\sqrt2}(0, \cos\phi\cos\theta+i\sin\phi,\sin\phi\cos\theta-i\cos\phi,-\sin\theta)\\
			\ket{p,-}=&\frac{1}{\sqrt2}(0, \cos\phi\cos\theta-i\sin\phi,\sin\phi\cos\theta+i\cos\phi,-\sin\theta)\\
			\ket{p,0}=&(\frac{|\vec{p}|}{m}, \frac{p^0}{m}\sin\theta\cos\phi,\frac{p^0}{m}\sin\theta\sin\phi,\frac{p^0}{m}\cos\theta)\label{eq:pwu-epsvector}
		\end{align}
		where $m=\sqrt{E^2-|\vec{p}|^2}$.

		This expansion suggests that we choose the Wigner D-matrix as an orthonormal
		basis for the $2$-particle phase space,
		\begin{align}
			H^{\alpha}_{jm}(\vec{x})=\sqrt{2j+1}D_{m \lambda_{\alpha}}^{j*}(\phi_a, \theta_a, 0)
		\end{align}
		As a result, in the scattering amplitude between two-particle states the
		corresponding amplitude $a$ becomes diagonal and
		depends only on one index:
		\begin{align}
			a_{jm,j^{\prime}m^{\prime}}^{\alpha\beta}=\delta_{jj^{\prime}}\delta_{mm^{\prime}}a_{j}^{\alpha\beta}\label{eq:pwu-at-relation}
		\end{align}
		where we introduce reduced $a$-coefficients $a_{j}^{\alpha\beta}$.

		Similarly, the $b$-coefficients can be reduced to a one-index version:
		\begin{align}
			b_{jm}^{\alpha\beta}
			= \sum_{j^{\prime}m^{\prime}}|a_{jm,j^{\prime}m^{\prime}}^{\alpha\beta}|^2=|a_j^{\alpha\beta}|^2
		\end{align}
		The set of unitarity conditions~\eqref{eq:pwu-constraints2} is thus reduced to
		\begin{align}
			|\mathrm{Re}\, a_j^{\alpha\alpha}|\le&\frac{1}{2}\\
			|\mathrm{Im}\, a_j^{\alpha\alpha}-\frac{1}{2}|\le&\frac{1}{2}\\
			\sum_{\beta\ne \alpha}b_j^{\alpha\beta}=
			\sum_{\beta\ne \alpha}|a_j^{\alpha\beta}|^2\le&\frac{1}{4}
		\end{align}
		These conditions are equivalent to those in
		Refs.~\cite{Baur:1987mt,Dahiya:2013uba,Corbett:2014ora}, 
		if only $2\to 2$ processes are considered.

		\subsection{$2\to n$ scattering: general idea}
		The unitarity conditions~\eqref{eq:pwu-constraints2} do not depend on the
		characteristics of the intermediate state~$c$, which may be any $n$-particle
		state.  We have made use of this fact by expressing the conditions in terms of
		$b$ coefficients, 
		\begin{align}
			b_{A}^{\alpha\gamma}\equiv&\frac{1}{4}\int\ud x_a\ud x_b \ud x_c H_A^{\alpha *}(x_a)H_A^{\alpha}(x_b)M^{\gamma\alpha *}(x_c,x_b)M^{\gamma\alpha}(x_c,x_a)\le\frac{1}{4}
		\end{align}
		which by construction are independent of the phase-space
		parameterization pertaining to $\Phi_c$.  We keep the dependency on the
		discrete quantum numbers $\gamma$ of the intermediate state~$c$.

		In analogy to the $2\to 2$ case above, we may use any orthonormal basis for the
		initial two-particle state $a$.  Choosing the same Wigner D-matrix expansion
		is most convenient, since due to angular-momentum conservation the
		$b$ coefficients only depend on one index,
		\begin{align}
			b^{\alpha\gamma}_{jm}\equiv b^{\alpha\gamma}_{jm^{\prime}}\equiv b^{\alpha\gamma}_{j}
		\end{align}
		independent of the complexity of the intermediate states $c$.

		At this point, we may discuss the connection to previous literature on the
		subject~\cite{Maltoni:2001dc,Dicus:2004rg,Yu:2013aca}.
		\begin{itemize}
			\item 
				In Refs.~\cite{Maltoni:2001dc,Dicus:2004rg}, unitarity constraints are
				formulated for the total cross section of $2\to n$ scattering
				under the assumption that the $j=0$ partial wave ($s$-wave) dominates.
				This assumption applies to some subset of the states that we consider here,
				but clearly is not justified for the generic case of polarized vector-boson
				scattering.

				In fact, with our notation, the cross section for $a\to c$ with discrete
				quantum numbers $\alpha,\gamma$ is given by:
				\begin{align}
					\sigma_{\alpha\gamma}(a\to c)
					= \frac{16\pi S_{\alpha}s}{[s-(m_{a1}+m_{a2})^2][s-(m_{a1}-m_{a2})^2]}
					\sum_{j}(2j+1)b_j^{\alpha\gamma}
					\label{eq:pwu-xs} 
				\end{align}
				where $b_j$ are the reduced $b$-coefficients after choosing the Wigner
				D-matrix as basis. 

				Assuming that the $j=0$ partial wave dominates in the high-energy limit, we
				obtain 
				\begin{align}
					\sigma_{\alpha\gamma}(a\to c)
					\approx \frac{16\pi S_{\alpha}}{s}
					b_0^{\alpha\gamma}\le \frac{4\pi S_{\alpha}}{s}
					\label{eq:pwu-xs-J0}
				\end{align}
				which is equivalent to the result of
				Ref.~\cite{Maltoni:2001dc,Dicus:2004rg}.  This inequality applies to any
				polarized cross section and could provide a stronger bound than its
				equivalent for an unpolarized cross section.

			\item 
				Ref.~\cite{Yu:2013aca} considers the more generic situation of $2\to n$
				scattering without $s$-wave dominance, but restricts the derivation to
				spin-less particles.  In that case, the Wigner D-matrix formalism collapses
				to the familiar formalism of Legendre polynomials and spherical harmonics.
				By the general relation
				\begin{align}
					P_l(\cos\theta_{ba}) = 
					\frac{4\pi}{2l+1}\sum_{m=-l}^{l}
					Y_l^{m}(\theta_{b},\phi_b)Y_l^{m*}(\theta_{a},\phi_a),
				\end{align}
				the relative polar angle $\theta_{ba}$ can be determined via
				\begin{align}
					\cos\theta_{ba} = 
					\cos\theta_b\cos\theta_a+\sin\theta_b\sin\theta_a\cos(\phi_b-\phi_a).
				\end{align}
				The Wigner D-matrix reduces to spherical harmonics as follows,
				\begin{align}
					D_{m0}^j(\phi,\theta,0)=\sqrt{\frac{4\pi}{2j+1}}Y_l^{m*}(\theta,\phi)
				\end{align}
				With these relations, it is easy to verify that our formulas reduce to the
				ones of Ref.~\cite{Yu:2013aca} in the spin-less case.
		\end{itemize}

		\subsection{Generalized $s$-wave}

		For some helicity combinations, the unitarity condition for $2\to n$
		scattering becomes independent of phase-space parameters in the high-energy
		limit.  This is the situation which was considered in
		Refs.~\cite{Maltoni:2001dc,Dicus:2004rg}.  In this subsection, we work out the
		details for our application.

		In the high-energy limit, effectively we can treat all external particles as
		massless, $p_i^2=0$.  The generalized $s$-wave condition for scattering $a\to
		c$ takes the form
		\begin{align}
			\bra{\gamma,\Phi_c}\mathcal{M}\ket{\alpha,\Phi_a}\approx \mathcal{C},
		\end{align}
		where $\mathcal{C}$ is a constant with respect to the kinematical parameters,
		for fixed total four-momentum.  In fact, in the EFT approximation, this
		situation occurs naturally for some of the terms since the leading
		contributions become polynomials of the Lorentz invariants.

		\begin{itemize}
			\item[(a)]
				For the case of inelastic scattering $\alpha\ne\gamma$, the
				$b$-coefficients with (multi-)index $A$ take the form
				\begin{align}
					b_{A}^{\alpha\gamma}
					&=
					\frac{1}{4}\int_{0}^{1}\ud\vec{x}_a\ud\vec{x_b}\,
					H^{\alpha*}_{A}(\vec{x}_a)\,H^{\alpha}_{A}(\vec{x}_b)\,
					J^{\frac{1}{2}}_{\alpha}(\vec{x}_a)\,J^{\frac{1}{2}}_{\alpha}(\vec{x}_b)\,
					\notag\\
					&\quad \times \int_0^1\ud \vec{x}_c 
					J_{\gamma}(\vec{x}_c)\,
					\bra{\gamma,\Phi_c}\mathcal{M}\ket{\alpha,\Phi_b}^{*}
					\bra{\gamma,\Phi_c}\mathcal{M}\ket{\alpha,\Phi_a}
					\notag\\
					&= \frac{1}{4} |\mathcal{C}|^2\Delta_{\gamma} |F_{A}^{\alpha}|^2
				\end{align}
				where the total phase-space volume $\Delta_{\gamma}$ is given
				by~\cite{Kleiss:1985gy,Platzer:2013esa}
				\begin{align}
					\Delta_{\gamma}\equiv
					\int_{0}^1\ud\vec{x_c}J_{\gamma}(\vec{x}_c)
					=\frac{1}{S_{\gamma}}\frac{1}{(2\pi)^{3n_{\gamma}-4}}
					\left(\frac{\pi}{2}\right)^{n_{\gamma}-1}
					\frac{s^{n_{\gamma}-2}}{(n_{\gamma}-1)!(n_{\gamma}-2)!}
				\end{align}
				and we define the function $F$ as
				\begin{align}
					F_{A}^{\alpha} 
					&=
					\int_{0}^{1}\ud \vec{x}_a H^{\alpha}_{A}(\vec{x}_a)\,
					J_{\alpha}^{\frac{1}{2}}(\vec{x}_a)
				\end{align}
				Using the Cauchy-Schwarz inequality, the orthonormality condition for
				the basis yields
				\begin{align}
					|F_{A}^{\alpha}|^2 
					&\le
					\int_{0}^{1}\ud \vec{x}_a |H_{\vec{l}_u}^{\alpha}(\vec{x}_a)|^2
					\int_{0}^{1}\ud \vec{x}_b J_{\alpha}(\vec{x}_b)
					= \Delta_{\alpha}
				\end{align}
				Therefore, we have
				\begin{align}
					b_{A}^{\alpha\gamma}
					\le \frac{1}{4}\Delta_{\alpha}\Delta_{\gamma} |\mathcal{C}|^2
					\label{eq:pwu-bineq}
				\end{align}
				The strongest bound is obtained if the
                                equals sign applies in
				Eq.~\eqref{eq:pwu-bineq}.  The inequality becomes
				\begin{align}
					\frac{1}{4}\Delta_{\alpha}\Delta_{\gamma}|\mathcal{C}|^2\le\frac{1}{4}
					\label{eq:pwu-1const-constraints}
				\end{align}
				To realize the optimal bound within a given phase-space
				parameterization, the following condition should be satisfied: 
				\begin{align}
					\frac{H^{\alpha}_{A}(\vec{x}_a)}{J^{\frac{1}{2}}_{\alpha}(\vec{x}_a)}
					= \text{constant}
				\end{align}
				The condition can be met if $H^{\alpha}_{A}(\vec{x}_a)$ and
				$J_{\alpha}(\vec{x}_a)$ are both constant.   Since a constant basis
				function is a member of commonly used orthonormal bases, the condition
				reduces to requiring a constant Jacobian determinant for the phase-space
				parameterization.  For an algorithm which achieves this,
				cf.\ Ref.~\cite{Platzer:2013esa}. 

				We observe that the bounds in Eq.~\eqref{eq:pwu-1const-constraints} are
				symmetric under the exchange $\alpha\leftrightarrow\gamma$, although the
				states $a$ and $c$ may differ in number or species of particles.  We may
				exploit this property by performing polarization sums to either the
				initial or final state, when applying the formalism to scattering
				processes.

			\item[(b)]
				In elastic scattering, i.e. $\alpha=\gamma$, the unitarity
				constraint may be expressed in terms of the $a$-coefficients rather than
				$b$-coefficients.  After an analogous derivation, we arrive at the
				following optimal constraint: 
				\begin{align}
					|\mathrm{Re}\,\frac{1}{2}\Delta_{\alpha}\mathcal{C}|\le\frac{1}{2}\\
					0\le|\mathrm{Im}\,\frac{1}{2}\Delta_{\alpha}\mathcal{C}-\frac{1}{4}|\le 1
				\end{align}
		\end{itemize}

		The above discussion can be also applied to the case
		that the independence of phase-space parameters results from summing over
		degenerate states (polarization, color, etc.).
		Explicitly, for a set of degenerate states $S$\footnote{We
			require all states in $S$ to have indentical particle numbers and symmetry
		factors.},
		\begin{align}
			\sum_{\gamma\in S}(\bra{\gamma,\Phi_c}\mathcal{M}\ket{\alpha,\Phi_a})^{*}\bra{\gamma,\Phi_c}\mathcal{M}\ket{\alpha,\Phi_b}=|\mathcal{C}_S|^2
		\end{align}
		where $|\mathcal{C}_S|^2$ is independent of the phase-space parameters
		$\vec{x}_a,\vec{x}_b,\vec{x}_c$. 
		With an optimal choice of kinematic variables and basis we obtain the bound
		\begin{align}
			\frac{1}{4}\Delta_{\alpha}\Delta_{\gamma}|\mathcal{C}_S|^2\le\frac{1}{4}
		\end{align}

		\subsection{Generic case: recursive kinematics}\label{App:pwu-rcm}
		For the concrete evaluation of unitarity bounds in the generic case where the
		phase-space parameter dependence remains nontrivial, we have to choose a
		specific phase-space parameterization.  In our calculations, we used the
		standard recursive generation of $2\to n$ phase space in terms of $2\to 2$
		scattering followed by a tree of $1\to 2$ momentum splittings.  The
		phase-space manifold ultimately is mapped to the the $3n-4$-dimensional unit
		hypercube, $\vec{x}\in[0,1]^{3n-4}$. Below, we review this construction and
		provide the detailed formulas.

		We denote the $n$-body phase-space element with total four-momentum $Q^{\mu}$ as
		$\ud\Phi_n\{Q^{\mu}\}$.
		\begin{enumerate}
			\item for $n>2$,
				this phase-space element is given by
				\begin{align}
					\ud\Phi_n\{Q^{\mu}\}
					&= \delta^{(4)}(\sum_{i=1}^n p_i^{\mu} -Q^{\mu})\,\ud\Phi_n
					\notag\\
					&=\frac{\ud^4p_n}{(2\pi)^3}\delta(p_n^2-m_n^2)\,
					\ud\Phi_{n-1}\{Q^{\mu}-p_n^{\mu}\}
				\end{align}
				Working in the c.m.\ frame of $Q^{\mu}$ where $Q^{\mu}_{\rm
				CM}=(\sqrt{Q^2},0,0,0)$, we obtain:
				\begin{align}
					\frac{\ud^4p_n}{(2\pi)^3}\delta(p_n^2-m_n^2)
					&= \frac{\ud^4p_{n,\rm CM}}{(2\pi)^3}\delta(p_{n,\rm CM}^2-m_n^2)
					\notag\\
					&= \frac{\rho^3(Q^2,m_n,\sum_{i=1}^{n-1}m_i) x_{3n-6}^2\sin \theta_n}
					{8\pi E_{n,\rm CM}}\ud x_{3n-6}\ud x_{3n-5}\ud x_{3n-4}
				\end{align}
				where the function $\rho$ is defined by
				\begin{align}
					\rho(s,m_1,m_2)=\sqrt{[s-(m_1+m_2)^2][s-(m_1-m_2)^2]}
				\end{align}
				and the four momentum $p_{n,\rm CM}$ satisfies
				\begin{align}
					p_{n,\rm CM}^{\mu}
					&= (E_{n,\rm CM}, \vec{p}_{n,\rm CM})\\
					\vec{p}_{n,\rm CM}
					&= x_{3n-6}\rho(Q^2,m_n,\sum_{i=1}^{n-1}m_i)
					(\sin\theta_n\cos\phi_n,\sin\theta_n\sin\phi_n,\cos\theta_n)\\
					E_{n,\rm CM}
					&= \sqrt{|\vec{p}_{n,\rm CM}|^2+m_n^2}\\
					\theta_n &= \pi x_{3n-5}\\
					\phi_n &= 2\pi x_{3n-4}
				\end{align}
				The corresponding four momentum in original frame can be obtained by a
				simple Lorentz boost: 
				\begin{equation}
					\begin{split}
						p_n^{\mu}
						&= \Lambda(Q_{\rm CM},Q,p_{n, \rm CM})\\
						&= p_{n,\rm CM}^{\mu}-2(Q_{\rm CM}^{\mu}+Q^{\mu})
						\frac{(Q_{\rm CM}+Q)\cdot p_{n,\rm CM}}{(Q_{\rm CM}+Q)^2}
						+2 Q^{\mu}\frac{Q_{\rm CM}\cdot p_{n,\rm CM}}{Q^2}
						\label{eq:pwu-boost}
					\end{split}
				\end{equation}

			\item 
				For $n=2$, working again in the c.m.\ frame, the formulas simplify
				accordingly:
				\begin{align}
					\ud\Phi_2(Q)
					&=
					\ud x_1 \ud x_2 \frac{\rho(Q^2,m_1,m_2)\sin\theta}{128\pi^4\sqrt{Q^2}}
				\end{align}
				with
				\begin{align}
					p_{1,\rm CM}^{\mu}
					&= (\sqrt{p_{CM}^2+m_1^2},-\vec{p}_{\rm CM})\\
					p_{2,\rm CM}^{\mu}
					&= (\sqrt{p_{CM}^2+m_2^2},\vec{p}_{\rm CM})\\
					\vec{p}_{\rm CM}
					&=\rho(Q^2,m_1,m_2)
					(\sin\theta_2\cos\phi_2,\sin\theta_2\sin\phi_2,\cos\theta_2)\\
					\theta_2 &= \pi x_1\\
					\phi_2 &= 2\pi x_2
				\end{align}
				Again, the corresponding four-momenta in the original frame can be obtained
				via the Lorentz boost given in Eq.~\eqref{eq:pwu-boost}. 

		\end{enumerate}

		\bibliographystyle{JHEP}
		\bibliography{ref}

		\end{document}